\documentclass[a4paper,fleqn]{cas-dc}

\usepackage[authoryear]{natbib}

\usepackage{algorithm}
\usepackage{algpseudocode}
\usepackage{booktabs}
\usepackage{listings}
\usepackage{graphicx}	
\usepackage{microtype}
\usepackage{epsfig}
\usepackage{caption}
\usepackage{float}
\usepackage{placeins}
\usepackage{color, colortbl}
\usepackage{stfloats}
\usepackage{enumitem}
\usepackage{xstring}
\usepackage{multirow}
\usepackage{xspace}
\usepackage{subcaption}
\usepackage{siunitx}
\usepackage{tabularx}
\usepackage{tcolorbox}
\usepackage{times}
\renewcommand{\ttdefault}{zi4}
\usepackage{url}
\usepackage{xcolor}
\usetikzlibrary{backgrounds,positioning,calc,fit,arrows.meta}
\usetikzlibrary{shapes.geometric}

\newcommand{\fref}[1]{Figure~\ref{#1}}
\newcommand{\tref}[1]{Table~\ref{#1}}
\newcommand{\eref}[1]{Equation~\ref{#1}}

\newcommand{\cref}[1]{Chapter~\ref{#1}}
\newcommand{\sref}[1]{Section~\ref{#1}}

\newcommand{\aref}[1]{Algorithm~\ref{#1}}

\definecolor{red}{rgb}{0.8,0,0}
\definecolor{purered}{rgb}{1,0,0}
\definecolor{pink}{rgb}{0.9,0,0.9}
\definecolor{darkred}{rgb}{0.6,0,0}
\definecolor{green}{rgb}{0.0,0.5,0}
\definecolor{blue}{rgb}{0,0,0.75}
\definecolor{darkblue}{rgb}{0,0,0.55}
\definecolor{lightcyan}{rgb}{0.5,0.7,0.7}
\definecolor{orange}{rgb}{0.9,0.3,0.1}
\definecolor{purple}{rgb}{0.6,0.0,0.6}
\definecolor{cyan}{rgb}{0.0,0.7,0.7}
\definecolor{darkgray}{rgb}{0.4,0.4,0.4}
\definecolor{bronze}{rgb}{0.7, 0.4, 0.18}
\definecolor{dorange}{rgb}{0.75, 0.4, 0.0}
\definecolor{darkgray}{rgb}{0.25,0.25,0.25}
\definecolor{black}{rgb}{0.0,0.0,0.0}

\definecolor{fkblue}{RGB}{53,92,125}
\definecolor{rsdpurple}{RGB}{122,92,153}
\definecolor{symgreen}{RGB}{93,138,102}
\definecolor{emorange}{RGB}{176,106,60}
\definecolor{refgray}{RGB}{77,77,77}

% ...

\newcommand{\R}[1]{{%
    \textbf{%
        \ifstrequal{#1}{1}{\textcolor{red}{R#1}}{%
        \ifstrequal{#1}{2}{\textcolor{blue}{R#1}}{%
        \ifstrequal{#1}{3}{\textcolor{magenta}{R#1}}{%
        \ifstrequal{#1}{4}{\textcolor{teal}{R#1}}{%
                           \textcolor{cyan}{R#1}%
        }}}}%
    }%
}}

\newlength{\oldintextsep}
\newlength{\oldcolumnsep}

\definecolor{goodbg}{HTML}{C6EFCE}
\definecolor{badbg}{HTML}{FFC7CE}

\newcommand{\good}[1]{\cellcolor{goodbg}#1}
\newcommand{\bad}[1]{\cellcolor{badbg}#1}

\newcommand{\xl}{\mathbf{x}_{l}} 
\newcommand{\xs}{\mathbf{x}_{s}} 
\newcommand{\xv}{\mathbf{x}_{v}} 
\newcommand{\lL}{\mathcal{L}} 
\newcommand{\lS}{\mathcal{S}} 

\newcommand{\fk}{{$f\text{--}k$ migration}}
\newcommand{\fkb}{{$f\text{--}k$}}
\newcommand{\rsd}{phasor-fields}
\newcommand{\Rsd}{Phasor-fields}

\definecolor{codegreen}{rgb}{0,0.6,0}
\definecolor{codegray}{rgb}{0.5,0.5,0.5}
\definecolor{codepurple}{rgb}{0.58,0,0.82}
\definecolor{backcolour}{rgb}{0.95,0.95,0.92}
\definecolor{grenate}{RGB}{200,0,0} 

\definecolor{codebg}{RGB}{250,250,250}
\definecolor{codeframe}{RGB}{198,198,198}
\definecolor{codekeyword}{RGB}{20,42,120}
\definecolor{codestring}{RGB}{140,38,28}
\definecolor{codecomment}{RGB}{34,120,58}
\definecolor{codelineno}{RGB}{150,150,150}

\lstdefinestyle{code_style}{
    backgroundcolor=\color{codebg},
    commentstyle=\color{codecomment},
    keywordstyle=\bfseries\color{codekeyword},
    numberstyle=\tiny\color{codelineno},
    stringstyle=\color{codestring},
    basicstyle=\ttfamily\scriptsize,
    frame=single,
    frameround=tttt,
    rulecolor=\color{codeframe},
    framesep=5pt,
    breakatwhitespace=false,
    breaklines=true,
    captionpos=b,
    keepspaces=true,
    numbers=left,
    numbersep=6pt,
    showspaces=false,
    showstringspaces=false,
    showtabs=false,
    tabsize=2,
    xleftmargin=16pt,
    framexleftmargin=11pt,
    linewidth=\linewidth,
    aboveskip=8pt,
    belowskip=4pt
}

\tcbset{
  mylisting/.style={
    colback=codebg,
    colframe=codeframe,
    boxrule=0.5pt,
    arc=3pt,
    outer arc=3pt,
    left=-2pt,
    right=4pt,
    top=4pt,
    bottom=4pt,
  }
}

\makeatletter
\newcommand{\CommentLine}[1]{%
  \Statex \hskip\ALG@thistlm {\color{codecomment}\# #1}%
}
\makeatother

\makeatletter
\newcommand\fs@nobottomruled{\def\@fs@cfont{\bfseries}\let\@fs@capt\floatc@ruled
  \def\@fs@pre{\hrule height.8pt depth0pt \kern2pt}%
  \def\@fs@post{}
  \def\@fs@mid{\kern2pt\hrule\kern2pt}%
  \let\@fs@iftopcapt\iftrue}
\makeatother
\floatstyle{nobottomruled}
\restylefloat{algorithm}

\makeatletter
\algnewcommand{\linedots}{\Statex \hskip\ALG@thistlm $\ldots$}
\makeatother

\algrenewcommand\alglinenumber[1]{{\ttfamily\scriptsize\color{codelineno}#1}}
\makeatletter
\renewcommand{\ALG@beginalgorithmic}{\setlength{\labelsep}{0.5em}}
\makeatother

\algrenewcommand\algorithmicfunction{\textbf{def}} 
\algrenewcommand\algorithmicprocedure{\textbf{def}} 

\algrenewcommand\algorithmicwhile{\textbf{while}}
\algrenewcommand\algorithmicdo{:}
\algrenewcommand\algorithmicif{\textbf{if}}
\algrenewcommand\algorithmicthen{:}
\algrenewcommand\algorithmicelse{\textbf{else:}}
\algrenewcommand\algorithmicforall{\textbf{for}}
\algrenewcommand\algorithmicreturn{\textbf{return}}

\algtext*{EndWhile}
\algtext*{EndIf}
\algtext*{EndFor}
\algtext*{EndFunction}
\algtext*{EndProcedure}

\algrenewtext{Function}[2]{\algorithmicfunction\ \textcolor{grenate}{\textbf{#1}}~(#2):}
\algrenewtext{Procedure}[2]{\algorithmicprocedure\ \textcolor{grenate}{\textbf{#1}}~(#2):}

\lstdefinelanguage{CUDA}{
  language=C++,
  morekeywords={
    __global__, __device__, __host__, __shared__, __constant__, __ldg,
    __syncthreads, __activemask, __match_any_sync, __shfl_sync, __ffs,
    __float2int_rz, __low2float, __high2float, __half22float2,
    threadIdx, blockIdx, blockDim, gridDim,
    dim3, cufftComplex, cudaMalloc, cudaMemcpy, cudaFree,
    cudaMemcpyHostToDevice, cudaMemcpyDeviceToHost,
    atomicAdd, make_cuFloatComplex, cuCmulf, make_float2, __half2
  }
}

\def\tsc#1{\csdef{#1}{\textsc{\lowercase{#1}}\xspace}}
\tsc{WGM}
\tsc{QE}
\tsc{EP}
\tsc{PMS}
\tsc{BEC}
\tsc{DE}

\begin{document}
\let\WriteBookmarks\relax
\def\floatpagepagefraction{1}
\def\textpagefraction{.001}

\shorttitle{Memory-efficient GPU pipelines for real-time non-line-of-sight reconstruction}

\shortauthors{A. López-Ruiz and D. Royo}

\title [mode = title]{Memory-efficient GPU pipelines for real-time non-line-of-sight reconstruction}                     

\author[1]{Alfonso López-Ruiz}[orcid=0000-0003-1423-9496]
\cormark[1]
\ead{alfonso.lopezr@unizar.es}

\author[1]{Diego Royo}[orcid=0000-0001-6880-322X]
\ead{droyo@unizar.es}

\affiliation[1]{organization={Universidad de Zaragoza--I3A},
    city={Zaragoza},
    country={Spain}}

\cortext[cor1]{Corresponding author}

\begin{abstract}
Non-line-of-sight (NLOS) imaging reconstructs scenes hidden around a corner from indirect light recorded by a single-photon avalanche diode (SPAD).
A single reconstruction is a large inverse problem: billions of photon timestamps must be binned, moved through memory, transformed and inverted.
As SPAD arrays raise acquisition throughput, reconstruction becomes the limiting stage.
We rebuild the GPU execution of two established wave-based algorithms, \fk{} and \rsd{}, for both streaming and offline processing.
On the phasor-fields side we assemble the ring-and-radius kernels of previous work once and offline, using the analytic Fourier transform of a ring, so the propagation kernel never exists in dense form at runtime, reducing the memory and bandwidth.
We reorganize the pipeline of both algorithms with fused kernels, warp-level photon binning, batched transforms, CUDA graph replay, and FP16 storage applied only where it reduces the actual bottleneck.
%
Our implementations are up to $42\times$ faster than the reference streaming pipeline and up to $14\times$ faster than the fastest published GPU baseline, all while using a fraction of the memory (down to 2.5\%), enabling vastly larger and finer reconstructions on the same hardware, or comparable ones within a much lower memory budget.
We report an ablation of each implementation choice and propose three
denoising strategies enabled by the resulting frame budget for next-generation NLOS video processing.
\end{abstract}



\maketitle

\section{Introduction}
\label{sec:intro}

\begin{figure*}
    \includegraphics[width=\textwidth]{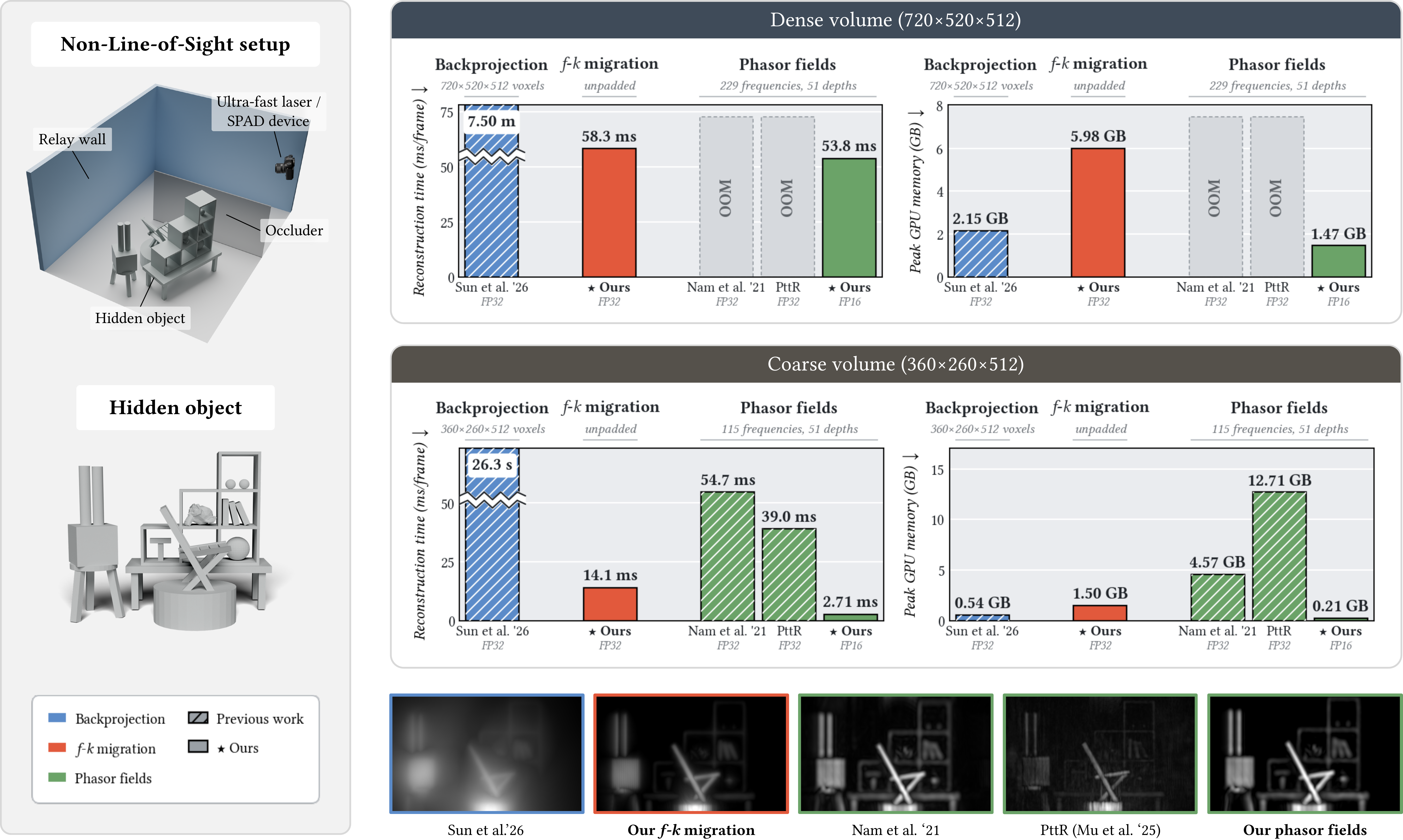}
    \caption{Performance overview on the office scene, reconstructed at two volume sizes: a dense $720\times520\times512$, and a coarser $360\times260\times512$ which therefore shows fewer details in the reconstruction. For each, the left panel reports reconstruction time per frame and the right panel peak video memory, with the methods grouped into three families: backprojection \citep{sun_cuda-accelerated_2026}, \fkb{} migration, and \rsd{} \citep{nam_low-latency_2021, mu_physics_2025}. Hatched bars are previous work and solid bars are ours. Backprojection is three to four orders of magnitude slower than the rest, so its bar runs off the top of the time panel. Against PttR, the fastest published GPU baseline, our \fkb{} is $2.7\times$ faster at the coarser volume and our \rsd{} about $14\times$, in a fraction of the memory; at the dense volume neither phasor-fields baseline fits in the $16$~\si{\giga\byte} of our GPU at all (\texttt{OOM}).}
    \label{fig:teaser}
\end{figure*}

Non-line-of-sight (NLOS) imaging reconstructs hidden objects by analyzing indirect light scattered by a visible \emph{relay wall}. 
In an active time-of-flight system, a pulsed laser illuminates the relay wall and a single-photon avalanche diode (SPAD) records, with picosecond resolution, the round-trip time of flight of each photon traveling to the hidden scene and back. Repeating this measurement across many wall positions yields a three-dimensional signal indexed by the two spatial coordinates of the wall and the photon time of flight, from which NLOS algorithms recover a volumetric reconstruction.

From a computing perspective, a single frame is therefore a large inverse problem whose cost grows with the sampling of the relay wall, the temporal resolution of the sensor, and the size of the reconstructed volume. In practice, it can comprise \emph{billions} of photon timestamps \citep{liu_non-line--sight_2019}, every one of which must be binned, moved through memory, transformed, and inverted.

With current technology, the main limitation is the low signal-to-noise ratio (SNR) inherent to three-bounce light transport: only a tiny fraction of the emitted light ever reaches the sensor, and the limited detection rate and noise level of the SPAD force setups to repeat the same measurement many times to accumulate a usable signal.
%
SPAD \emph{arrays} relieve this acquisition bottleneck by recording photon events in parallel \citep{schwartz_single-photon_2008}. Still, current technology forces a trade-off between pixel count and timing resolution. Arrays with sufficient timing resolution for NLOS reconstruction ($\sim\!\!10\text{ps}$), only reach $16\times16$ pixels \citep{conca_design_2019}. On the other hand, newer 3D-stacked architectures reach far more pixels at coarser timing, such as the $256\times128$, $60\text{ps}$ sensor of \citet{efe_characterization_2026} built for LiDAR. As 3D stacking technology advances, arrays of $128\times128$ pixels and beyond with precise timing would raise pixel-level capture parallelism by nearly two orders of magnitude over current high-timing-resolution arrays.


Our work does not address the SNR problem. We observe instead that every credible remedy for low SNR produces more data, and that reconstruction has already become a limiting factor in its own right. The throughput of NLOS imaging algorithms determines the size and resolution of the volumes that can be reconstructed at all, and the SPAD arrays now under development will only push that demand further.

Previous work has accelerated NLOS reconstruction with GPU rasterization \citep{arellano_fast_2017}, CUDA \citep{nam_low-latency_2021, sun_cuda-accelerated_2026}, field-programmable gate array (FPGA) hardware \citep{liao_fpga_2022}, and high-level GPU frameworks such as PyTorch \citep{mu_physics_2025}. We use two main baselines: the streaming \rsd{} pipeline of \cite{nam_low-latency_2021}, which can capture and reconstruct up to five frames per second, and Physics to the Rescue (PttR) \citep{mu_physics_2025}, the fastest published GPU baseline for preprocessed data, built using PyTorch.

In this work, we take two established algorithms, \fk{} \citep{lindell_wave-based_2019} and \rsd{} \citep{liu_non-line--sight_2019}, and rebuild their execution on the GPU. We combine existing algorithmic optimizations with new ones, most notably an offline construction of the ring-and-radius kernels of \cite{jiang_ring_2022} that builds them from the analytic Fourier transform of a ring and so removes dense kernel reconstruction from the runtime entirely, and three strategies to reduce background noise in continuous reconstruction. Their workload is highly parallel, yet not automatically fast on a GPU: we analyze how performance and memory scale with capture resolution, reconstruction size, and float precision. We identify the dominant bottlenecks
and accelerate the reconstructions with CUDA. As a result, on streaming reconstruction we are $42\times$ faster than that reference pipeline using $8.2\%$ of its video memory ($0.27$ against $3.34$~\si{\giga\byte}), and on full multi-plane volumes we are roughly an order of magnitude faster than PttR ($7.7\times$ and $14.0\times$) using only $2.5\%$ of its peak memory on average ($0.22$--$0.29$ against $8.9$--$12.7$~\si{\giga\byte}). The two gains compound: the memory savings lift the ceiling on what can be reconstructed at all, letting us handle reconstruction volumes far larger than prior work, or comparable ones in commodity GPUs.

We evaluate two operating modes: \emph{streaming} reconstruction, where one frame is processed while the next is captured, and \emph{offline} reconstruction, where all transient data is already available. Streaming tests whether the method can keep up with future capture hardware, while offline reconstruction tests scalability for high-resolution experiments. We will release our implementation upon acceptance to make these comparisons reproducible.

In summary, our contributions are:
\begin{itemize}
    \item High-throughput, low-memory GPU pipeline design for \fk{} and \rsd{}.
    \item Algorithmic optimizations, notably an offline construction of the ring-and-radius kernel of \cite{jiang_ring_2022} from the analytic Fourier transform of a ring, which removes the runtime rebuild.
    \item CUDA implementations and benchmarking for both \emph{streaming} and \emph{offline} settings.
    \item An analysis of performance and memory usage across reconstruction sizes, precision modes, and padding settings, and against previous accelerated NLOS imaging implementations.
    \item Three strategies for removing noise in streamed reconstructions.
\end{itemize}

\section{Related Work}
\label{sec:related}

\subsection{Time-of-flight NLOS imaging}

We focus on active NLOS imaging, where a laser source illuminates the scene. The first NLOS imaging method relied on expensive streak cameras to capture time-resolved indirect light \citep{velten_recovering_2012, velten_femto-photography_2013}. The introduction of SPADs \citep{buttafava_non-line--sight_2015, schwartz_single-photon_2008}, far more affordable than streak cameras, broadened access to NLOS imaging and incentivized the development of different reconstruction methods \citep{faccio_non-line--sight_2020}. Broadly, there are two possible capture configurations: confocal, where the illuminated and measured points on the relay wall are co-located, and non-confocal, which relaxes this restriction. In our work, we accelerate two wave-based methods: the phasor-fields formulation \citep{liu_non-line--sight_2019} and the \fk{} approach \citep{lindell_wave-based_2019}. Our implementation supports both confocal and non-confocal capture.

\paragraph{Algorithmic optimizations for NLOS imaging.} Wave-based NLOS methods let us bring tools from conventional optics into the NLOS domain. For example, some works implement convolutional versions of plane-to-plane light propagation operators \citep{liu_phasor_2020}, use approximations targeting efficiency \citep{ahn_convolutional_2019} or reducing memory usage \citep{luesia-lahoz_zone_2023}, or even use the phase information of light to achieve similar quality with fewer operations \citep{luesia-lahoz_zero-phase_2025}. Most relevant to our work, \cite{jiang_ring_2022} exploit radial symmetry in propagation operators and propose efficient ring-based representations.
%
%
Their formulation only keeps a single radial profile and rebuilds the full propagation kernel at reconstruction time, which recovers the memory but pays for it in bandwidth. We take this representation to the GPU and, additionally, remove the need to rebuild. The Fourier transform of a ring is analytic, so the radial kernel can be assembled once offline and the propagation kernel then samples it directly (\sref{sec:method:rsd_ring}).
The kernel never exists in dense form at runtime, so in our implementation the memory saving and the traffic saving apply at the same time. On a different topic, \cite{nam_low-latency_2021} reduce noise in consecutive frames of streamed NLOS reconstruction using depth-dependent averages. Other works have explored learning-based techniques \citep{ye_plug-and-play_2024}, but show limited generalization capabilities. In our work, we propose two novel denoising techniques enabled by our high-throughput capabilities by combining different reconstructions of the same frame.


\paragraph{Hardware acceleration for NLOS imaging.} \cite{arellano_fast_2017} used GPU rasterization to accelerate backprojection by tessellating the ellipsoids used for reconstruction. \cite{nam_low-latency_2021} achieved NLOS imaging at five frames per second, introducing several optimizations that accelerate capture and improve reconstruction quality, but their main speed-up comes from their CUDA implementation of a \rsd{} method.
\cite{liao_fpga_2022} reported real-time \rsd{} reconstruction on FPGA hardware, but their implementation is not publicly available. Their accelerator implements the ring-and-radius formulation of \cite{jiang_ring_2022} in hardware, including its kernel reconstruction at runtime, so the rebuild cost that our offline construction removes applies to this baseline as well. More recently, \cite{sun_cuda-accelerated_2026} implemented CUDA backprojection for irregular relay surfaces. Still, all these approaches target relatively small reconstruction volumes, and do not scale well.

Our closest point of comparison is Physics to the Rescue \citep{mu_physics_2025} (PttR), which implements \rsd{} in PyTorch and reports very high throughput, but for \emph{single-plane} reconstruction. However, reconstruction of an entire volume requires multiple planes at different depths, especially when the depth of the hidden object is unknown. Their learned feature embedding pipeline avoids this explicit depth sweep and reports real-time performance at 11.8 FPS, but it solves the problem with a learned model. We achieve a $40\%$ speed-up with respect to PttR on \emph{single-plane} reconstruction. Still, our gains are much more apparent in high-throughput \emph{multiple-plane} (volumetric) reconstruction, where we reconstruct $7.7\times$ faster on the \emph{Z} scene and $14.0\times$ faster on the office scene while using on average $2.5\%$ of its peak video memory (\tref{table:perf_related_work}). 

\paragraph{Baselines.} In summary, we compare against the baseline NLOS imaging works listed in \tref{tab:comparison_baselines}. We distinguish between methods designed for real-time \emph{streaming} and \emph{offline} references. Across all these works the reported numbers differ in reconstruction algorithm, hardware platform, and depth settings, so our evaluation separates experiments we can rerun on our hardware from values reported in the literature.

\begin{table*}[t]
    \centering
    \small
    \setlength{\tabcolsep}{4pt}
    \caption{Previous NLOS imaging methods considered in our evaluation. We distinguish between baselines rerun in our hardware and methods included as reported literature values. \emph{Partially modified} denotes code changes required to run the method on our data, mainly for loading \texttt{HDF5} capture datasets. Every method is run with its own published reconstruction operator.}
    \resizebox{\linewidth}{!}{
        \begin{tabular}{l l l l}
            \toprule
            \textbf{Compared work} & \textbf{Model} & \textbf{Comparison use} & \textbf{Main limitation for comparison} \\
            \midrule
            Fast backprojection \citep{arellano_fast_2017} & Backprojection & Rerun (partially modified) & Different reconstruction model; costly for many ellipsoids \\
            CUDA irregular-relay NLOS \citep{sun_cuda-accelerated_2026} & Backprojection & Rerun & Different reconstruction model \\
            NLOS at 5 FPS \citep{nam_low-latency_2021} & Phasor-fields & Rerun & Real-time but low-resolution setting \\
            FPGA-accelerated NLOS \citep{liao_fpga_2022} & Phasor-fields & Reported only & Specialized hardware; code unavailable \\
            Physics to the Rescue \citep{mu_physics_2025} & Phasor-fields / learned & Rerun & Highest FPS reported for single-depth reconstruction \\
            \cmidrule{1-4}
            \fk{} \citep{lindell_wave-based_2019} & \fk{} & Offline reference & Original offline implementation \\
            Phasor-fields \citep{liu_non-line--sight_2019} & Phasor-fields & Offline reference & Original offline implementation \\
            \bottomrule
        \end{tabular}
    }
    \label{tab:comparison_baselines}
\end{table*}

\subsection{GPU pipelines for streaming and memory-bound applications}

The bottlenecks we observe in previous NLOS reconstruction implementations (many short kernel launches, keeping the GPU fed, volumes larger than available memory, etc.) are not specific to NLOS imaging. They appear in any system where an instrument feeds a GPU faster than software can consume the data \citep{magro_real-time_2014}, and most of our solutions are taken from that literature. Our kernels are short, so launch overhead is a significant fraction of the frame time. We record the per-frame sequence into a CUDA graph and replay it, following \citet{ekelund_boosting_2025}, who use graphs for iterative solvers, and \citet{li_set_2026}, who add event chaining across
streams. To keep the device busy we overlap capture and compute in a
producer-consumer front end \citep{yuan_3dpipe_2026, xue_cpu-gpu_2024}. \citet{badawood_enhanced_2026} argue that pipelines of this kind should be evaluated on latency and not only on throughput, so we report frame periods rather than kernel times. Memory capacity is a more serious constraint. Volumes that do not fit in device memory are usually staged in tiles \citep{jurado_out--core_2022, jaros_out--core_2026}, or the access pattern is restructured \citep{liu_mapo_2026}. We never build the propagation kernel
densely, so there is nothing to stage. We treat precision the same way and
lower it only where it buys performance, as in mixed-precision solvers that
choose the precision of each component \citep{tsai_three-precision_2023},
batching the small operations that remain \citep{luszczek_batched_2024}.



\section{Background: non-line-of-sight imaging}
\label{sec:background}

This section formalizes the active time-of-flight NLOS imaging problem, and introduces the two wave-based NLOS imaging methods used in this work: \fk{} \citep{lindell_wave-based_2019} and \rsd{}-based reconstruction \citep{liu_non-line--sight_2019}, focusing on the parts that are relevant for our GPU implementation.

\fref{fig:histogramming} illustrates a typical NLOS imaging setup. A laser emits ultrashort pulses toward points $\xl \in \lL$ on a visible \emph{relay} wall. A small fraction of the light scatters from the wall toward the hidden scene and then returns to the relay wall after interacting with the hidden object. An ultrafast sensor records the indirect illumination at points $\xs \in \lS$, yielding a time-resolved impulse response $H(\xl, \xs, t)$. Here, $t$ denotes the time of flight from $\xl$ to $\xs$ through the hidden scene. NLOS reconstruction then operates on the impulse response $H(\xl, \xs, t)$ to estimate a hidden-scene response $f(\xv)$ at points $\xv$ in the reconstruction volume.


There are two capture modalities: \emph{confocal capture}, where the laser and sensor are directed to the same relay-wall point (thus $\xl = \xs$), and \emph{non-confocal capture}, where illumination and sensing points can differ. 
Confocal capture is slower, since laser and sensor must move together, but it leads to a more structured reconstruction problem, since every light path starts and ends at the same point.

The standard \fk{} formulation assumes confocal data, whereas \rsd{}-based reconstruction is formulated for both modalities. Still, non-confocal measurements can also be approximately converted to a confocal form through temporal shifts of $H(\xl, \xs, t)$ \citep{lindell_wave-based_2019}. Thus, one can apply both reconstruction methods under any capture modality. Next, we briefly review the two methods.

\subsection{NLOS imaging methods}
\label{sec:background:methods}

\paragraph{$f\text{--}k$ migration.} \citet{lindell_wave-based_2019} interpret the hidden scene as a wave field $\Psi(x, y, z, t)$ in which each scene point emits a spherical wave at $t = 0$. The \emph{confocal} impulse response $H(\xs, \xs, t)$ samples this wave field at relay-wall positions $\xs = (x_s, y_s, z_s)$, with the wall plane placed at $z_s = 0$, and at later times $t > 0$.

The reconstruction can then be written as a boundary-value problem for this wave field. Stolt interpolation is used in the frequency domain to migrate the measured field from the relay wall, where $z = 0$, back to the time slice $t = 0$, which corresponds to the hidden scene:
\begin{align}
    \Psi(x, y, z = 0, t) \;\;\; &\stackrel{\text{Stolt}}{\implies} \;\;\; \Psi(x, y, z, t = 0);\\
    f(\xv) &= \left| \Psi(x_v, y_v, z_v, t = 0) \right|^2
    \label{eq:fk-2}
\end{align}

\paragraph{Phasor-fields method.} \cite{liu_non-line--sight_2019} treat the relay wall as the virtual aperture of a camera with line-of-sight toward the hidden scene. They modulate an illumination signal $\mathcal{P}(\xl, t)$ on top of the impulse response, yielding $\hat{\mathcal{P}}_\omega(\xl, \xs) = \mathcal{F}\left\{ \mathcal{P}(\xl, t) \ast_t H(\xl, \xs, t)\right\}(\omega)$ which represents out-of-focus light waves at the relay wall of angular frequency $\omega$ via a convolution $\ast_t$ and Fourier transform $\mathcal{F}$. Then, NLOS reconstruction is formulated as a plane-to-plane propagation from the relay wall to the hidden scene, analogous to how a camera focuses light, to generate an image:
\begin{equation}
    \hat{\mathcal{P}}_\omega(\xv) = \int_\lS e^{i\omega t_s} \int_\lL e^{i\omega t_l} \, \hat{\mathcal{P}}_\omega(\xl, \xs) \, \mathrm{d}\xl \mathrm{d}\xs,
    \label{eq:phasor-fields_1}
\end{equation}
where $t_l = \vert \xl - \xv \vert / c$ and $t_s = \vert \xv - \xs \vert / c$ represent times of flight, with $c$ the speed of light. The resulting volumetric reconstruction $f(\xv)$ is the sum over all frequencies $f(\xv) = \int_{-\infty}^{+\infty} \hat{\mathcal{P}}_\omega(\xv) \, \mathrm{d}\omega$.

When the relay wall and reconstruction planes are parallel, we can apply two optimizations. \cite{liu_phasor_2020} formulate \eref{eq:phasor-fields_1} using a spatial convolution. When both planes are also aligned, the times of flight $t_l$ and $t_s$ have radial symmetry. This structure enables more efficient \rsd{} formulations based on ring- or radius-wise kernels \citep{jiang_ring_2022}, reducing computation and memory use while preserving the same propagation model. Since the two-dimensional Fourier transform of a radial function is itself radial, a single one-dimensional profile carries the same information as a full two-dimensional map.

\section{Proposed GPU implementation}
\label{sec:method}

In this section, we describe the design of our NLOS imaging implementations. We first discuss general processing steps (\sref{sec:method:processing}), and later present our optimizations for \fk{} (\sref{sec:method:fk}) and \rsd{} (\sref{sec:method:rsd_ring}) methods. Next, we present several techniques to combine consecutive NLOS reconstructions to reduce background noise (\sref{sec:method:noise_removal}). Most of the section is written from the \emph{streaming} point of view, where frames arrive repeatedly and launch overhead matters. Finally, we describe how the \emph{offline} path differs (\sref{sec:offline_recs}).


Unless stated otherwise, the results presented throughout this section are generated using the dataset of \cite{nam_low-latency_2021}, with dimensions $190 \times 190 \times 3002$ ($x \times y \times t$). The 3002 temporal bins are transformed into 208 frequency components and propagated over 52 depth planes.

\begin{figure*}
\centering
\includegraphics[width=.8\linewidth]{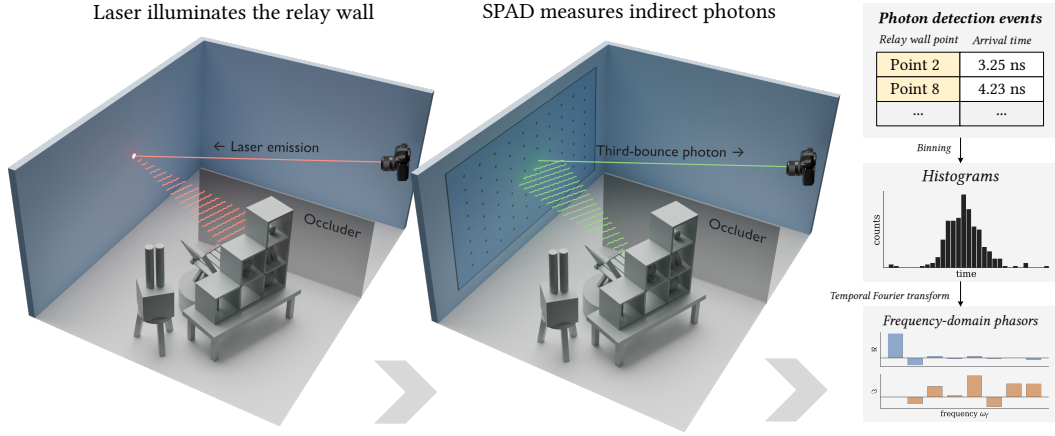}
\caption{Illustration of the capture setup and the collected data. An ultrafast laser emits a pulse onto the visible relay wall, from which light scatters toward the hidden scene. The scattered light then returns to the relay wall and is captured by the SPAD array as third-bounce photons. The recorded measurements consist of individual photon arrival events, which can be aggregated into transient histograms or directly accumulated into temporal-frequency phasors for wave-based reconstruction methods such as \rsd.}
\label{fig:histogramming}
\end{figure*}

\begin{figure}
    \centering
    \includegraphics[width=\linewidth]{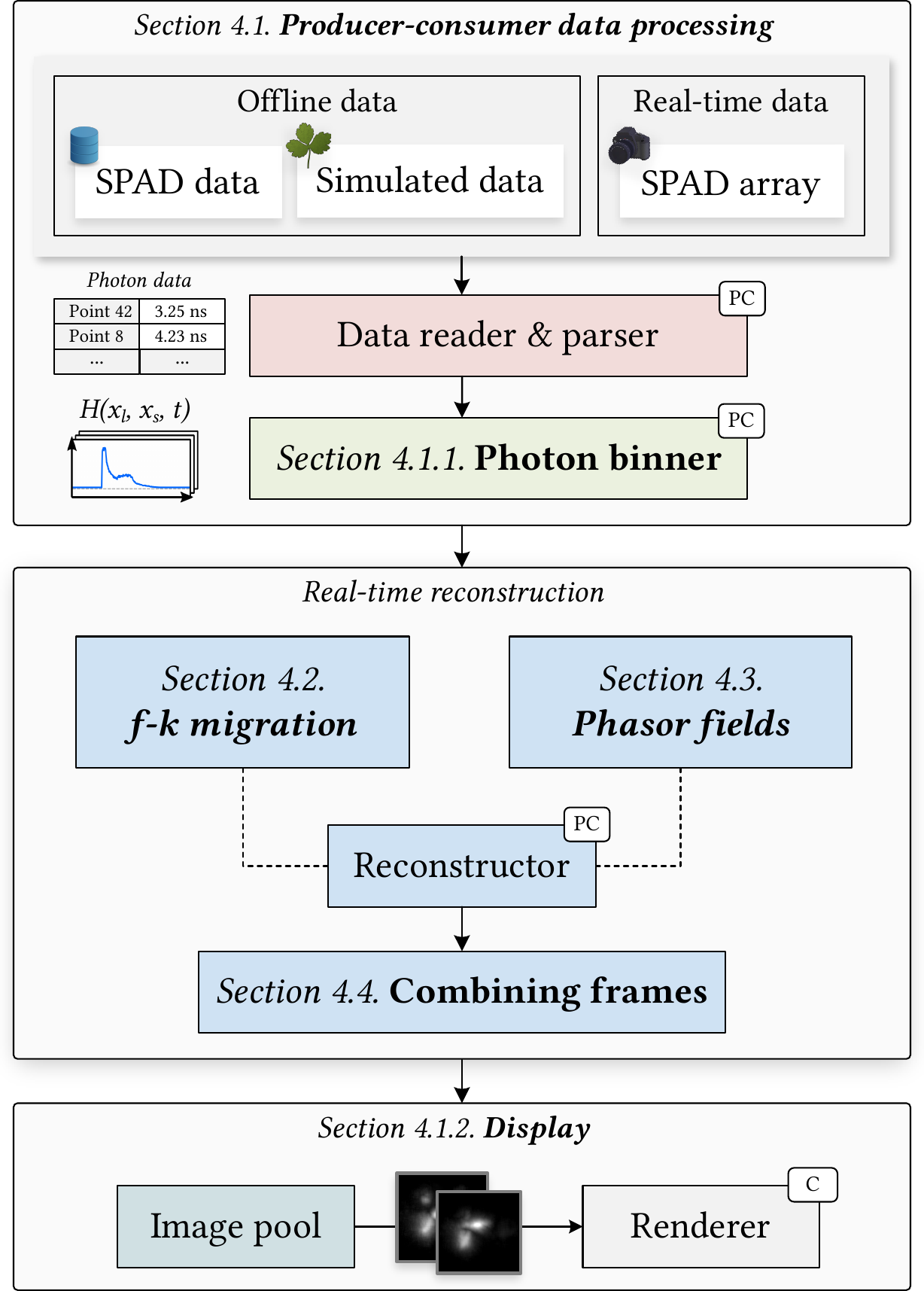}
    \caption{Overview of our producer-consumer pipeline. Labels \textbf{P} and \textbf{C} indicate whether a module is a producer, a consumer, or both.}
    \label{fig:producer_consumer}
\end{figure}

\subsection{Producer-consumer data processing}
\label{sec:method:processing}  

SPAD devices output a list of timestamps corresponding to photon arrival times, which are processed to form a transient histogram $H$ and later used for reconstruction (see \fref{fig:histogramming}). Our work accomplishes this following the producer-consumer (PC) structure from \fref{fig:producer_consumer}. The PC pattern is well suited for simultaneously solving the multiple tasks required by a SPAD asynchronously in a multi-threaded environment. Briefly, PC systems are composed of threads that (i) \textbf{P}roduce data and push them into a queue, (ii) pop data for \textbf{C}onsuming, or (iii) perform both operations (\textbf{PC}). In a PC system, multiple instances of each worker may exist; however, for simplicity in this explanation, we assume a single instance of each component. 

We adapted the work of \cite{nam_low-latency_2021} to implement the PC data processing. Our implementation supports reading data from three different sources. These include photon timestamps read from a SPAD device (\emph{live data}), or read from a binary file (\emph{photon disk data}). We can also read existing NLOS datasets \citep{galindo_dataset_2019, lindell_wave-based_2019}, where photons already come pre-binned into histograms (\emph{preprocessed disk data}).


The \emph{Data reader \& parser} module reads raw photon data and streams the complete record of photons captured by the SPAD array after iterating over every relay wall target. Next, the \emph{Photon binner} module accumulates the photons into histograms, one per relay wall target, yielding the impulse response $H$. For our \rsd{} pipeline, this stage can also perform a time-domain Fourier transform to compute the phasor field required by the method.
The \emph{Reconstructor} works either over the impulse response $H$ (\fk{}) or its corresponding frequency version (\rsd{}), and writes the result into a texture from the \emph{Image pool}. For \emph{preprocessed disk data}, there is no \emph{Data reader \& parser}. The \emph{Photon binner} instead copies the already available transient data into the layout expected by the selected reconstruction path: index-major temporal bins for \fk{}, and index-major compact phasors for our \rsd{} pipeline.

\subsubsection{Photon binner}

Processing data from the SPAD requires binning raw photon records before reconstruction. We perform this operation on the GPU, whereas \cite{nam_low-latency_2021} accumulate the photon events directly into a Fourier-domain histogram on the CPU using multi-threading. As discussed in the supplementary material, this stage becomes increasingly heavy when the photon count grows. In our default path with large photon counts, photon batches are copied into pinned host memory allocated with CUDA. This memory is locked in physical RAM and can be transferred efficiently through Direct Memory Access (DMA). 


\begin{figure}
    \centering
    \includegraphics[width=\linewidth]{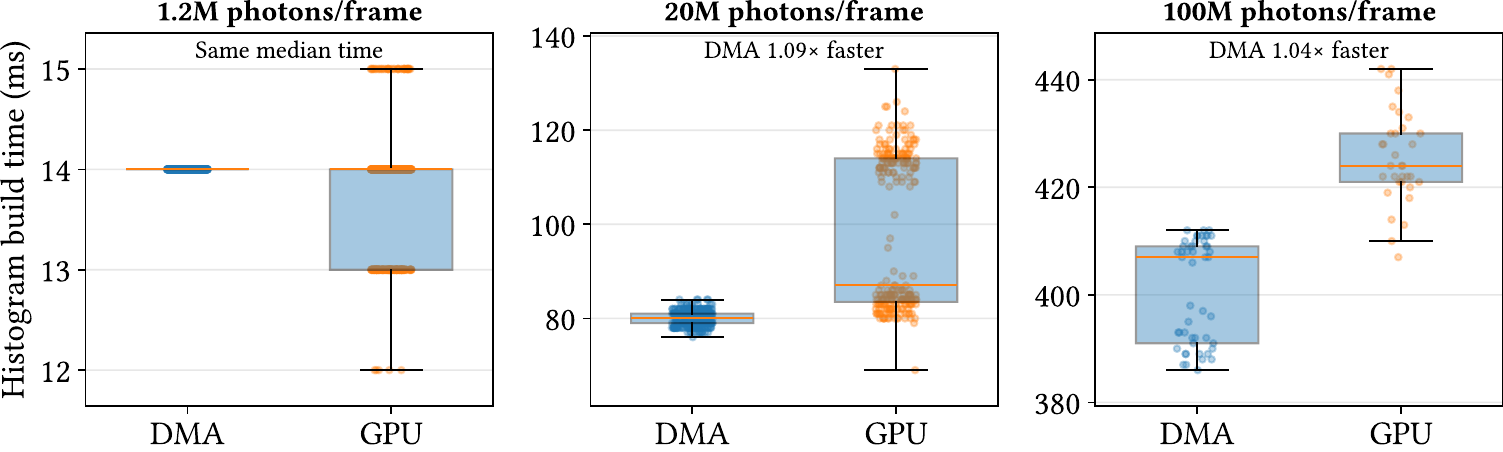}
    \caption{Histogram construction time for DMA and GPU-resident buffers using different photon counts. By default, histograms are built with 1.2M photons, matching the binary datasets of \cite{nam_low-latency_2021}; the 20M and 100M cases are generated by uniformly oversampling this baseline.}
    \label{fig:dma_vs_gpu_photon_count}
\end{figure}

\fref{fig:dma_vs_gpu_photon_count} compares the performance recorded when using DMA and GPU buffers. At low photon counts, the extra indirection during computations through host-pinned memory can outweigh the transfer benefit. At higher counts, the DMA path is more robust because moving the photon batch becomes a larger part of the frame. We therefore use it as the default for high-throughput streaming.

Another important speed-up comes from organizing phasor accumulation at the warp level. Histogramming photons into phasors still requires atomic additions to the output histogram, because many photons may contribute to the same relay-wall point and frequency. Instead of assigning a thread to every photon-frequency pair, our phasor path assigns a warp to one photon and lets the lanes cover different temporal frequencies, as illustrated in \autoref{lst:warp_phasor_binning}. This keeps the writes atomic, but reduces scheduling overhead and produces a more regular access pattern for the compact phasor layout.

\begin{lstlisting}[
  language=CUDA,
  caption={Warp-per-photon phasor binning. One warp owns one photon, and its lanes stride over temporal frequencies before atomically updating the compact phasor histogram.},
  label={lst:warp_phasor_binning}
]
uint thread = blockIdx.x * blockDim.x + threadIdx.x;
uint warpId = thread >> 5;      	// / 32
uint lane = threadIdx.x & 31;		  // % 32

// Omitted indexing and data recovery lines

for (uint freq = lane; freq < numFreqs; freq += 32)
{
  float2 sinCos = &cosSinLUT[lutIndex];

  atomicAdd(&(hist[histIndex].x), sinCos.x);
  atomicAdd(&(hist[histIndex].y), sinCos.y);
}
\end{lstlisting}

\subsubsection{Display}

The purpose of the \emph{Image pool} and \emph{Renderer} modules in \fref{fig:producer_consumer} is to enable interoperability between \emph{OpenGL} (rendering) and \emph{CUDA} (compute). The \emph{Image pool} maintains a set of buffers that are written by the reconstructor and read by the \textit{Renderer}. The pipeline operates as follows:
(i) the reconstructor waits for an available image buffer; once written,
(ii) the image is pushed into the presentation queue and removed from the writing queue.
Asynchronously,
(iii) the renderer waits for an image in the presentation queue; once displayed,
(iv) the image is returned to the writing queue for reuse.
The \emph{Renderer} writes the contents of an available image into a CUDA surface, which can be both written from CUDA kernels and sampled from OpenGL shaders.

\subsubsection{Optimization of CUDA kernels}
\label{sec:method:graphs}

In streaming mode, the same reconstruction pipeline is executed once per frame. Kernel launch overhead therefore becomes visible, especially after the main kernels have been optimized. CUDA graphs let us record a fixed sequence of operations and replay it with lower scheduling overhead. The captured work includes \texttt{cuFFT} calls, custom kernels, reductions, normalization, and optional band-pass filtering. Presentation through the CUDA/OpenGL display path remains outside this reconstruction graph. \sref{sec:results:ablations} measures how graph replay affects performance.

\subsection{\fk\ optimizations}
\label{sec:method:fk}

The \fk{} method consists of four main stages: (i) input scaling and padding to measure the wave field $\Psi(x, y, z, t)$ at $z=0,\;t>0$, (ii) a 3D Fast Fourier Transform (FFT), (iii) Stolt interpolation, and (iv) an inverse 3D FFT to finally obtain $\Psi$ at $z>0,\;t=0$ which corresponds to the reconstructed volume $f(\xv)$. The Fourier transforms dominate runtime, and the buffers they pass between stages dominate memory. That footprint grows cubically with the side length of the padded reconstruction volume. Existing implementations therefore become too expensive when they allocate separate arrays for padding, remapping coordinates, shifted spectra, and FFT (see the supplementary material). In our implementation, the input transient volume is stored with index-major layout, thus keeping temporal samples adjacent and giving \texttt{cuFFT} a regular 3D volume to process.

\subsubsection{Video memory footprint}

We reduce the overall memory footprint by solving the Stolt interpolation on the fly, avoiding auxiliary remapping buffers. Each thread handles one output voxel and evaluates its source position directly from its discrete frequency coordinates. In our implementation, the spatial coordinates map back to shifted Cartesian grid locations, so only the temporal/depth coordinate requires interpolation. The remapping kernel therefore reads two neighboring samples along the contiguous third dimension and performs a linear interpolation. \autoref{lst:fk_interp} shows the interpolation and write step of the CUDA kernel. We also evaluated a shared-memory variant that pre-caches the block data into shared memory, but we observed marginal gains, as shown later in \sref{sec:results:ablations}.

\begin{lstlisting}[
  language=CUDA,
  caption={Partial CUDA code of Stolt interpolation.},
  label={lst:fk_interp}
]
uint z0u = static_cast<uint>(z0);
uint z1u = static_cast<uint>(z1);

uint idx0 = getVolumeIdx(x0, y0, z0u, fftRes);
uint idx1 = getVolumeIdx(x0, y0, z1u, fftRes);

cufftComplex c0 = &H[idx0], c1 = &H[idx1];
cufftComplex res = complexLerp(c0, c1, dz);

uint outZ = wrapShiftedIndex(z, shift.z, fftRes.z);
uint outIdx = getVolumeIdx(x0, y0, outZ, fftRes);

float scale = fabs(fz) * safeRCP(sqrt_term);
result[outIdx] = complexMulScalar(res, scale);
\end{lstlisting}

\begin{figure}
    \centering
    \includegraphics[width=\linewidth]{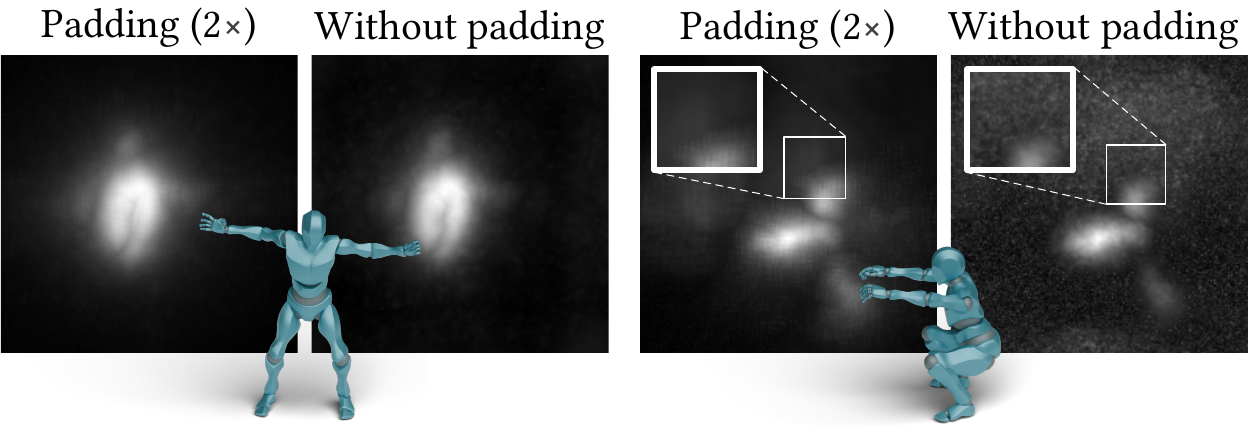}
    \caption{Two scenes reconstructed with and without padding using \fkb. The reconstruction on the left side does not show artifacts without padding, whereas the right one does; however, the hidden scene remains recognizable and the versions without padding were computed at higher speed.}
    \label{fig:fft_artifacts}
\end{figure}

Moreover, our \fkb{} algorithm uses two working FFT buffers, $\Psi^{(0)}$ and $\Psi^{(1)}$, in addition to the input transient volume $\Psi \in \mathbb{R}^{N_x \times N_y \times N_t}$ and the reconstructed image $f(\xv)$. Each processing stage alternately reads from one working buffer and writes to the other. In the padded configuration used in our experiments, the working buffers have size $2N_x \times 2N_y \times 2N_z$. This zero-padding reduces FFT wrap-around artifacts, as illustrated in \fref{fig:fft_artifacts}, but also increases memory use and runtime. We therefore include an unpadded variant as a faster alternative. This trade-off is summarized quantitatively in \fref{fig:fps_vram_padding_fp}, where \textsc{Pad $\times 1$} denotes the unpadded configuration and \textsc{Pad $\times 2$} denotes $\times 2$ zero-padding. Prior real-time approaches~\citep{nam_low-latency_2021} have likewise omitted padding, presumably for similar efficiency reasons.

\subsubsection{Data quantization}

Although \fkb{} consumes much less video memory than standard \rsd{}, its complex-valued working data can still be stored in reduced precision. In the \texttt{FP32} path, $\Psi^{(0)}$ and $\Psi^{(1)}$ are stored as 64-bit complex values. In the \texttt{FP16} path, they are stored as 32-bit complex values and transformed with \texttt{cuFFT Xt}; however, \texttt{cuFFT Xt} operates over data with power-of-two dimensions, so non-power-of-two padded volumes are rounded up before planning the FFT, which can cancel the memory saved by quantization. 

\begin{figure}
    \centering
    \includegraphics[width=\linewidth]{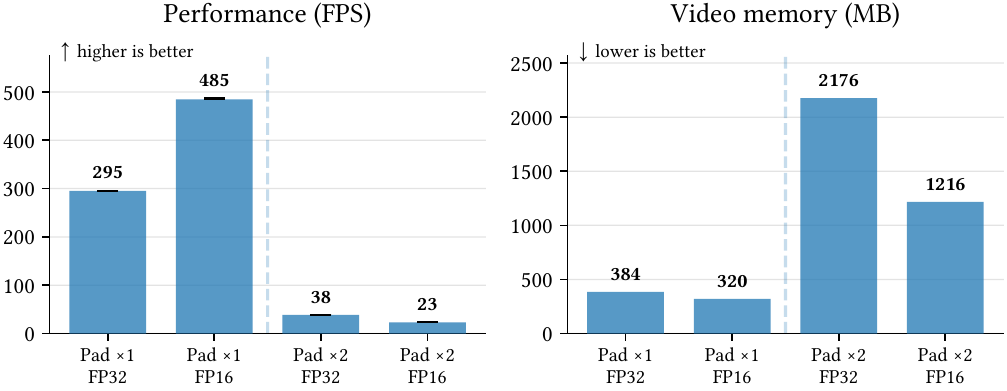}
    \caption{Performance and video memory footprint of \fkb\ with different padding and precision configurations, measured on a confocal dataset shaped as $256 \times 256 \times 256$.}
    \label{fig:fps_vram_padding_fp}
\end{figure}

\texttt{FP16} also has a limited numeric range, so the input is scaled before the transform and the spectrum is scaled before the inverse FFT to keep the reconstruction stable. As shown in \fref{fig:fps_vram_padding_fp}, \texttt{FP16} reduces video memory consumption, but its performance depends on the padding configuration and the conversion overhead. For instance, the \textsc{\texttt{FP16}-Pad ($\times 2$)} setting exhibited worse performance due to \texttt{FP32}-\texttt{FP16} conversions, and could have been even worse if the dataset was not shaped with power-of-two dimensions. Therefore, we do not treat reduced precision as a general optimization. In practice, \fkb{} benefits most from reducing temporary allocations and avoiding unnecessary remapping buffers.

\subsubsection{Algorithm details}

\begin{figure}
    \centering
    \resizebox{0.65\linewidth}{!}{%
        \begin{tikzpicture}[
    >=Latex,
    font=\small,
    node distance=0.45cm,
    oplight/.style={
        draw,
        rounded corners=2pt,
        fill=blue!8,
        minimum width=0.83\columnwidth,
        minimum height=1.2cm,
        align=center
    },
    opmed/.style={
        draw,
        rounded corners=2pt,
        fill=blue!20,
        minimum width=0.83\columnwidth,
        minimum height=1.2cm,
        align=center
    },
    opheavy/.style={
        draw,
        rounded corners=2pt,
        fill=blue!35,
        minimum width=0.83\columnwidth,
        minimum height=1.2cm,
        align=center
    },
    ann/.style={
        font=\scriptsize,
        inner sep=1.5pt,
        fill=none,
        draw=none
    },
    line/.style={->, thick}
]

\node (input) {\strut Input transient volume $\Psi$};

\node[oplight, below=of input] (pad)   {Distance falloff compensation + zero-padding};
\node[opheavy, below=of pad]   (fft)   {3D FFT};
\node[opmed, below=of fft]     (stolt) {Stolt remapping + cyclic shift};
\node[opheavy, below=of stolt] (ifft)  {3D inverse FFT};
\node[oplight, below=of ifft]  (maxz)  {Max over depth + crop padded region};

\node[below=of maxz] (output) {\strut Final image $f(\xv)$};

\draw[line] (input) -- (pad);
\draw[line] (pad) -- (fft);
\draw[line] (fft) -- (stolt);
\draw[line] (stolt) -- (ifft);
\draw[line] (ifft) -- (maxz);
\draw[line] (maxz) -- (output);

\node[ann, anchor=north west] at ($(pad.north west)+(0.06,-0.06)$)
    {$x \times y \times t$};
\node[ann, anchor=north west] at ($(fft.north west)+(0.06,-0.06)$)
    {$2x \times 2y \times 2t$};
\node[ann, anchor=north west] at ($(stolt.north west)+(0.06,-0.06)$)
    {$2x \times 2y \times 2f$};
\node[ann, anchor=north west] at ($(ifft.north west)+(0.06,-0.06)$)
    {$2x \times 2y \times f$ written};
\node[ann, anchor=north west] at ($(maxz.north west)+(0.06,-0.06)$)
    {$2x \times 2y \times 2z$};

\node[ann, anchor=south east] at ($(pad.south east)+(-0.06,0.06)$)
    {$2x \times 2y \times 2t$};
\node[ann, anchor=south east] at ($(fft.south east)+(-0.06,0.06)$)
    {$2x \times 2y \times 2f$};
\node[ann, anchor=south east] at ($(stolt.south east)+(-0.06,0.06)$)
    {$2x \times 2y \times 2f$};
\node[ann, anchor=south east] at ($(ifft.south east)+(-0.06,0.06)$)
    {$2x \times 2y \times 2z$};
\node[ann, anchor=south east] at ($(maxz.south east)+(-0.06,0.06)$)
    {$x \times y$};

\node[font=\scriptsize, anchor=north] at ($(output.south)+(0,-0.15)$) {Relative runtime cost};

\node[draw, fill=blue!8, minimum width=0.35cm, minimum height=0.25cm]
    at ($(output.south)+(-1.6,-0.7)$) {};
\node[anchor=west, font=\scriptsize]
    at ($(output.south)+(-1.35,-0.7)$) {lower};

\node[draw, fill=blue!20, minimum width=0.35cm, minimum height=0.25cm]
    at ($(output.south)+(-0.2,-0.7)$) {};
\node[anchor=west, font=\scriptsize]
    at ($(output.south)+(0.05,-0.7)$) {medium};

\node[draw, fill=blue!35, minimum width=0.35cm, minimum height=0.25cm]
    at ($(output.south)+(1.3,-0.7)$) {};
\node[anchor=west, font=\scriptsize]
    at ($(output.south)+(1.55,-0.7)$) {higher};

\end{tikzpicture}
    }
    \caption{Overview of our optimized \fk\ pipeline with $2 \times$ padding. Each stage shows the corresponding input and output dimensions. Stolt remapping only writes the active half of the padded spectrum, while the final stage selects the maximum response along depth for each spatial location. The third dimension is denoted as $t$, $f$, or $z$ depending on the domain, but all three have the same resolution.}
    \label{fig:algorithm_overview}
\end{figure}
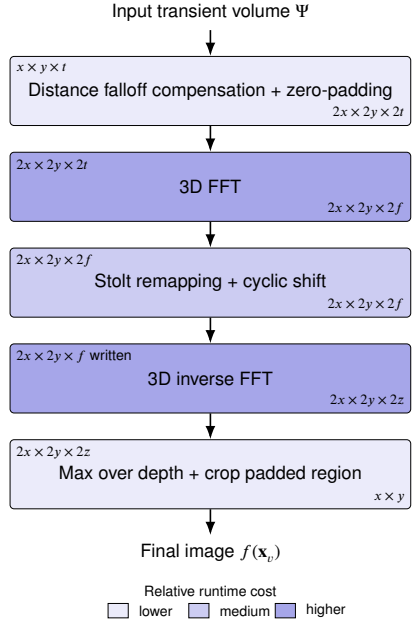

A key optimization arises from the observation that the original \fkb{} formulation keeps only one half of the interpolated spectrum (see the supplementary material). As illustrated in \fref{fig:algorithm_overview}, this reduces the active Stolt domain to half of the padded frequency volume. The destination buffer is first cleared, and the remapping kernel writes only this active half. Additional acceleration comes from fusing operations that were originally separate. Input scaling, which multiplies each temporal bin by its normalized time index, is fused with zero padding. The cyclic shift is applied directly during the remapping write (see \autoref{lst:fk_interp}). After the inverse FFT, a warp-level reduction computes the maximum response along depth for each spatial location while discarding the padded region. The resulting image is then normalized and optionally band-pass filtered.

\subsection{Phasor-fields optimizations}
\label{sec:method:rsd_ring}

The phasor-field framework of \cite{liu_non-line--sight_2019} derives several reconstruction methods, each modelled after a line-of-sight imaging system. We focus on the one modelled after a confocal camera. The name refers to the \emph{virtual} imaging system the operator emulates, not to the capture configuration: as \eref{eq:ring_decomposition} makes explicit through $(\sigma_s, \sigma_t)$, the same operator serves confocal and non-confocal captures. As explained before, preprocessing performs a time-domain Fourier transform over photon timestamps to obtain a phasor field $\hat{\mathcal{P}}_\omega(\xl, \xs) \in \mathbb{C}^{N_x \times N_y \times N_f}$. The algorithm consists of three main stages: (i) batched 2D spatial FFT, (ii) convolution with plane-to-plane propagation operators, and (iii) batched 2D inverse FFT. In this case, runtime and memory footprint are both dominated by propagation operators. For this reason, real-time \rsd{} implementations precompute a complex propagation kernel $K \in \mathbb{C}^{N_x \times N_y \times N_f \times N_d}$ and keep it for the whole session. Kernel size grows with padding, temporal frequencies and depths from plane-to-plane propagations.


Additional optimizations exploit the radial symmetry of the \rsd{} kernel in the Fourier domain. \cite{jiang_ring_2022} described this idea through ring and radius-based kernel representations. Our implementation follows the same intuition, but stores a representation more suitable for GPU runtime. We precompute a unit-ring Fourier basis analytically, then assemble a compact radial kernel for each frequency-depth pair. The supplementary material covers the radial sampling, cutoff, and interpolation used by the implementation. The stored layout is $[\rho,f,d]$, where $\rho$ is the radial coordinate in the Fourier domain, $f$ is temporal frequency, and $d$ is propagated depth. During propagation, each Cartesian Fourier pixel accesses this radial kernel through a precomputed pixel-to-radius map. \fref{fig:rsd_pipeline_comparison} compares our pipeline with the real-time \rsd{} implementations of \cite{nam_low-latency_2021} and \cite{mu_physics_2025}.

\begin{figure}
    \centering
    \includegraphics[width=.8\linewidth]{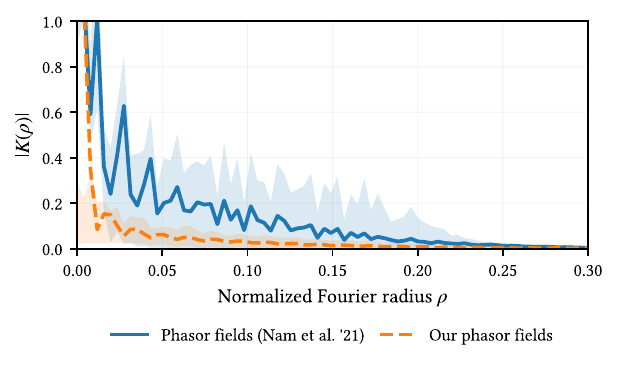}
    \caption{Mean radial magnitude of the dense \rsd{} Fourier-domain kernel and of our radial approximation, averaged over temporal frequencies and depths. The comparison is one of shape rather than of exact values.}
    \label{fig:radial_comparison_symmetry_rsd}
\end{figure}

\begin{figure}
\centering
\resizebox{\linewidth}{!}{%
    \input{figs/latex/rsd_comparison}
}
\caption{Comparison of three \rsd{} reconstructions. \cite{nam_low-latency_2021}'s \rsd{} requires many explicit GPU stages, \cite{mu_physics_2025}'s \rsd~utilizes many individual tensor operations, and our \rsd{} implementation compresses the main work into fewer stages. Darker boxes indicate heavier stages. Dimensions are annotated at the top-left and bottom-right of each stage, with $f$ temporal frequencies, $d$ propagated depths and $n$ the frames used by multi-frame enhancement.}
\label{fig:rsd_pipeline_comparison}
\end{figure}


Dense 2D kernel slices are therefore never constructed at runtime. Constructing those slices would either require allocating a dense buffer again or would reduce concurrency by constructing only one slice at a time. \cite{jiang_ring_2022}'s algorithm rebuilds the dense Cartesian kernel from the compact basis before every use, through an index map that assigns each Cartesian pixel its nearest stored sample. They report this mapping as the heaviest function in their implementation. Instead, we never rebuild it by storing the final radial kernels directly and only perform radius lookup, interpolation, low-pass weighting as well as multiply and accumulate operations. This makes our \rsd{} implementation especially useful in memory-constrained real-time settings. The approximation does not exactly reproduce the dense \rsd{} kernel (see \fref{fig:radial_comparison_symmetry_rsd}). The continuous operator between parallel planes is exactly radial, as \eref{eq:ring_decomposition} later makes explicit, but the dense kernel discretizes it on the finite rectangular grid of the relay wall, whose window and Cartesian sampling break that symmetry; the discretized kernel is therefore only approximately radial, and a radial representation cannot reproduce it pointwise by construction. What the radial form discards is mostly angular structure in the low-energy tail, which is why the reconstructions of \fref{fig:humanoid_dynamic_quality} remain comparable.

\subsubsection{Video memory footprint}

The dense kernel $K$ described at the start of this section dominates the memory budget of the original \rsd{}. Our compact representation consists of: (i) the Fourier-domain phasor data, (ii) the compact radial kernel, (iii) the pixel-to-radius map, and (iv) the working buffers needed for accumulation and inverse FFT. This reduces memory usage substantially and removes dense kernel reconstruction from runtime.

\subsubsection{Data quantization}

The most direct way to further reduce video memory is to quantize the stored propagation data, which here means the compact radial kernel rather than a dense Cartesian buffer. Radial kernels are accumulated in \texttt{FP32} during precomputation and only downcast when they are stored, which keeps precomputation stable. The FFT planning and the main reconstruction stages are left unchanged.

\subsubsection{Our improved ring-and-radius construction}

\cite{jiang_ring_2022} keep a single radial profile and rebuild the dense kernel at reconstruction time. We remove that rebuild: since the Fourier transform of a ring is analytic (\eref{eq:ring_basis}), the radial kernel can be assembled once offline and sampled directly during propagation.

After padding and forward FFTs, the input phasors are represented as $\hat{\mathcal{P}}_\omega(u, v, f) \in \mathbb{C}^{N_x \times N_y \times N_f}$, where $(u,v)$ index the relay-wall spatial frequencies produced by the batched 2D FFT. This is the phasor field of \sref{sec:background:methods}, sampled after the spatial transform rather than at the wall. Let $k(\cdot,d,f)$ denote the \rsd{} propagation kernel that takes the relay wall to a plane at depth $d$ for temporal frequency $f$, written in the spatial domain; the dense kernel $K$ introduced above is its two-dimensional Fourier transform, $K = \mathcal{F}_{2D}\{k\}$. For parallel relay-wall and reconstruction planes it depends only on the wall coordinates through the radius $r = \sqrt{x^2 + y^2}$, so it can be written as a weighted sum of unit rings $\delta_r$ of radius $r$,
\begin{equation}
\begin{aligned}
    k(\cdot,d,f) &= e^{\,i \kappa_f (\sigma_t d + \Delta)} \sum_r A(r,d,f)\, \delta_r, \\[2pt]
    A(r,d,f) &= e^{\,i \sigma_s \kappa_f \sqrt{r^2 + d^2}},
\end{aligned}
\label{eq:ring_decomposition}
\end{equation}
where $\kappa_f$ is the wavenumber of temporal frequency $f$ and $\Delta$ a calibration offset of the optical path. The pair $(\sigma_s, \sigma_t)$ selects the capture modality: $(2, 0)$ for confocal data, where illumination and detection share a wall point and the light covers the same distance twice, and $(1, 1)$ for non-confocal data, where only one leg is propagated and the other is approximated by the plane depth. The kernel carries phase only: we drop the $1/\sqrt{r^2+d^2}$ amplitude falloff, which varies slowly across the aperture; discarding it mildly reweights contributions across the aperture rather than adding structure, and the per-frame normalization before display absorbs the remaining overall scale. The two-dimensional Fourier transform of a unit ring is analytic,
\begin{equation}
    B(r,\rho) = 2\pi r\, J_0(2\pi r \rho),
    \label{eq:ring_basis}
\end{equation}
where $\rho$ is the sampled radial coordinate in the Fourier domain and $J_0$ is the zeroth-order Bessel function of the first kind. Only the zeroth order appears because a unit ring is constant in the angular coordinate, so every harmonic $n \neq 0$ carries a factor $\int_0^{2\pi} e^{i n \theta}\, \mathrm{d}\theta = 0$ and drops out. \eref{eq:ring_basis} is therefore exact rather than a truncated expansion, which leaves no accuracy parameter to tune at this step. Integrating the ring coefficients of \eref{eq:ring_decomposition} against this basis yields the stored radial kernel
\begin{equation}
\begin{aligned}
    K_r(\rho,f,d) &= e^{\,i \kappa_f (\sigma_t d + \Delta)} \int A(r,d,f)\, B(r,\rho)\, \mathrm{d}r \\[2pt]
    &\approx e^{\,i \kappa_f (\sigma_t d + \Delta)} \sum_j q_j\, A(r_j,d,f)\, B(r_j,\rho),
\end{aligned}
\label{eq:radial_kernel}
\end{equation}
where $q_j$ are trapezoidal weights over the sampled radii $r_j$. The modality phase does not depend on $r$, so it leaves the quadrature untouched and is applied once per $(f,d)$ pair after the sum. We compute this integral offline just once for each $(f,d)$ pair. This precomputation introduces two numerical approximations: using a discrete sum (quadrature) and stopping at a maximum radius. Our precomputed kernel differs from the exact discrete kernel in two other ways: we ignore the amplitude falloff (as noted above), and we enforce perfect radial symmetry on the square relay-wall grid.


At runtime, each Cartesian Fourier pixel $(u,v)$ is associated with a continuous radial coordinate $\rho(u,v)$ through a precomputed lookup map. The propagation stage can then be written as
\begin{equation}
    \widehat{\mathcal{I}}(u, v, d) =
    \sum_{f=1}^{N_f}
    \hat{\mathcal{P}}_\omega(u, v, f)\, w(f)\,
    \widetilde{K}_r(\rho(u,v), f, d),
\end{equation}
where $\widetilde{K}_r$ denotes the sampled radial kernel evaluated at the radius associated with pixel $(u,v)$.

In practice, $\rho(u,v)$ usually does not fall exactly on an integer radial sample, so the kernel is evaluated by linearly interpolating between neighboring radial bins to avoid the extra artifacts that would appear with a nearest-neighbor lookup. The access pattern is shown in \autoref{lst:radial_rsd_lookup}: one lane computes the radius-dependent indices and weights, broadcasts them to the lanes covering neighboring depths, and each lane then loads the corresponding \texttt{FP16} radial-kernel values.

\begin{lstlisting}[
  language=CUDA,
  caption={Symmetry-aware \rsd{} lookup. Lane zero computes the radial interpolation and cutoff terms, broadcasts them across the warp, and each lane loads the two \texttt{FP16} radial-kernel samples for its depth.},
  label={lst:radial_rsd_lookup}
]
uint lane = threadIdx.x;
uint fullMask = 0xffffffffu;
int i0 = 0, i1 = 1;
float a = 0.0f, cutoffWeight = 1.0f;
int outsideCutoff = 0;

if (lane == 0)
{
  float rf = radiusMap[xy];
  cutoffWeight = lowPassWeight(rf, rhoCutoffCoord, ringSize);
  outsideCutoff = (cutoffWeight <= 0.0f) ? 1 : 0;

  int lastBase = max(0, min(int(rhoCutoffIndex) - 1, int(ringSize) - 2));
  i0 = max(0, min(int(floorf(rf)), lastBase));
  i1 = min(i0 + 1, int(rhoCutoffIndex));
  a = rf - static_cast<float>(i0);
}

i0 = __shfl_sync(fullMask, i0, 0);
i1 = __shfl_sync(fullMask, i1, 0);
a = __shfl_sync(fullMask, a, 0);
cutoffWeight = __shfl_sync(fullMask, cutoffWeight, 0);
outsideCutoff = __shfl_sync(fullMask, outsideCutoff, 0);

if (outsideCutoff || !validDepth)
  return;

__half2 h0 = rsdKernel[(i0 * numFreqs + wave) * numDepths + depth];
__half2 h1 = rsdKernel[(i1 * numFreqs + wave) * numDepths + depth];
float2 k0 = __half22float2(h0);
float2 k1 = __half22float2(h1);
float2 k = make_float2(
  cutoffWeight * fmaf(a, k1.x - k0.x, k0.x),
  cutoffWeight * fmaf(a, k1.y - k0.y, k0.y));
\end{lstlisting}

A frequency-parallel implementation would require atomics on the output buffer. Instead, our propagation kernel assigns one thread to each output element $(u,v,d)$ and accumulates all temporal frequencies locally. Threads sharing the same Fourier pixel broadcast the radius lookup, interpolation weights, input phasors, and frequency weights across the warp while neighboring lanes cover adjacent depths. Once all temporal frequencies have been accumulated, the depth slices are transformed back through a batched inverse FFT, performing $N_d$ two-dimensional inverse FFTs in parallel. 


\subsection{Post-processing for background noise removal}
\label{sec:method:noise_removal}

The raw output of NLOS reconstruction often shows small artifacts, especially on the background region. We present and evaluate three noise removal strategies. Two exploit temporal consistency across nearby frames: an extension of the depth-dependent average (DDA) of \cite{nam_low-latency_2021}, and a new averaging strategy driven by our coherence metric. The third, our Mixture of Experts (MoE), instead combines several reconstructions of the same frame, which is only practical because our pipeline can reconstruct each one in a fraction of the frame budget.


\subsubsection{Depth-Dependent Average (DDA)}
\label{sec:method:dda}

This algorithm follows the idea of \cite{nam_low-latency_2021}, but we adapt it to our data and enable a variable number of frames, instead of only three. The Depth-Dependent Average (DDA) stores the last $N_w$ reconstructed depth volumes in a circular buffer and computes a weighted average according to the depth, where the window length $N_w$ is a runtime parameter. This allows temporal denoising to operate before the 3D information is collapsed into a 2D image. The weights vary with depth. After temporal averaging, the rest of the pipeline remains as in previous sections; the output is reduced over depth, normalized for display, and optionally band-pass filtered.

\begin{figure}
    \centering
    \includegraphics[width=\linewidth]{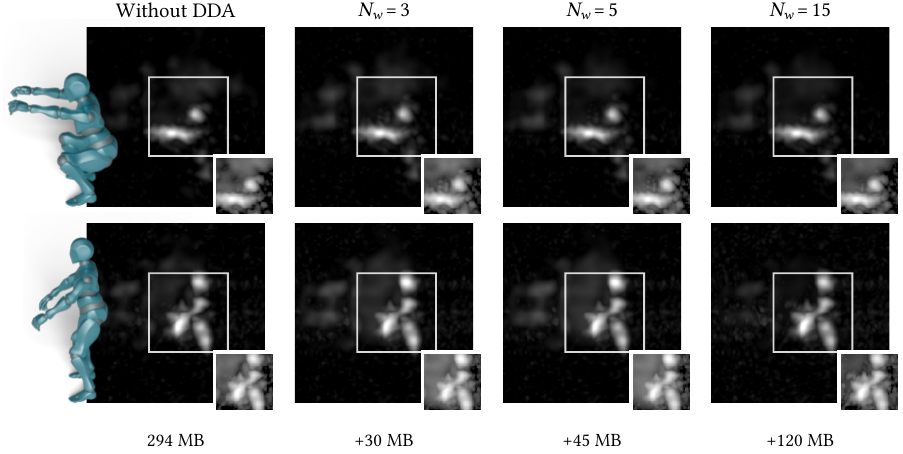}
    \caption{Denoising results from DDA on a dynamic capture, for temporal windows of $N_w$ frames. Each row is one frame of the sequence, with the corresponding pose rendered at the left; the leftmost column reconstructs that frame without denoising. All panels of a row show the same frame, and the inset magnifies the same region in every panel.}
    \label{fig:dda}
\end{figure}

\fref{fig:dda} illustrates the trade-off introduced by temporal averaging. The effect is mostly temporal rather than spatial. On a moving sequence, a short window leaves the per-frame appearance nearly unchanged while reducing frame-to-frame flicker, without visibly smearing the moving object. The benefit saturates quickly with window length, so we keep short windows by default; longer ones mainly add smoothing and runtime overhead.

\begin{figure}
    \centering
    \includegraphics[width=\linewidth]{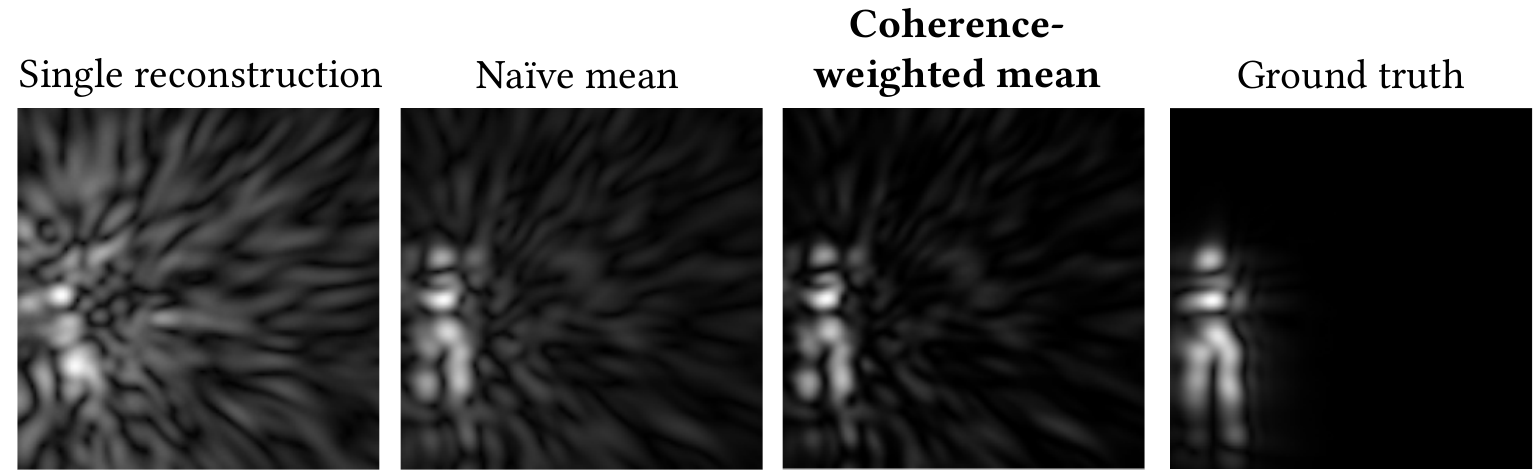}
    \caption{Merging reconstructed frames in the complex domain on a simulated scene. Weighting
    the coherent mean by $C_i(\xv)$ attenuates the incoherent background speckle while
    preserving the hidden surface, whose phase is consistent across the window.}
    \label{fig:frame_merging}
\end{figure}

\subsubsection{Coherence-weighted merging}
\label{sec:method:coherence}

Both of our reconstruction operators are wave-based, so the volume they return is complex valued and each voxel carries a phase as well as an amplitude. The depth-dependent average works on $\lvert f_i(\xv)\rvert$ and discards that phase. It therefore cannot tell whether the successive estimates of a voxel actually agree, or merely have similar magnitudes. Signal from a static surface keeps a consistent phase from frame to frame, whereas reconstruction noise does not, so the phase is exactly the quantity that separates them.

We exploit this by measuring, over a window of $N_w$ frames, how coherently the complex
estimates add up,
\begin{equation}
    C_i(\xv) =
    \frac{\bigl\lvert \sum_{j=i-N_w+1}^{i} f_j(\xv) \bigr\rvert}
         {\epsilon + \sum_{j=i-N_w+1}^{i} \lvert f_j(\xv) \rvert},
    \label{eq:coherence}
\end{equation}
with $\epsilon$ a small constant for numerical stability. The ratio compares the magnitude of the coherent sum with the sum of magnitudes, so $0 \le C_i(\xv) < 1$; the upper bound is approached but never reached, since $\epsilon > 0$. It approaches one where the contributions share a phase, and collapses toward zero where they cancel. The merged frame is the coherent mean scaled by this factor,
\begin{equation}
    \bar{f}_i(\xv) =
    C_i(\xv)\,
    \frac{1}{N_w} \Bigl\lvert \sum_{j=i-N_w+1}^{i} f_j(\xv) \Bigr\rvert.
    \label{eq:coherent_merge}
\end{equation}

\fref{fig:frame_merging} shows the effect on a simulated scene: the background speckle, which is incoherent across the window, is strongly attenuated, while the surface of the hidden object survives the weighting. The assumption behind \eref{eq:coherence} is that the underlying signal is static over the window.

\subsubsection{Mixture of experts (MoE)}
\label{sec:method:moe}

Given the low latency of our reconstruction methods, a direct approach for reducing background noise is to mix complementary reconstructions and keep the structures on which they agree. Intuitively, \rsd{} recovers finer detail at the price of more background noise, while \fk{} trades resolution for a cleaner background. Our goal is to keep high-frequency detail while retaining the noise suppression of the low-pass response.

Our Mixture of experts (MoE) shares part of the spectral work between the two branches. First, the input is padded into a shared frequency-minor layout and a batched spatial FFT is applied once over the relay-wall dimensions. The \rsd{} branch reads this shared spectrum, applies its temporal-frequency transform and propagates it with the radial kernel representation, and the resulting depth volume is reduced to an image. The \fk{} branch expands the same shared spectrum into the padded \fk{} volume, performs the temporal FFT, Stolt remapping, inverse FFT and depth maximum. Since both branches only read the shared spectrum and write to disjoint buffers, the captured CUDA graph executes them concurrently on separate streams with dedicated FFT work areas, joining right before mixing. 

\begin{figure}
\centering
\includegraphics[width=\linewidth]{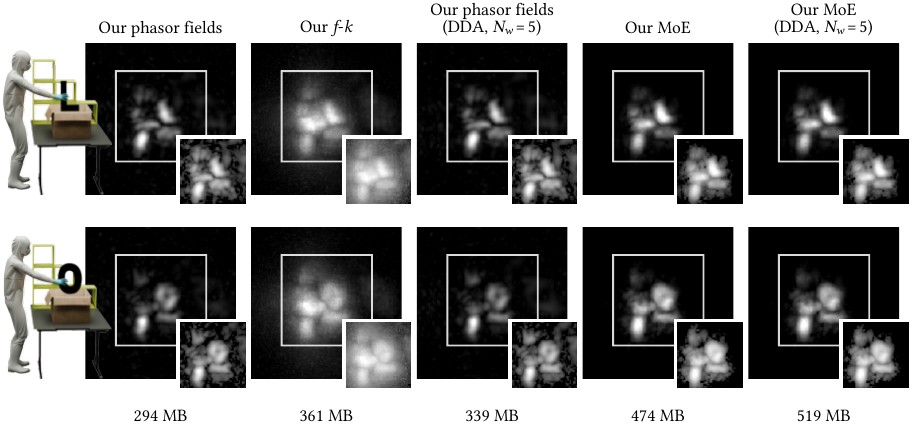}
\caption{Merging \rsd~and~\fk~with MoE. The inset magnifies the same region through a display tone curve ($\gamma = 2.2$).}
\label{fig:expert_mixture_fk_rsd}
\end{figure}

\fref{fig:expert_mixture_fk_rsd} compares each branch with the mixture. The two are not redundant, even though both are filtered backprojections of the same measurement. What separates them is the filter each applies along the temporal-frequency axis \citep{marco_comprehensive_2026}. \fk{} implements the exact inverse, so it keeps the full band and implicitly favors high frequencies. The \rsd{} branch instead applies a narrow band-pass centered on the virtual wavelength, with almost no response at zero frequency. Neither filter is preferable everywhere in a scene, so we combine them rather than choosing one:
\begin{equation}
    f(\xv) =
    f_{\textit{f-k}}(\xv)\,
    {\bigl(f_{\textit{\rsd}}(\xv)+\epsilon\bigr)}^{\beta},
    \label{eq:moe}
\end{equation}
with $\epsilon$ a small constant for numerical stability. Formally it is a product of experts, in which \fk{} supplies the amplitude and \rsd{} acts as a confidence gate of weight $\beta$. Setting $\beta = 1$ recovers the plain product of the two reconstructions. We use $\beta = 0.3$; larger values keep suppressing background but start to erode dim geometry with it. The mixed result is normalized and optionally band-pass filtered.

\subsection{Offline reconstructions}
\label{sec:offline_recs}

Offline reconstruction uses separate implementations from the streaming path. Since each offline reconstruction is typically executed only once for a given input volume, it does not benefit from CUDA graphs. 


For \fk{}, the offline implementation follows the same main ideas as the streaming version: fused input scaling and padding, on-the-fly Stolt remapping, two FFT work buffers, and optional \texttt{FP16} execution. For \rsd{}, however, we do not use the \emph{streaming} \rsd{} implementation. Instead, for fair comparison with existing offline baselines, we accelerate the \rsd{} implementation released by \cite{lindell_wave-based_2019}, which follows work from \cite{liu_non-line--sight_2019}. This offline path precomputes the plane-to-plane propagation operators and the transform matrices for the given volume, uses \texttt{cuBLAS}/\texttt{cuFFT} for the main operations, and can use many more temporal samples and propagated depths than the real-time configuration. 

\section{Results and evaluation}
\label{sec:results}

We evaluate the proposed implementations from two viewpoints. First, we measure the complete \emph{streaming} path used for real-time NLOS imaging, where photon records are binned while the previous frame is reconstructed. This is the setting in which frame rate and GPU memory pressure matter at the same time. Second, we evaluate \emph{offline} reconstruction, where all the captured data is already available and the experiments mainly reflect reconstruction time. We then isolate the main implementation choices: padding, reduced precision, CUDA graphs, photon binning, and Stolt interpolation.

\paragraph{Execution hardware.} All experiments were run on a desktop system with an Intel(R) Core(TM) i7-14700KF CPU at 3.40~\si{\giga\hertz}, 64~GB RAM, an NVIDIA RTX 4080 SUPER GPU with 16~GB VRAM, and Windows 11. The implementation is written in C++23 with CUDA 13.0 and OpenMP for CPU multi-threading. Real-time visualization uses OpenGL 4.6 together with CUDA/OpenGL interoperability.

\paragraph{Adaptation to real SPAD arrays.} Our experiments process raw photon records captured by a SPAD array, but they do not stream directly from live acquisition hardware. Instead, we read the recorded photons from disk as fast as possible. This lets us stress the processing pipeline without being limited by the data interface of a particular camera. The supplementary material examines how our frame rate varies with the photon budget, and finds that the pipeline stays above interactive rates at photon counts far larger than those of the dynamic capture. We will therefore release our implementation with the hope that future SPAD devices and data interfaces will accommodate our faster pipeline.

\subsection{Real-time performance}
\label{sec:results:realtime}

Our real-time comparison uses the dynamic SPAD data released by \citet{nam_low-latency_2021}. The data consists of raw photon timestamps, so both implementations start from the same input and include photon processing before reconstruction. \textbf{The reported frame times are not isolated kernel timings, but the times needed by the streaming pipeline to produce one reconstructed frame}. \tref{table:realtime_comparison} reports average frame time, frame rate, peak GPU memory, and peak CPU memory. We also include an unpadded \fk{} variant because prior real-time systems commonly avoid FFT padding to reduce cost.

The fastest configuration is unpadded \fk{}, which reaches 919.9 FPS on this dataset. This number should be interpreted together with the padding discussion in \sref{sec:method:fk}: removing padding is a speed-memory trade-off and may introduce wrap-around artifacts. With padding enabled, \fk{} still reaches 81.4 FPS. Our \rsd{} variant reaches 448.5 FPS in 0.27~GB of video memory, the least of any configuration here. The reference \rsd{} runs at 10.7 FPS and allocates 3.34~GB, an order of magnitude more, mainly because it uses more intermediate buffers.

\begin{table}
\centering
\mbox{} \hfill
\caption{Average frame time and peak RAM and VRAM usage on \emph{dynamic} data, for our implementations and \cite{nam_low-latency_2021}. The capture is a raw SPAD recording binned frame by frame, so it measures whether a method keeps interactive rates under frame-by-frame streaming processing. PttR \citep{mu_physics_2025} is absent here because it does not handle dynamic captures. Dataset dimensions are shown above the results.}
\label{table:realtime_comparison}
\resizebox{\linewidth}{!}{
\begin{tabular}{|l|r|l|r|r|}
\toprule
\multicolumn{5}{|c|}{\textbf{Dynamic data}}\\ 
\multicolumn{5}{|c|}{$190 \times 190 \times 3002 \Rightarrow$ 208 frequencies, 52 depths, \texttt{nlosbox1} dataset}\\
\cmidrule{1-5}
Approach & $\downarrow$ Frame time & $\uparrow$ FPS & $\downarrow$ Peak VRAM & $\downarrow$ Peak RAM\\
\cmidrule{1-5}
Our \fkb\ & $12.282 \pm 0.475$ \si{\milli\second} & 81.42 & $1,104.1$ \si{\mega\byte} & \multirow{4}{*}{$1,154.5$ \si{\mega\byte}}\\
Our \fkb\ (w/o padding) & \good{$1.087 \pm 0.061$ \si{\milli\second}} & \good{919.85} & $302.1$ \si{\mega\byte} & \\
\cmidrule{1-4}
Phasor-fields \citep{nam_low-latency_2021} & \bad{$93.769 \pm 4.951$} \si{\milli\second} & \bad{10.66} & \bad{$3,425.3$ \si{\mega\byte}} & \\
Our \rsd\ & $2.230 \pm 0.126$ \si{\milli\second} & 448.48 & \good{$280.4$} \si{\mega\byte} & \\
\bottomrule
\end{tabular}}
\end{table}

The supplementary material breaks this capture down across every precision and padding configuration, and shows the same general trend.

\begin{figure}
    \centering
    \begin{subfigure}[t]{0.85\linewidth}
        \includegraphics[width=\linewidth]{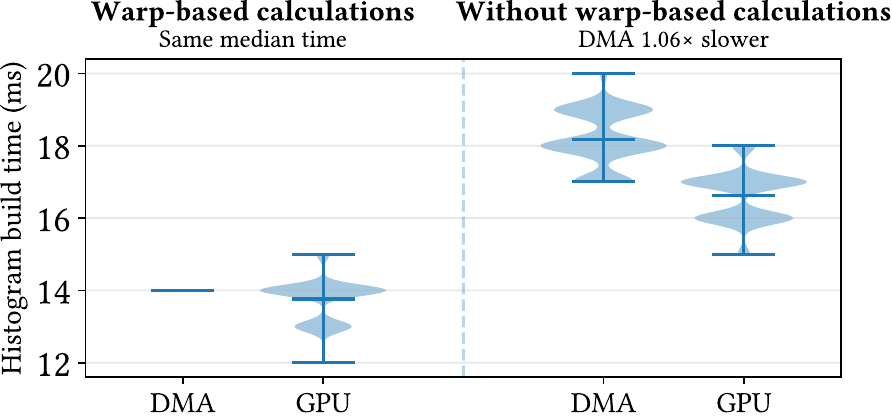}
        \caption{Photon binning}
    \end{subfigure}\\
    \begin{subfigure}[t]{0.85\linewidth}
        \includegraphics[width=\linewidth]{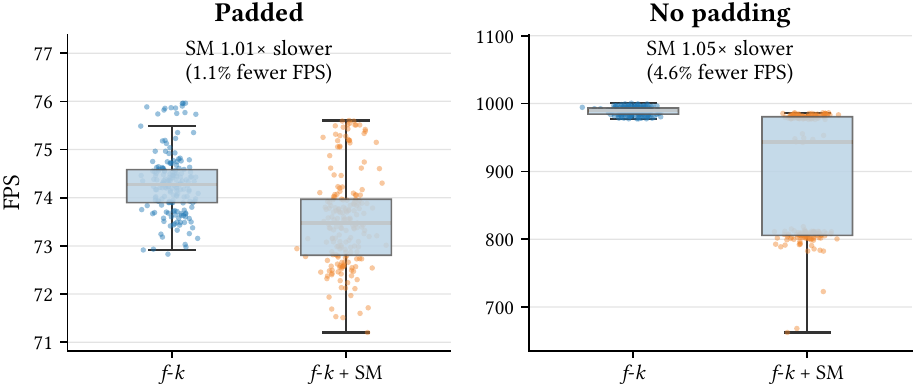}
        \caption{Stolt shared-memory variant of \fk{}}
    \end{subfigure}\\
    \begin{subfigure}[t]{0.85\linewidth}
        \includegraphics[width=\linewidth]{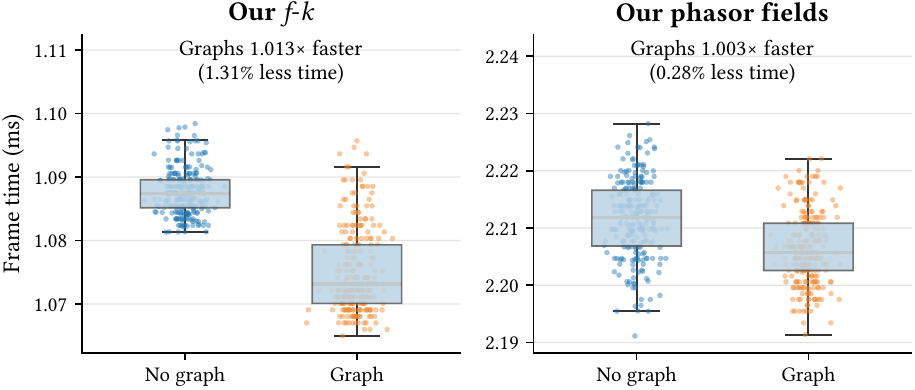}
        \caption{CUDA graph replay}
        \label{fig:graph_effect}
    \end{subfigure}

    \caption{Ablation of implementation decisions: (a) the photon-binning path, (b) the shared-memory Stolt-remapping variant of \fk{}, and (c) replaying a captured CUDA graph instead of launching each operation separately. 
    }
    \label{fig:implementation_ablation}
\end{figure}

\subsection{Implementation choices and ablations}
\label{sec:results:ablations}

\textbf{Photon binning.} Photon binning becomes visible in the frame time once the number of photons grows. Subfigure~(a) of \fref{fig:implementation_ablation} compares the relevant variants. The warp-per-photon phasor path improves throughput by distributing the frequency work for one photon across a warp, while still using atomic additions for the final histogram updates. DMA through pinned host memory is not always faster for small batches, but \fref{fig:dma_vs_gpu_photon_count} shows why it is useful in the larger-photon-count regime: transfer cost eventually dominates, and avoiding an additional copy through pageable memory becomes worthwhile.

\textbf{Stolt interpolation.} Subfigure~(b) shows a case where a more complex memory scheme is not automatically better. Preloading the interpolation neighborhood into shared memory reduces some global reads, but the optimized Stolt kernel already performs only a simple two-sample interpolation along the contiguous third dimension. The extra synchronization therefore cancels most of the expected gain.


\textbf{CUDA graphs.} Graphs address a different source of overhead. Once the main kernels are short, the launch sequence itself becomes a measurable part of the frame time. \fref{fig:graph_effect} shows the effect of replaying a captured reconstruction graph instead of launching each operation separately: $1.3\%$ for \fkb{} and $0.3\%$ for our \rsd{}. The gain is small because our reconstruction submits only about fifteen operations per frame, so there is little launch overhead once the kernels have been fused; reported speed-ups of $1.4\times$ and above come from iterative solvers that launch hundreds of small kernels per step \citep{ekelund_boosting_2025, li_set_2026}. The benefit is most relevant for repeated \emph{streaming} reconstruction, where the same sequence of FFTs and kernels is executed every frame.

\subsection{Comparison with other methods}
\label{sec:results:comparison}

We next compare throughput against other fast NLOS reconstruction approaches. \tref{table:perf_related_work} reports millions of voxels processed per second, average reconstruction time, and peak video memory. The CPU and GPU baselines include Fast Backprojection~\citep{arellano_fast_2017}, CUDA-based backprojection~\citep{sun_cuda-accelerated_2026}, PttR~\citep{mu_physics_2025}, and the \rsd{} pipeline of \citet{nam_low-latency_2021}. PttR is the relevant reference here: it is the fastest published GPU pipeline on preprocessed data, whereas \citet{nam_low-latency_2021} is our reference for streaming reconstruction and is included for completeness. We also include the FPGA result of \citet{liao_fpga_2022}; since no source code is available, we report their timing for their stated configuration instead of re-running it.

The comparison mainly shows where each family of methods scales. Backprojection methods avoid the dense Fourier-domain kernel, but their throughput remains low for the tested volumes. Our \rsd{} variant gives the best throughput in the reported real-time comparisons because it keeps the propagation kernel small. \fkb{} remains the most stable option across larger temporal dimensions, since its cost is controlled only by FFT volume size.

\begin{table}[t]
\centering
\scriptsize
\setlength{\tabcolsep}{3pt}
\caption{Performance of fast reconstruction algorithms. Throughput is the processed data volume divided by reconstruction time: for the wave-based methods this is the input transient cube of each block header, of which the \rsd{} family propagates the listed frequency and depth counts, while backprojection uses its reconstructed volume, $256 \times 256 \times 491$ voxels on Z and $360 \times 260 \times 512$ on office. The frequency counts are chosen from the virtual wavelength and capped at the Nyquist limit of the temporal sampling.
Green and red cells mark the best and worst value per metric within each dataset.}
\label{table:perf_related_work}
\resizebox{\linewidth}{!}{%
\begin{tabular}{@{}lrrr@{}}
\toprule
Approach & Mvox/s $\uparrow$ & Time (s) $\downarrow$ & VRAM (GB) $\downarrow$\\
\midrule
\multicolumn{4}{@{}l}{\textbf{RTX 4080 SUPER - Z}: $256^2 \times 512$, $255$ frequencies, $30$ depths}\\
\addlinespace[0.25em]
Our \fkb{} & 4,846.80 & $0.0069 \pm 0.0000$ & 0.750\\
Our \rsd{} & \good{\strut 9,577.03} & \good{\strut $0.0035 \pm 0.0000$} & \good{\strut 0.293}\\
\addlinespace[0.3em]
Phasor fields \citep{nam_low-latency_2021} & 521.67 & $0.0643 \pm 0.0042$ & 4.389\\
Phasor fields \citep{mu_physics_2025} & 1,242.38 & $0.0270 \pm 0.0023$ & \bad{\strut 8.851}\\
CUDA backprojection & 1.21 & $26.593 \pm 0.377$ & 0.366\\
Fast backprojection & \bad{\strut 0.37} & \bad{\strut $91.238 \pm 0.781$} & 0.689\\
\addlinespace[0.6em]
\midrule
\multicolumn{4}{@{}l}{\textbf{RTX 4080 SUPER - office}: $360 \times 260 \times 512$, $115$ frequencies, $51$ depths}\\
\addlinespace[0.25em]
Our \fkb{} & 3,406.01 & $0.0141 \pm 0.0001$ & 1.496\\
Our \rsd{} & \good{\strut 17,745.27} & \good{\strut $0.0027 \pm 0.0000$} & \good{\strut 0.219}\\
\addlinespace[0.3em]
Phasor fields \citep{nam_low-latency_2021} & 855.57 & $0.0560 \pm 0.0029$ & 4.565\\
Phasor fields \citep{mu_physics_2025} & 1,269.18 & $0.0378 \pm 0.0023$ & \bad{\strut 12.707}\\
CUDA backprojection & 1.74 & $27.586 \pm 0.229$ & 0.538\\
Fast backprojection & \bad{\strut 0.10} & \bad{\strut $467.194 \pm 3.895$} & 4.164\\
\addlinespace[0.6em]
\midrule
\multicolumn{4}{@{}l}{\textbf{Stratix 10 FPGA}: $128^2 \times t$ (N.A.), $69$ frequencies, $51$ depths}\\
\addlinespace[0.25em]
FPGA-acc. \citep{liao_fpga_2022} & N.A. & 0.04 & N.A.\\
\bottomrule
\end{tabular}}
\end{table}

The memory column is what separates the two \rsd{} implementations. Dense propagation kernels grow with the number of retained frequencies, so on the office scene PttR needs $12.7$~\si{\giga\byte} of video memory at the $115$ frequencies of its block, and \cite{nam_low-latency_2021} $4.6$~\si{\giga\byte}, against $0.22$~\si{\giga\byte} for our symmetry-aware variant. The gap is a property of the representation rather than of the operating point: a dense kernel is stored per spatial pixel, ours per radius.

\subsubsection{Memory scalability}
\label{sec:results:mem_scale}

\tref{tab:memory_scalability_pad1} and \tref{tab:memory_scalability_pad2} report peak GPU memory for increasingly large reconstruction volumes. These measurements do not include the CPU memory needed to load the full dataset, which is common to all methods. To keep the tables readable, the input volume is listed separately from the \rsd{} frequency count $\Omega_{\mathrm{PF}}$ and the \rsd{} depth count $d_{\mathrm{PF}}$. There is no corresponding column for \fk{}: it migrates the entire wave field rather than selecting a band of frequencies, so its footprint is governed directly by the volume in the first column. The tables use the reduced-precision configuration that is most relevant for memory pressure: our \rsd{} and MoE store their propagation data in \texttt{FP16}, while \fkb{} remains in \texttt{FP32} in these runs because the \texttt{FP16} \texttt{cuFFT} path imposes power-of-two constraints.

Padding increases the memory footprint of all methods. MoE remains the largest consumer because it runs both branches, allocating the padded \fk{} volume alongside the \rsd{} working buffers; however, since its \rsd{} expert shares the radial kernel representation of \sref{sec:method:rsd_ring}, its stored propagation data no longer grows with the spatial dimensions of the relay wall. Our phasor-fields implementation needs 4.5~GB in the largest padded configuration and is the only method that completes it: the padded \fkb{} transform exceeds what \texttt{cuFFT} accepts at that size, and the MoE runs out of memory. That headroom is the practical reason for adopting the radial kernel representation in the real-time pipeline despite the slight loss in visual quality it introduces.

\begin{table}[t]
\centering
\small
\setlength{\tabcolsep}{3.8pt}
\caption{Peak GPU memory usage for our reconstruction methods \textbf{without padding}. $\Omega_{\mathrm{PF}}$ and $d_{\mathrm{PF}}$ are the number of temporal frequencies and propagated depths used by the phasor-fields runs. Missing entries are reported as \emph{OOM}. Precision: \fkb{} in \texttt{FP32}; phasor fields and MoE in \texttt{FP16}.}
\label{tab:memory_scalability_pad1}
\resizebox{\linewidth}{!}{%
\begin{tabular}{@{}lllrrr@{}}
\toprule
 &  &  & \multicolumn{3}{c}{Peak GPU memory usage (\si{\giga\byte})} \\
\cmidrule(l){4-6}
Volume & $\Omega_{\mathrm{PF}}$ & $d_{\mathrm{PF}}$ & \fkb{} & Phasor fields & MoE \\
\midrule
$90\times65\times128$ & 29 & 13 & 0.031 & \textbf{0.005} & 0.035 \\
$180\times130\times256$ & 58 & 26 & 0.188 & \textbf{0.026} & 0.234 \\
$360\times260\times512$ & 115 & 51 & 1.496 & \textbf{0.208} & 1.872 \\
$720\times520\times1024$ & 229 & 103 & 11.960 & \textbf{1.658} & 14.976 \\
\bottomrule
\end{tabular}%
}
\end{table}

\begin{table}[t]
\centering
\small
\setlength{\tabcolsep}{3.8pt}
\caption{Peak GPU memory usage for our reconstruction methods \textbf{with padding}. $\Omega_{\mathrm{PF}}$ and $d_{\mathrm{PF}}$ are the number of temporal frequencies and propagated depths used by the phasor-fields runs. Missing entries are reported as \emph{OOM}. Precision: \fkb{} in \texttt{FP32}; phasor fields and MoE in \texttt{FP16}.}
\label{tab:memory_scalability_pad2}
\resizebox{\linewidth}{!}{%
\begin{tabular}{@{}lllrrr@{}}
\toprule
 &  &  & \multicolumn{3}{c}{Peak GPU memory usage (\si{\giga\byte})} \\
\cmidrule(l){4-6}
Volume & $\Omega_{\mathrm{PF}}$ & $d_{\mathrm{PF}}$ & \fkb{} & Phasor fields & MoE \\
\midrule
$90\times65\times128$ & 29 & 13 & 0.133 & \textbf{0.009} & 0.143 \\
$180\times130\times256$ & 58 & 26 & 1.064 & \textbf{0.072} & 1.147 \\
$360\times260\times512$ & 115 & 51 & 8.477 & \textbf{0.567} & 9.170 \\
$720\times520\times1024$ & 229 & 103 & \textsc{oom} & \textbf{4.530} & \textsc{oom} \\
\bottomrule
\end{tabular}%
}
\end{table}

\subsection{Decomposition of stage time}
\label{sec:results:stage_time}

\begin{figure}
    \centering
    \includegraphics[width=\linewidth]{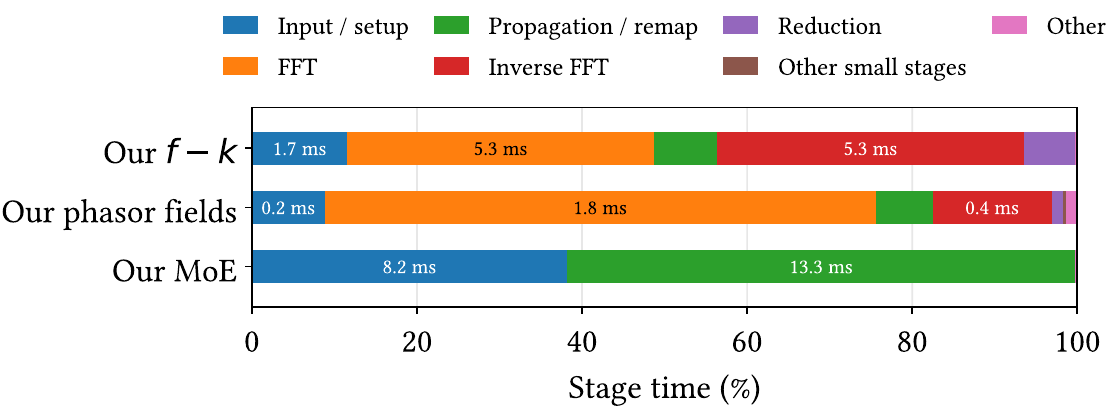}
    \caption{Stage-time decomposition for the main reconstruction pipelines. }
    \label{fig:stage_time}
\end{figure}

\fref{fig:stage_time} breaks the frame time into the processing stages. This view is useful because the fastest algorithm is not always limited by the same part of the pipeline. In \fk{}, the FFT and inverse FFT stages dominate. For our \rsd{} variant, the compact kernel reduces memory traffic in propagation, so launch overhead and common preprocessing steps become more visible. This explains why there is not a single optimization workflow for all the reconstruction methods.

\subsection{Data quantization}
\label{sec:results:quantization}

\begin{figure}
    \centering
    \includegraphics[width=\linewidth]{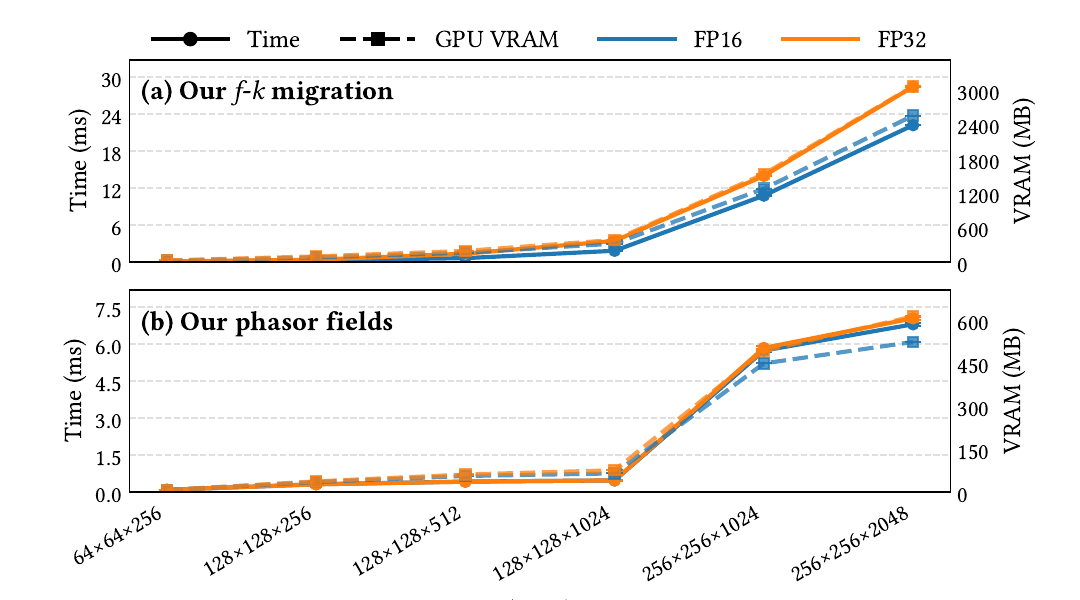}
    \caption{Frame time and peak GPU memory for \texttt{FP16} and \texttt{FP32} configurations. }
    \label{fig:quantization_results}
\end{figure}

Reduced precision is useful only when it reduces the actual bottleneck. \fref{fig:quantization_results} therefore reports both time and GPU memory for \texttt{FP16} and \texttt{FP32} variants. For our \rsd{}, \texttt{FP16} mainly affects the stored propagation kernel, so the expected benefit is lower VRAM usage with limited impact on the FFT stages. For \fkb{}, the trade-off is less direct: \texttt{FP16} reduces element size, but \texttt{cuFFT Xt} requires power-of-two dimensions. As a result, \texttt{FP16} is not automatically faster for \fk{} even when it reduces memory. This is why the memory-scalability tables use \texttt{FP16} storage for our \rsd{} propagation data but keep \fkb{} in \texttt{FP32} for large data volumes.

\subsection{Equivalence with reference implementations}

We compare our reconstructions using the implementation from \cite{nam_low-latency_2021} as a reference to show that our results are practically equivalent across quantization levels, propagation kernel formulations and consecutive reconstructed frames.

\begin{figure}
    \centering
    \begin{subfigure}[t]{\linewidth}
        \includegraphics[width=\linewidth]{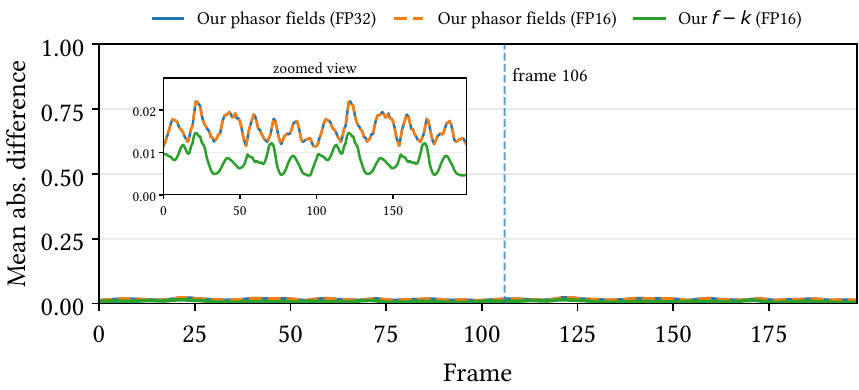}
        \caption{Sequence-wide visual error.}
        \label{fig:humanoid_quality_over_time}
    \end{subfigure}

    \vspace{0.45em}

    \begin{subfigure}[t]{1\linewidth}
        \centering
        \includegraphics[width=\linewidth]{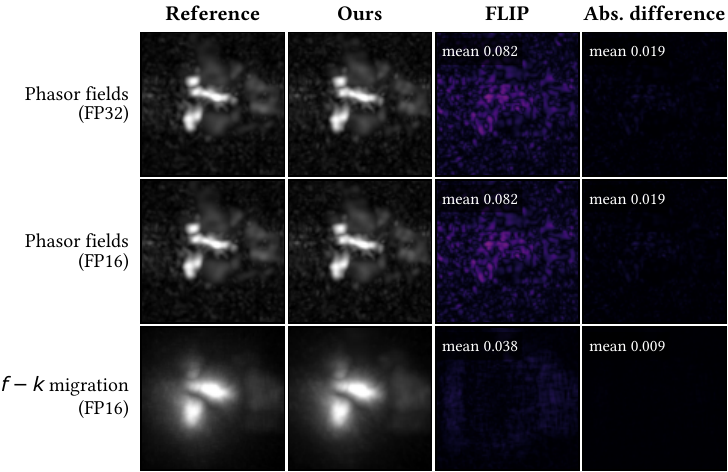}
        \caption{Representative frame.}
        \label{fig:humanoid_representative_frame}
    \end{subfigure}

    \caption{Equivalence with reference implementation on the dynamic capture of \cite{nam_low-latency_2021}. The sequence plot reports the mean absolute difference over time, while the frame panel shows the reference image, our reconstruction, FLIP metric and absolute difference. The reference for \rsd{}-based methods is \cite{nam_low-latency_2021}'s reconstruction; for \fkb{}, the reference is \fkb{} using \texttt{FP32}. The two \rsd{} curves overlap.}
    \label{fig:humanoid_dynamic_quality}
\end{figure}

\fref{fig:humanoid_dynamic_quality} illustrates this on a dynamic dataset. The top panel tracks the mean absolute difference for each frame, and the bottom panel shows the corresponding reference, reconstruction, and error map for one representative frame. We show \emph{FLIP} error \citep{andersson_flip_2020}, and the absolute difference between reference and candidate images. For the tested methods, the mean error remains low throughout the sequence, and our reconstruction qualitatively looks very similar to the reference frame. Over the whole sequence the mean absolute difference stays at $0.016$ for our \rsd{} and $0.008$ for our \fkb{}. The \texttt{FP16} and \texttt{FP32} \rsd{} curves overlap in the plot. 
The residual difference with respect to \cite{nam_low-latency_2021} is therefore attributable to the radial approximation of \sref{sec:method:rsd_ring} rather than to reduced-precision storage, which is consistent with the kernel-domain comparison of \fref{fig:radial_comparison_symmetry_rsd}.

\subsection{Background noise suppression}
\label{sec:results:background}

\begin{figure}
    \centering
    \includegraphics[width=.65\linewidth]{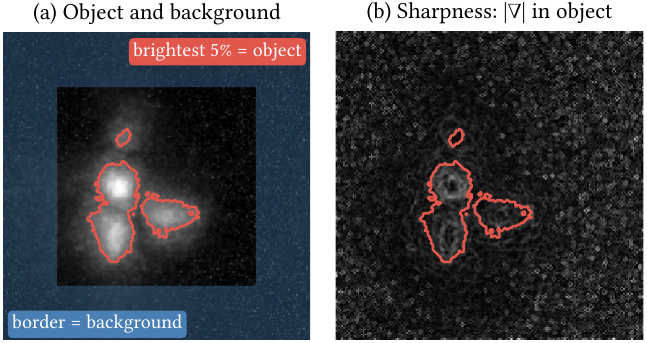}
    \caption{The regions used in \tref{tab:background_suppression}. (a) The brightest 5\% of pixels form the \textbf{object} (red contour), and the outer border forms the \textbf{background} (blue band). Their ratio is the \textbf{contrast}. (b) \textbf{Sharpness} is the mean gradient magnitude inside the same object contour.}
    \label{fig:background_metrics}
\end{figure}

The mixture described in \sref{sec:method:moe} is meant to reduce background response without removing scene content, so we measure both. Every row of \tref{tab:background_suppression} applies the same single normalization and band-pass step. The object level is the mean of the brightest 5\% of pixels, and the background level is the mean over an outer border where the hidden object never appears. Their ratio is the contrast, and the mean gradient magnitude over the object serves as a sharpness proxy. \fref{fig:background_metrics} draws both regions on a measured frame, so each column of the table can be read against the part of the image it summarizes. Values are averaged over the full 100-frame sequence of the \texttt{nlosbox1} capture.

\begin{table}
\centering
\caption{Background suppression on \texttt{nlosbox1}, in \texttt{FP32}. Each method is reported with and without padding. Green and bold mark the best value in each column. The sweep over the pooling weight $\beta$ is reported in the supplementary material.}
\label{tab:background_suppression}
\resizebox{\linewidth}{!}{
\begin{tabular}{|l|c|r|r|r|r|}
\toprule
Method & Padding & $\uparrow$ Object & $\downarrow$ Background & $\uparrow$ Contrast & $\uparrow$ Sharpness\\
\cmidrule{1-6}
\multirow{2}{*}{Our \fkb{}} & no & \good{\textbf{0.615}} & 0.0379 & 16.2 & 0.043\\
 & yes & \good{\textbf{0.656}} & 0.0475 & 13.8 & 0.037\\
\cmidrule{1-6}
\multirow{2}{*}{Our \rsd{}} & no & 0.405 & 0.0115 & 35.1 & \good{\textbf{0.056}}\\
 & yes & 0.410 & 0.0079 & 51.6 & \good{\textbf{0.056}}\\
\cmidrule{1-6}
\multirow{2}{*}{Our MoE ($\beta = 0.3$)} & no & 0.449 & \good{\textbf{$7.4\cdot10^{-6}$}} & \good{\textbf{60,981}} & 0.053\\
 & yes & 0.479 & \good{\textbf{0.0009}} & \good{\textbf{512}} & 0.052\\
\bottomrule
\end{tabular}
}
\end{table}

Unpadded, the mixture leaves a background of $7.4\cdot10^{-6}$, three orders of magnitude below either branch alone, while keeping 73\% of the \fkb{} object level and more than the \rsd{} branch retains on its own.

Padding changes the picture in a way worth stating. It raises every object level slightly, but it also admits real signal into the outer border, so the \fkb{} and mixture backgrounds grow and the mixture's contrast falls from $60{,}981$ to $512$. The ordering between methods holds: \rsd{} is the exception, its background falls and its contrast in fact improves with padding, from 35.1 to 51.6, whereas \fkb{} loses both contrast and sharpness.

We also compared the two temporal strategies of \sref{sec:method:dda} and \sref{sec:method:coherence} on real dynamic captures, using the same contrast measure. The subject of \texttt{nlosbox1} moves slowly relative to the frame rate. There, merging in the complex domain over $N_w = 3$ frames raises contrast from 36 to 48 and halves the frame-to-frame flicker, well ahead of the depth-dependent average. The coherence weight of \eref{eq:coherence} is not what produces this gain. Applying it lowers contrast to 42, and it degrades further as the window grows. On a faster capture the ordering reverses altogether, with complex merging dropping contrast from 27 to 22 while the depth-dependent average raises it to 32. Coherence weighting assumes a static signal over the window, and a moving surface violates it, so we use the depth-dependent average by default and use complex merging for slow or static content.

\subsection{Offline reconstruction}
\label{sec:results:offline}

\begin{figure}
    \centering
    \begin{subfigure}[t]{\linewidth}
        \includegraphics[width=\linewidth]{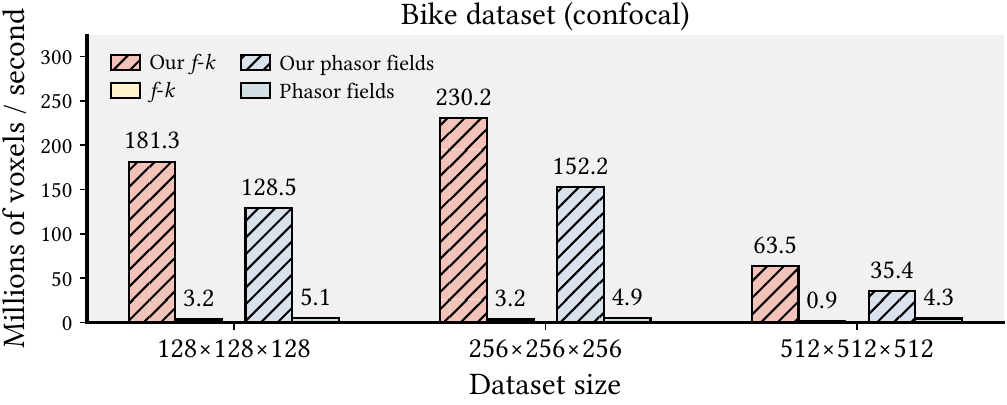}
    \end{subfigure}
    
    \vspace{0.5em}
    
    \begin{subfigure}[t]{\linewidth}
        \includegraphics[width=\linewidth]{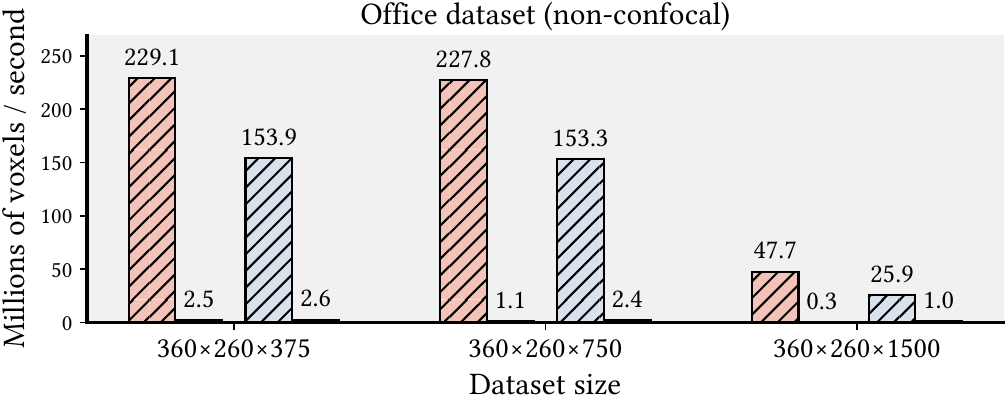}
    \end{subfigure}
    
    \vspace{0.5em}
    
    \begin{subfigure}[t]{\linewidth}
        \includegraphics[width=\linewidth]{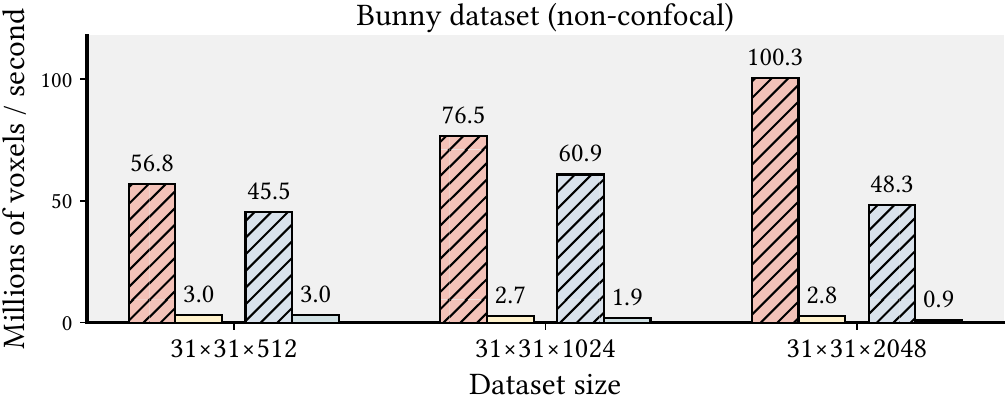}
    \end{subfigure}

    \caption{Offline throughput, in millions of voxels processed per second, for representative confocal and non-confocal datasets. Hatched bars are our implementations and solid bars the original MATLAB ones. Full timing and memory results are reported in the supplementary material.}
    \label{fig:mvox_second}
\end{figure}

\tref{table:offline_summary} reports one representative configuration per dataset, and \fref{fig:mvox_second} summarizes offline throughput in millions of voxels processed per second. Both are extracts: the complete sweep over every resolution is in the supplementary material. We evaluate datasets from \citet{lindell_wave-based_2019} and \citet{galindo_dataset_2019}, including confocal and exhaustive captures. Exhaustive captures are first converted to confocal measurements using Normal Moveout Correction. In this setting, the input transient volume is already available, so the reported time focuses on reconstruction rather than on streaming photon records. \textbf{The reported time is not the reconstruction kernels alone. It covers the whole per-run sequence: allocating and zeroing the output volume, reconstructing, and releasing every intermediate buffer.} We measure it this way because those stages are a real cost of running a reconstruction once.

Across most experiments, the optimized \fkb{} implementation gives the highest throughput. The optimized offline \rsd{} path is faster by a few milliseconds in only a few tests, and whether it wins depends on the selected dimensions. Both optimized implementations are notably faster than their original MATLAB counterparts, which run entirely on the CPU. Across the full offline benchmark table, reconstruction time is reduced by 98.3\% for \fkb{} and 96.1\% for \rsd{} on average, and peak memory by 52.5\%. The curves in \fref{fig:mvox_second} are not strictly monotonic: small datasets are affected by allocation and setup overhead, medium-sized datasets make better use of the GPU, and the largest datasets become limited by FFT size and memory traffic. In our measurements, increasing spatial resolution is more expensive than increasing the number of temporal bins, because the dominant FFT and inverse FFT stages grow over the spatial dimensions as well.

\begin{table*}
\scriptsize
\centering
\caption{Offline comparison against the original MATLAB implementations, showing the largest configuration measured for each dataset. This is a representative extract: every dataset was also evaluated at lower spatial and temporal resolutions, and those results, together with the per-resolution trends, are reported in full in the supplementary material. Throughput is the reconstructed volume divided by reconstruction time. For exhaustive captures, the dimensions column lists the original measurement grid before normal moveout correction; normal moveout correction produces a $31\times31\times T$ confocal transient, which is the volume used for the throughput calculation. Max. tracked memory reports the peak footprint of each implementation's main buffers: host RAM for the original MATLAB code, and the peak of all tracked CUDA allocations for ours, which counts device buffers and pinned host staging together and is therefore not a pure video-memory figure.}
\label{table:offline_summary}
\begin{tabular}{|l|l|l|l|l|l|l|l|l|}
\toprule
& & & \multicolumn{2}{|c|}{$\downarrow$ Time (\si{\second})} & \multicolumn{2}{c|}{$\uparrow$ Throughput (Mvox/\si{\second})} & \multicolumn{2}{c}{$\downarrow$ Max. tracked memory (\si{\giga\byte})}\\
\cmidrule{4-9}
Dataset & Dimensions & Algorithm & Ours & Original & Ours & Original & Ours & Original \\
\cmidrule{1-9}
\multicolumn{9}{|c|}{Confocal -- \fk\ dataset \citep{lindell_wave-based_2019}}\\
\cmidrule{1-9}
\multirow{2}{*}{bike} & \multirow{2}{*}{512$\times$512$\times$512} & \fk & \textbf{2.114 $\pm$ 0.03} & 146.624 $\pm$ 8.47 & \textbf{63.50} & 0.92 & \textbf{17.000} & 52.430 \\
& & Phasor fields & \textbf{3.791 $\pm$ 0.03} & 30.872 $\pm$ 0.67 & \textbf{35.40} & 4.35 & \textbf{22.001} & 33.579 \\
\cmidrule{1-9}
\multirow{2}{*}{teaser} & \multirow{2}{*}{512$\times$512$\times$512} & \fk & \textbf{2.016 $\pm$ 0.01} & 154.730 $\pm$ 14.77 & \textbf{66.59} & 0.87 & \textbf{17.000} & 52.433 \\
& & Phasor fields & \textbf{3.916 $\pm$ 0.08} & 31.458 $\pm$ 2.65 & \textbf{34.28} & 4.27 & \textbf{22.001} & 33.579 \\
\cmidrule{1-9}
\multirow{2}{*}{statue} & \multirow{2}{*}{512$\times$512$\times$512} & \fk & \textbf{2.018 $\pm$ 0.01} & 152.567 $\pm$ 8.22 & \textbf{66.51} & 0.88 & \textbf{17.000} & 52.430 \\
& & Phasor fields & \textbf{3.758 $\pm$ 0.02} & 32.195 $\pm$ 1.86 & \textbf{35.72} & 4.17 & \textbf{22.001} & 33.579 \\
\cmidrule{1-9}
\multicolumn{9}{|c|}{Confocal -- Zaragoza dataset \citep{galindo_dataset_2019}}\\
\cmidrule{1-9}
\multirow{2}{*}{usaf} & \multirow{2}{*}{256$\times$256$\times$1964} & \fk & \textbf{1.747 $\pm$ 0.00} & 139.592 $\pm$ 22.22 & \textbf{73.70} & 0.92 & \textbf{16.303} & 52.429 \\
& & Phasor fields & \textbf{4.172 $\pm$ 0.01} & 31.655 $\pm$ 1.26 & \textbf{30.85} & 4.07 & \textbf{21.112} & 33.621 \\
\cmidrule{1-9}
\multirow{2}{*}{bunny} & \multirow{2}{*}{256$\times$256$\times$1964} & \fk & \textbf{1.734 $\pm$ 0.00} & 138.419 $\pm$ 8.19 & \textbf{74.24} & 0.93 & \textbf{16.303} & 52.429 \\
& & Phasor fields & \textbf{4.169 $\pm$ 0.02} & 33.418 $\pm$ 0.67 & \textbf{30.88} & 3.85 & \textbf{21.112} & 33.621 \\
\cmidrule{1-9}
\multirow{2}{*}{z} & \multirow{2}{*}{256$\times$256$\times$1964} & \fk & \textbf{1.780 $\pm$ 0.08} & 142.930 $\pm$ 6.27 & \textbf{72.29} & 0.90 & \textbf{16.303} & 52.429 \\
& & Phasor fields & \textbf{4.208 $\pm$ 0.02} & 32.743 $\pm$ 1.42 & \textbf{30.58} & 3.93 & \textbf{21.112} & 33.636 \\
\cmidrule{1-9}
\multicolumn{9}{|c|}{Tal-generated dataset \citep{royo_diegoroyotal_2024}, inspired by a Phasor Fields scene \citep{liu_phasor_2020}}\\
\cmidrule{1-9}
\multirow{2}{*}{office} & \multirow{2}{*}{360$\times$260$\times$1500} & \fk & \textbf{2.945 $\pm$ 0.05} & 507.808 $\pm$ 47.68 & \textbf{47.67} & 0.28 & \textbf{17.784} & 75.938 \\
& & Phasor fields & \textbf{5.414 $\pm$ 0.02} & 138.281 $\pm$ 13.87 & \textbf{25.93} & 1.02 & \textbf{23.023} & 66.424 \\
\cmidrule{1-9}
\multicolumn{9}{|c|}{Exhaustive -- Zaragoza dataset \citep{galindo_dataset_2019}}\\
\cmidrule{1-9}
\multirow{2}{*}{concavities} & \multirow{2}{*}{16$\times$16$\times$16$\times$16$\times$2048} & \fk & \textbf{0.018 $\pm$ 0.00} & 0.588 $\pm$ 0.04 & \textbf{107.02} & 3.21 & \textbf{0.352} & 0.770 \\
& & Phasor fields & \textbf{0.027 $\pm$ 0.00} & 2.107 $\pm$ 0.01 & \textbf{68.74} & 0.90 & \textbf{0.436} & 0.558 \\
\cmidrule{1-9}
\multirow{2}{*}{t (in a box)} & \multirow{2}{*}{16$\times$16$\times$16$\times$16$\times$2048} & \fk & \textbf{0.015 $\pm$ 0.00} & 0.744 $\pm$ 0.05 & \textbf{122.69} & 2.53 & \textbf{0.351} & 0.773 \\
& & Phasor fields & \textbf{0.024 $\pm$ 0.00} & 4.263 $\pm$ 0.15 & \textbf{78.85} & 0.44 & \textbf{0.436} & 0.562 \\
\cmidrule{1-9}
\multirow{2}{*}{bunny} & \multirow{2}{*}{16$\times$16$\times$16$\times$16$\times$2048} & \fk & \textbf{0.019 $\pm$ 0.00} & 0.685 $\pm$ 0.10 & \textbf{100.31} & 2.76 & \textbf{0.352} & 0.769 \\
& & Phasor fields & \textbf{0.039 $\pm$ 0.02} & 2.187 $\pm$ 0.03 & \textbf{48.33} & 0.86 & \textbf{0.436} & 0.560 \\
\bottomrule
\end{tabular}
\end{table*}

\section{Conclusions and future work}
\label{sec:conclusion}

We have presented CUDA implementations of wave-based NLOS reconstruction pipelines for both streaming and offline processing. The proposed system reorganizes their execution around GPU dataflow: fused kernels, compact memory layouts, batched transforms, CUDA graphs, and reduced intermediate allocations. On the phasor-fields side, building the ring-and-radius kernels of \cite{jiang_ring_2022} offline, from the analytic Fourier transform of a ring, keeps the propagation data compact and removes dense kernel reconstruction from the runtime. On the dynamic SPAD dataset of \citet{nam_low-latency_2021}, the padded \fkb{} path reaches 81.4~FPS and our \rsd{} variant reaches 448.5~FPS in 0.27~GB of video memory, compared with 10.7~FPS in 3.34~GB for the reference streaming \rsd{} pipeline of \citet{nam_low-latency_2021}: $42\times$ the frame rate in $8.2\%$ of the memory. Against the fastest published GPU baseline, Physics to the Rescue \citep{mu_physics_2025}, we reconstruct full multi-plane volumes roughly an order of magnitude faster ($7.7\times$ on the \emph{Z} scene and $14.0\times$ on the office scene) while using on average $2.5\%$ of its peak video memory; on the single-plane setting it is built for, the margin is $1.4\times$. In the offline setting, where the reference implementations are CPU-bound MATLAB code, reconstruction time is reduced by 98.3\% for \fk{} and 96.1\% for \rsd{} on average, and by 99.4\% and 96.1\% on the largest office dataset.

From the reported results we can conclude that no single optimization dominates all setups. Unpadded \fk{} gives the highest frame rate on small dynamic inputs, but padding remains an important quality and robustness trade-off. Our \rsd{} variant replaces the dense propagation kernel with a radial representation, reducing memory traffic and video-memory usage while introducing a small visual error. Similarly, \texttt{FP16} is useful when it targets our stored \rsd{} propagation data, but it is not a universal replacement for \texttt{FP32}; for \fkb{}, \texttt{cuFFT} planning constraints and conversion overhead can even outweigh the reduced element size.

Our streaming experiments process raw photon records from disk rather than from a live SPAD interface, so future systems should evaluate the same reconstruction path together with camera-side transfer constraints. Photon binning and data movement also remain important costs at high photon counts, motivating more asynchronous uploads and better integration between acquisition and reconstruction. Finally, both \fk{} and \rsd{} still depend heavily on Fourier-domain operators, so future improvements will likely come from better transform scheduling, more memory-efficient propagation representations, and denoising strategies that preserve temporal detail while exploiting the high frame rates enabled by the GPU pipeline.

\section*{Declaration of generative AI and AI-assisted technologies in the manuscript preparation process}

During the preparation of this work the authors used Claude to revise and improve parts of the manuscript text and to generate the scripts that produce the figures and tables from human-measured data. The authors reviewed and edited the output as needed and take full responsibility for the content of the published article.




\bibliographystyle{cas-model2-names}

\bibliography{references}

\clearpage
\newpage
\appendix
\section{Integration in a SPAD array}
\label{sec:photon_experiment}

In this section, we evaluate the potential results that could be achieved by running our code on current and next-generation live SPAD arrays. The experiments still use recorded photons rather than a physical camera stream, but they answer two practical questions: the number of photons our work can process every frame, and the image quality to expect from these photons.


These experiments use the dynamic dataset released by \citet{nam_low-latency_2021} for streamed reconstruction. Each frame contains approximately $1.2 \cdot 10^{6}$ photon records. Reading those records from disk lets us stress the processing pipeline without being limited by the bandwidth or buffering behavior of a specific acquisition device. With this, we take the opposite perspective: instead of fixing the photon count and reporting the resulting frame rate, we fix the frame rate and determine how many photons our system can process per second.

\begin{figure}
    \centering
    \includegraphics[width=\linewidth]{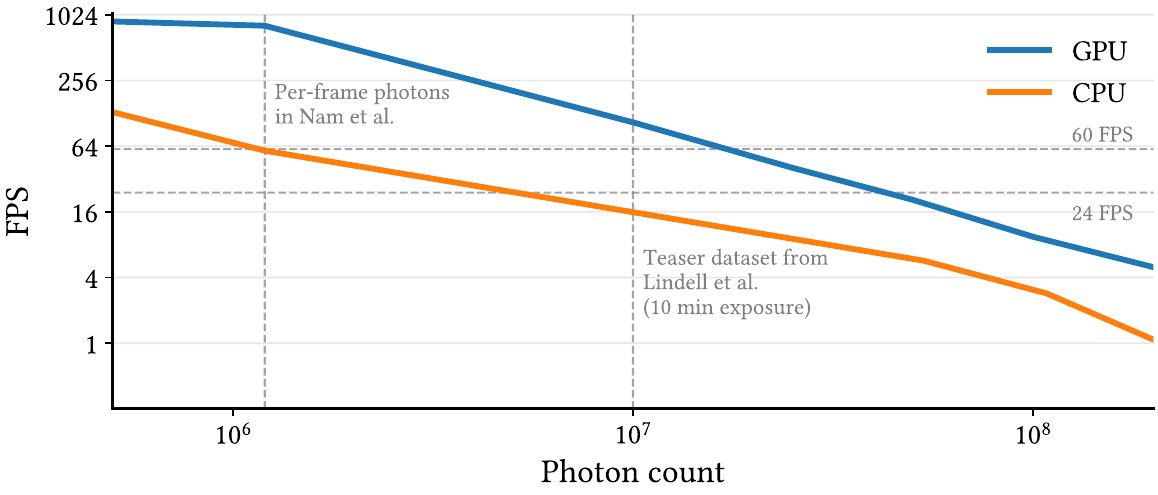}
    \caption{Photon binning time for a dynamic NLOS capture of shape $190\times190\times3002$ ($x \times y \times t$), whose photons are accumulated directly into $208$ temporal frequencies. The photon count varies slightly from frame to frame and typically stays near $1.2\cdot10^6$.}
    \label{fig:photon_binning_nam_dataset}
\end{figure}

\fref{fig:photon_binning_nam_dataset} shows that our pipeline remains above 60~FPS for photon counts up to approximately 47M per frame, and above 24~FPS up to approximately 96M photons per frame. These counts are much larger than those of the dynamic dataset used in the main streamed comparison.

\begin{figure}
    \centering
    \includegraphics[width=\linewidth]{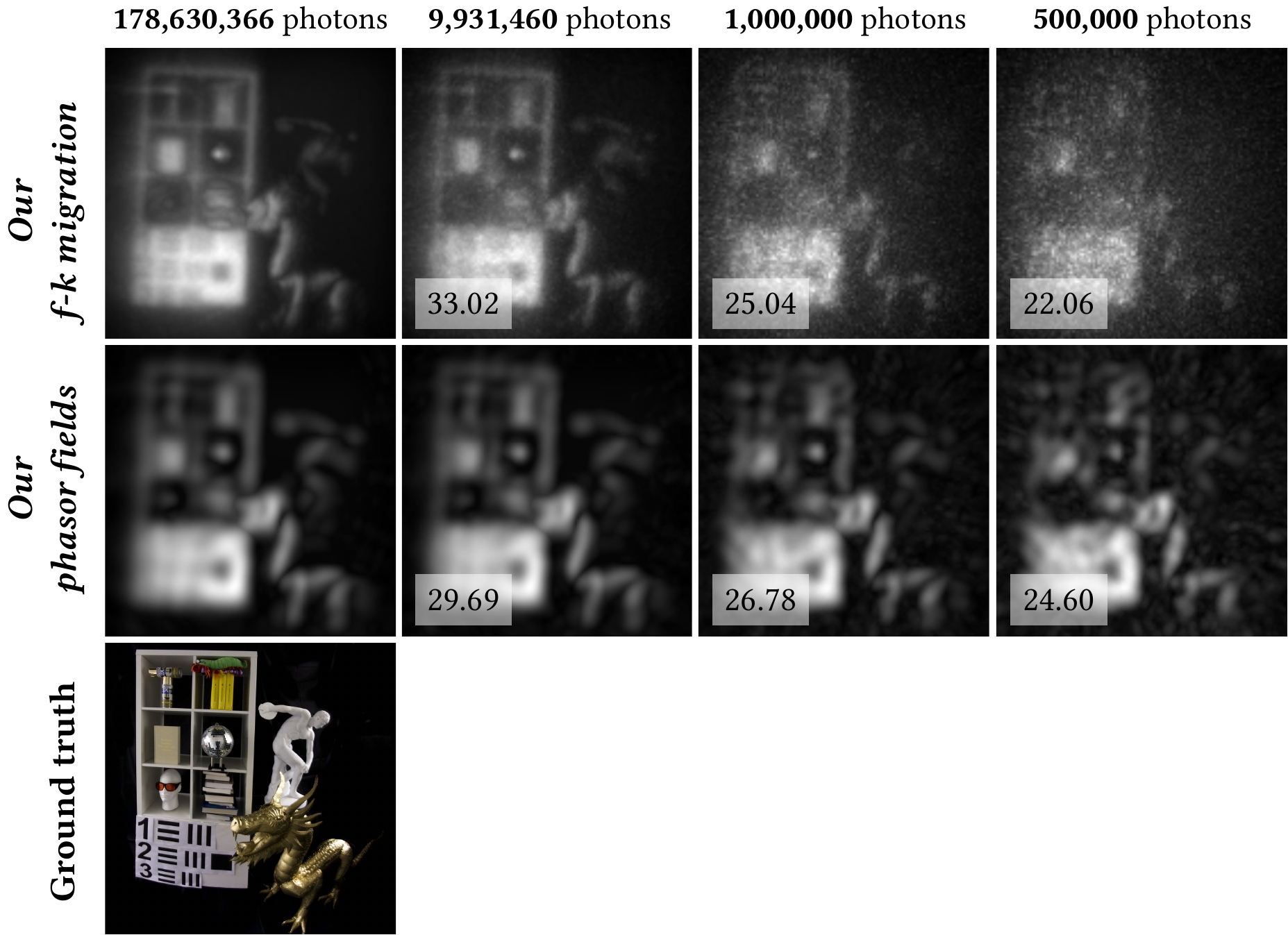}
    \caption{Teaser dataset from \citep{lindell_wave-based_2019}. The first and second columns correspond to measurements of 180 and 10 minutes, respectively. We downsample the former to $10^6$ and $5 \cdot 10^5$ photons to show that the hidden scene remains recognizable with fewer captured photons. The first and second rows show our \fkb{} and \rsd{} reconstructions, respectively. The numbers within each reconstruction indicate the Peak Signal-to-Noise Ratio with respect to the most informed reconstruction, shown in the first column.}
    \label{fig:exp_time_variation}
\end{figure}

While our work handles several million photons until surpassing the 60~FPS threshold, we also aim to demonstrate that, for a lower number of photons, the hidden scene remains recognizable using both \fkb{} and \rsd{} algorithms. \fref{fig:exp_time_variation} shows reconstructions of a scene with progressively fewer photons.
Both our \fkb{} and \rsd{} preserve the main hidden shape at reduced counts, although \rsd{} is more robust to noise in this example. The third column uses fewer photon records than a typical frame of the dynamic dataset from \citet{nam_low-latency_2021}, and the fourth column uses less than half of that amount.

For context, the dynamic dataset of \citet{nam_low-latency_2021} is processed at approximately 400~FPS in this benchmark. The teaser dataset of \citet{lindell_wave-based_2019}, which contains 178M photons captured over 180~\si{\minute}, is processed at slightly below 16~FPS. The reconstructions with fewer photons in \fref{fig:exp_time_variation} suggest that much shorter effective exposures or higher acquisition rates can still be useful for interactive operation.

Finally, we estimate the acquisition data rate. \citet{nam_low-latency_2021} report photon events streamed from the hardware queue over USB~3.0, whose speed rate is 5~\si{\giga\bit\per\second}, i.e.\ at most about 500~\si{\mega\byte\per\second} of payload~\citep{west_real-time_2023}. Since each photon record occupies 4~bytes~\citep{nam_low-latency_2021}, the link is bounded at about 125M photon events per second. With the photon counts of the dynamic sequence, that bandwidth gives an ideal upper bound of about 104 frames per second; protocol and device overheads probably place the practical rate below that. If fewer photons are sufficient for the target scene, as in \fref{fig:exp_time_variation}, the acquisition-side frame rate can be higher.

\section{Further implementation details}
\label{sec:appendix_section}

\subsection{\textit{f--k} migration implementation}

The following algorithms show the pseudocode of the original \fk\ implementation (\aref{alg:original_fk}) and our version (\aref{alg:f_k}). We mainly use them to highlight the dataflow difference. The reference pipeline allocates separate arrays for scaling, padding, Fourier shifts, and Stolt remapping. Our CUDA path keeps the input volume and uses two complex working buffers of size $(2N_x, 2N_y, 2N_z)$ for the FFT-domain stages. The initial time-bin scaling is fused with padding, cyclic shifts are handled inside the kernels, and Stolt remapping is evaluated on the fly instead of being stored as an auxiliary coordinate volume. Finally, after the inverse FFT we take the maximum magnitude over depth, rather than the squared amplitude of Equation~2 of the main paper, which in our experiments improves the tone mapping of the final NLOS image. 

\begin{algorithm}
\begin{tcolorbox}[mylisting]
\footnotesize
\begin{algorithmic}[1]
\Function{\fkb\_original}{$\Psi, N_x, N_y, N_z, d_{\max}, \mathrm{apt_{width}}$}
    \State $(X_w, Y_w, Z_w) \gets \mathrm{mgrid}[-N_x{:}N_x,\,-N_y{:}N_y,\,-N_z{:}N_z]$
    \State $(X_w, Y_w, Z_w) \gets (X_w/N_x,\; Y_w/N_y,\; Z_w/N_z)$
    \State $\mathcal{T} \gets \mathrm{tile}(\mathrm{linspace}(0,1,N_z),\,(N_x, N_y, N_z))$
    \Statex
    \State $\Psi \gets \Psi \odot \mathcal{T}$
    \Comment{Initial scaling and zero-padding}
    \State $\Psi' \gets \mathrm{zeros}(2N_x, 2N_y, 2N_z)$
    \State $\Psi'[{:}N_x,{:}N_y,{:}N_z] \gets \Psi$
    \Statex
    \State $\widehat{\Psi} \gets \mathcal{F}\{\mathrm{fftshift}(\Psi')\}$
    \Comment{Forward FFT}

    \State $s \gets N_x \cdot d_{\max} \,\big/\, \left(4N_z \cdot (\mathrm{apt_{width}}/2)\right)$
    \State $Z_w' \gets \sqrt{s^2(X_w^2 + Y_w^2) + Z_w^2}$
    \Comment{Stolt remapping}
    \State $\Psi' \gets \mathrm{interpn}(X_w,Y_w,Z_w,\;\widehat{\Psi},\;(X_w,Y_w,Z_w'))$

    \State $\widehat{\Psi} \gets \Psi' \odot \mathbf{1}_{Z_w>0}$
    \Comment{Spectral filtering/compensation}
    \State $\widehat{\Psi} \gets \widehat{\Psi} \odot \frac{|Z_w|}{\max(Z_w')}$

    \State $\Psi \gets \mathrm{ifftshift}(\mathcal{F}^{-1}\{\widehat{\Psi}\})$
    \Comment{Inverse FFT}
    \State $f(\xv) \gets \mathrm{max\_z}(|\Psi^2|)$
    \Statex     
    \State \Return $f(\xv)$
    \Comment{Final reconstruction}
\EndFunction
\end{algorithmic}
\normalsize
\end{tcolorbox}
\caption{\textbf{\fk\ \citep{lindell_wave-based_2019}.} Reconstruction pipeline following prior work. Pseudocode shown in a Python-like style. $N_x$, $N_y$, and $N_z$ denote the volume dimensions. $d_{\max}$ refers to the maximum distance, whereas $\mathrm{apt_{width}}$ is the physical size in meters of the scanned relay wall.}
\label{alg:original_fk}
\end{algorithm}

\begin{algorithm}
\begin{tcolorbox}[mylisting]
\footnotesize
\begin{algorithmic}[1]
\Function{\textit{our}\_\fkb{}}{$\Psi,N_x,N_y,N_z,d_{\max},\mathrm{apt_{width}}$}
    \State $\widehat{\Psi}(\cdot, \cdot, \cdot) \gets 0$
    \State $\widehat{\Psi} \gets \mathrm{scale\_pad\_fftshift}(\Psi)$
    \Comment{Time-bin scaling fused with padding}
    \State $\widehat{\Psi} \gets \mathcal{F}\{\widehat{\Psi}\}$
    \State $s \gets N_x \cdot d_{\max}/\left(4N_z \cdot \mathrm{apt_{width}}/2\right)$
    \State $\Psi' \gets \mathrm{stolt}(\widehat{\Psi}, s)$
    \Comment{On-the-fly remapping}
    \State $\widehat{\Psi} \gets \mathcal{F}^{-1}\{\Psi'\}$
    \State $f(\xv) \gets \mathrm{ifftshift\_max\_magnitude}(\widehat{\Psi})$
    \State \Return $f(\xv)$
\EndFunction
\end{algorithmic}
\normalsize
\end{tcolorbox}
\caption{Our CUDA-based \textbf{\fk} pipeline. }
\label{alg:f_k}
\end{algorithm}

\subsection{\Rsd{} implementation}

Similarly to the previous section, Algorithm~\ref{alg:original_rsd} presents the reference \rsd\ implementation used by \cite{nam_low-latency_2021}. In the reference implementation, the initial Fourier transforms are not batched: although they could be executed as a group, they are instead handled sequentially. The convolved volume
$\mathcal{C}$ is also processed sequentially, weighting each slice $\mathcal{C}(\cdot, \cdot, c)$ by $w_c$. Moreover, the Fourier-transformed matrices are not padded, which reduces memory consumption at the expense of introducing artifacts.


\begin{algorithm}
\begin{tcolorbox}[mylisting]
\footnotesize
\begin{algorithmic}[1]
\Function{\textit{phasor\_fields}\_\textit{nam}}{$P, K, N_x, N_y, N_f, d_{\min}, d_{\max}, \Delta d, w$}
    \For{$c \gets 1$ \textbf{to} $N_f$}
        \Comment{2D forward FFTs}
        \State $\widehat{P}_c \gets \mathcal{F}_{2D}\{P(\cdot,\cdot,c)\}$
    \EndFor

    \State $i \gets 0$
    \State $P(\cdot, \cdot, \cdot) \gets 0$
    \For{$d \gets d_{\min}$ \textbf{to} $d_{\max}$, \textbf{step} $\Delta d$}
        \State $\mathcal{C} \gets \widehat{P} \odot K(\cdot, \cdot, \cdot, i)$
        \Comment{Convolution over the whole frequency stack}
        \For{$c \gets 1$ \textbf{to} $N_f$}
            \State $P(\cdot, \cdot, i) \gets P(\cdot, \cdot, i) + w_c\mathcal{C}(\cdot, \cdot, c)$
        \EndFor
        \State $P(\cdot, \cdot, i) \gets \mathcal{F}^{-1}_{2D}\{P(\cdot, \cdot, i)\}$
        \Comment{Inverse FFT}
        \State $\widehat{P}(\cdot, \cdot, i) \gets |P(\cdot, \cdot, i)|$
        \Comment{Magnitude at depth $i$}
        \State $i \gets i + 1$
    \EndFor
    \State $f(\xv) \gets \mathrm{max\_z}(\widehat{P})$
    \State \Return $f(\xv)$
\EndFunction
\end{algorithmic}
\normalsize
\end{tcolorbox}
\caption{\textbf{\Rsd{} as implemented by \cite{nam_low-latency_2021}.} Reconstruction pipeline following prior work. $P$ holds the sampled phasor field $\hat{\mathcal{P}}_\omega$ and $K$ the dense propagation kernel, both as defined in Section~4.3 of the main paper; $\widehat{P}$ denotes the stack of transformed slices $\widehat{P}_c$ produced by the first loop. $N_x$, $N_y$, and $N_f$ denote the relay-wall spatial dimensions and the number of temporal-frequency slices; $d_{\min}$, $d_{\max}$, and ${\Delta}d$ define the propagated-depth range, and $w$ weights each frequency slice.}
\label{alg:original_rsd}
\end{algorithm}


\subsubsection{Our improved ring-and-radius construction}
\label{sec:appendix:symmetry_rsd}

Our symmetry-aware \rsd{} variant replaces the dense Cartesian propagation kernel used in streamed reconstructions with a compact radial representation.
We store the radial kernel as a 3D array indexed by radial Fourier bin $\rho$, temporal frequency $f$, and depth $d$. A precomputed lookup table maps each Cartesian Fourier pixel to its continuous radius. At runtime, the GPU linearly interpolates this radial representation and accumulates all temporal frequencies locally for each $(u,v,d)$ output voxel.

We evaluate the analytic Bessel function representation only during precomputation. For a circular ring of spatial radius $r$, the angular component of its 2D Fourier transform depends solely on the Fourier radius $\rho$, reducing to the zeroth-order Hankel term:
\begin{equation}
    B(r,\rho) = 2\pi r\,J_0(2\pi r\rho).
\end{equation}
This expression is exact, as noted in Section~4.3 of the main paper. Because a unit ring is uniform across the angular coordinate, all higher-order harmonics ($n \neq 0$) integrate to zero. This formulation is exact and involves no series truncation. The only approximations are in the finite trapezoidal quadrature integration over sampled radii, the high-frequency radial cutoff described below, and enforcing radial symmetry on the discrete, rectangular relay-wall grid. During precomputation, we evaluate this basis once over a grid of spatial and Fourier radii. The pipeline then multiplies these real-valued basis samples by the complex propagation phase for each $(f,d)$ pair, integrates across the sampled spatial rings, and stores the resulting radial kernel $K_r(\rho,f,d)$.

The sampling is chosen in two stages, both derived from the capture geometry. Let $\Delta x$ and $\Delta y$ be the relay-wall sample spacings along $x$ and $y$, and let $N_u$ and $N_v$ be the padded FFT grid extents along $y$ and $x$ respectively, so that $N_v \Delta x$ and $N_u \Delta y$ are the physical side lengths of the padded aperture. The largest spatial radius covered by that grid and the radial Nyquist limit of the Cartesian spectrum are then:
\begin{equation}
\begin{aligned}
    r_{\max} &= \sqrt{\left(\tfrac{1}{2}N_v \Delta x\right)^2 + \left(\tfrac{1}{2}N_u \Delta y\right)^2}, \\[2pt]
    \rho_{\max} &= \sqrt{\left(\tfrac{1}{2\Delta x}\right)^2 + \left(\tfrac{1}{2\Delta y}\right)^2}.
\end{aligned}
\label{eq:supp_radial_extent}
\end{equation}

First, spatial radii are sampled uniformly over $[0, r_{\max}]$. To resolve the phase oscillations of the \rsd{} kernel $A(r,d,f)$ (Equation~4 of the main paper), we differentiate its phase to obtain the instantaneous radial frequency $\sigma_s r / (\lambda \sqrt{r^2+d^2})$. This quantity is upper-bounded by $\sigma_s/\lambda_{\min}$ cycles per meter, where $\lambda_{\min}$ is the shortest reconstructed wavelength and $\sigma_s$ is the modality factor ($2$ for confocal, $1$ for non-confocal). Rounding this Nyquist sampling limit up to the nearest power of two determines the number of rings and their uniform spacing:
%
\begin{equation}
    N_r = 2^{\left\lceil \log_2\left( \left\lceil 2 \sigma_s r_{\max}/\lambda_{\min} \right\rceil + 1 \right) \right\rceil},
    \quad
    \Delta r = \frac{r_{\max}}{N_r - 1},
\label{eq:supp_ring_count}
\end{equation}
floored at $N_r = 2$. The radial integral of Equation~6 of the main paper uses trapezoidal weights over these samples, which for uniform spacing are:
\begin{equation}
    q_0 = q_{N_r-1} = \tfrac{1}{2}\Delta r,
    \qquad
    q_j = \Delta r \quad (0 < j < N_r-1).
\label{eq:supp_trapezoid}
\end{equation}

Second, the radial spectral domain spans $[0, \rho_{\max}]$, using the same Nyquist criterion to similarly determine the sample count:
\begin{equation}
    N_\rho = 2^{\left\lceil \log_2\left( \left\lceil 2 r_{\max} \rho_{\max} \right\rceil + 1 \right) \right\rceil}.
\label{eq:supp_rho_count}
\end{equation}
Because discrete spatial rings cannot resolve arbitrarily high spatial frequencies, the ring sampling rate $f_s = (N_r-1)/r_{\max}$ caps the usable bandwidth at:
\begin{equation}
    \rho_{\mathrm{cutoff}} = \min\!\left(\rho_{\max},\; \tfrac{1}{2} f_s\right).
\label{eq:supp_cutoff}
\end{equation}
Samples beyond $\rho_{\mathrm{cutoff}}$ are set to zero during precomputation. The runtime lookup works in the discrete bin coordinate of the stored axis:
\begin{equation}
    \tilde{\rho} = \rho\,\frac{N_\rho - 1}{\rho_{\max}}.
\label{eq:supp_bin_coord}
\end{equation}
To prevent boundary artifacts, we apply a smoothstep taper over the final $\Delta_{\mathrm{taper}} = 8$ bins below this cutoff:
\begin{equation}
    a_{\mathrm{cutoff}}(\tilde{\rho}) =
    \begin{cases}
        1, & \tilde{\rho} \le \tilde{\rho}_{\mathrm{cutoff}} - \Delta_{\mathrm{taper}},\\[2pt]
        t^2\,(3 - 2t), & \tilde{\rho}_{\mathrm{cutoff}} - \Delta_{\mathrm{taper}} < \tilde{\rho} < \tilde{\rho}_{\mathrm{cutoff}},\\[2pt]
        0, & \tilde{\rho} \ge \tilde{\rho}_{\mathrm{cutoff}},
    \end{cases}
\label{eq:supp_taper}
\end{equation}
with $t = (\tilde{\rho}_{\mathrm{cutoff}} - \tilde{\rho})/\Delta_{\mathrm{taper}}$, where $\tilde{\rho}_{\mathrm{cutoff}}$ is $\rho_{\mathrm{cutoff}}$ expressed in the same bin coordinate.
For the $190\times190$ non-confocal capture of \citet{nam_low-latency_2021} used throughout this appendix, where $\sigma_s = 1$, these expressions give $N_r = 128$ rings (from a Nyquist minimum of $92$) and $N_\rho = 256$ stored radii (from a minimum of $191$), with $\rho_{\max} = \SI{70.34}{\per\meter}$ and $\rho_{\mathrm{cutoff}} = \SI{47.02}{\per\meter}$. A confocal capture of the same geometry doubles the ring count through $\sigma_s = 2$.

\begin{figure}
    \centering
    \resizebox{0.42\textwidth}{!}{%
        \begin{tikzpicture}[
    >=Latex,
    font=\small,
    node distance=0.48cm,
    inputnode/.style={
        draw,
        rounded corners=2pt,
        fill=gray!8,
        minimum width=0.42\textwidth,
        text width=0.38\textwidth,
        minimum height=1.02cm,
        align=center,
        inner sep=3pt
    },
    storednode/.style={
        draw,
        rounded corners=2pt,
        fill=blue!10,
        minimum width=0.42\textwidth,
        text width=0.38\textwidth,
        minimum height=1.25cm,
        align=center,
        inner sep=3pt
    },
    opnode/.style={
        draw,
        rounded corners=2pt,
        fill=blue!22,
        minimum width=0.42\textwidth,
        text width=0.38\textwidth,
        minimum height=1.25cm,
        align=center,
        inner sep=3pt
    },
    line/.style={->, thick}
]

\node[inputnode] (o_input) at (0,-1.35)
    {Sampled spatial radii $r$\\[-1pt]
    {\scriptsize uniform spacing, trapezoidal weights, sized to the shortest wavelength used}};

\node[opnode, below=of o_input] (o_basis)
    {Unit ring basis\\[-1pt]
    $B(r,\rho)=2\pi r\,J_0(2\pi r\rho)$\\[-1pt]
    {\scriptsize evaluated over spatial radius $r$ and Fourier radius $\rho$}};

\node[opnode, below=of o_basis] (o_integrate)
    {Multiply by propagation phase $(f,d)$\\[-1pt]
    and integrate over rings};

\node[storednode, below=of o_integrate] (o_output)
    {Stored radial kernel $K_r(\rho,f,d)$\\[-1pt]
    {\scriptsize size $\rho \times f \times d$; samples beyond the radial cutoff are zero}};

\draw[line] ($(o_input.south)+(0,-0.05)$) -- ($(o_basis.north)+(0,0.05)$);
\draw[line] ($(o_basis.south)+(0,-0.05)$) -- ($(o_integrate.north)+(0,0.05)$);
\draw[line] ($(o_integrate.south)+(0,-0.05)$) -- ($(o_output.north)+(0,0.05)$);

\end{tikzpicture}
    }
    \caption{Offline precomputation of our symmetry-aware \rsd{} radial kernel. Spatial radii are sampled and turned into a Bessel-function ring basis $B(r,\rho)$, which is multiplied by the propagation phase and integrated over rings to form the stored radial kernel $K_r(\rho,f,d)$; both are defined in Section~4.3 of the main paper. This stage runs once, ahead of runtime. \sref{sec:appendix:symmetry_rsd} describes how the runtime lookup then samples this kernel per pixel.}
    \label{fig:rsd_ring_basis}
\end{figure}
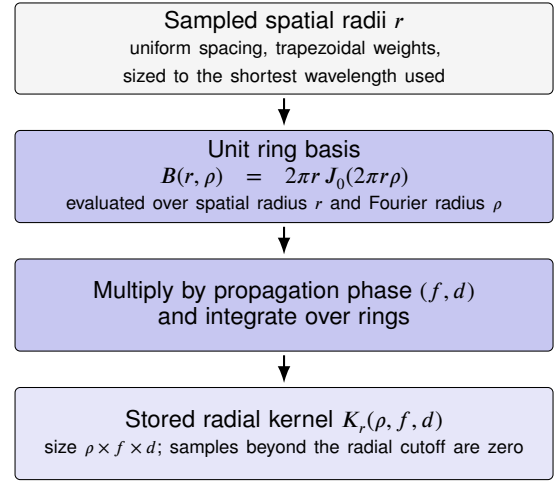

At runtime, each Cartesian FFT pixel reads a precomputed continuous radius from the pixel-to-radius map, applies the taper weight of \eref{eq:supp_taper}, and linearly interpolates between the two neighboring $\rho$ bins before accumulating over temporal frequencies. \fref{fig:rsd_ring_basis} summarizes this offline precomputation phase. 

\begin{algorithm}
\begin{tcolorbox}[mylisting]
\footnotesize
\begin{algorithmic}[1]
\Function{\textit{ours}\_\textit{phasor\_fields}}{$P, K_r, R, N_x, N_y, N_d, N_f, w$}
    \State $\widehat{P} \gets \mathcal{F}\{\mathrm{pad}_{\mathrm{tl}}(P)\}$
    \Comment{Top-left padding and batched 2D FFTs}
    \ForAll{$(u,v,d)$ \textbf{in parallel}}
        \State $\mathcal{C}(u,v,d) \gets 0$
        \State $\rho \gets R(u,v)$
        \For{$f \gets 1$ \textbf{to} $N_f$}
            \State $a \gets a_{\mathrm{cutoff}}(\tilde{\rho})$
            \Comment{Taper of \eref{eq:supp_taper}}
            \State $h \gets a \cdot \mathrm{linear\_sample}(K_r(\cdot,f,d), \rho)$
            \State $\mathcal{C}(u,v,d) \gets \mathcal{C}(u,v,d) + w(f)\,\widehat{P}(u,v,f)\,h$
        \EndFor
    \EndFor
    \State $\widehat{P} \gets \mathcal{F}^{-1}\{\mathcal{C}\}$
    \Comment{Batched inverse FFTs over depth}
    \State $P' \gets |\widehat{P}|$
    \State $f(\xv) \gets \mathrm{max\_z}(P')$
    \State \Return $f(\xv)$
\EndFunction
\end{algorithmic}
\normalsize
\end{tcolorbox}
\caption{Our symmetry-aware \textbf{\rsd} dataflow. The dense kernel $K$ from Algorithm~\ref{alg:original_rsd} is replaced by a radial kernel $K_r$ and a pixel-to-radius map $R$.}
\label{alg:symmetry_rsd}
\end{algorithm}

Algorithm~4 details the runtime execution loop for this streamed radial \rsd{} kernel. For each temporal frequency, $a_{\mathrm{cutoff}}$ evaluates the low-pass taper (\eref{eq:supp_taper}) near the radial sampling boundary, while $\texttt{linear\_sample}$ interpolates between adjacent radial bins. The taper factor $a$ operates independently of the frequency weights $w(f)$, which are applied directly during local accumulation.  By bypassing the generation of dense 2D kernel slices at runtime, this approach substantially reduces VRAM consumption and memory bandwidth. 



\subsubsection{Offline \rsd{} simplification}

This subsection discusses our separate offline \rsd{} implementation used for the preprocessed experiments, different from our streamed pipeline from Section~4.3.
We use as a baseline the implementation by \citet{lindell_wave-based_2019}, which is slightly different from that of \citet{nam_low-latency_2021} and follows the algorithm of \citet{liu_non-line--sight_2019}. We use this baseline as reference for all our offline reconstruction experiments.

The first memory issue we addressed was the convolution of the propagation kernel $K$ of Section~4.3 of the main paper, which acts here as the point spread function of the plane-to-plane propagation, with the signal's phasor representation. The original implementation allocates two real-valued buffers for the phasor representation, $P_{\mathrm{cos}}$ and $P_{\mathrm{sin}}$, which hold the cosine and sine components of the phasor field $\hat{\mathcal{P}}_\omega$ of the main paper. In this formulation the two buffers are treated as if they were independent signals: each one is Fourier-transformed, multiplied by $K$ in the frequency domain, and then inversely transformed, and the final complex result is obtained by combining both outputs.

However, these two components are in fact the real and imaginary parts of a single complex phasor. Since the Fourier transform, the pointwise multiplication by $K$, and the inverse transform are all linear operations, applying them separately to $P_{\mathrm{cos}}$ and $P_{\mathrm{sin}}$ is equivalent to applying them once to their complex combination. Therefore, we found that the following two expressions are equivalent:
\begin{align}
    P' &= \mathcal{F}^{-1}\{\widehat{P}_{\mathrm{cos}} \cdot K\}
      + i\,\mathcal{F}^{-1}\{\widehat{P}_{\mathrm{sin}} \cdot K\} \\
        &= \mathcal{F}^{-1}\{(\widehat{P}_{\mathrm{cos}} + i\,\widehat{P}_{\mathrm{sin}})\cdot K\}
\end{align}
with $\widehat{P}_{\mathrm{cos}} = \mathcal{F}\{P_{\mathrm{cos}}\}$ and $\widehat{P}_{\mathrm{sin}} = \mathcal{F}\{P_{\mathrm{sin}}\}$.

Additionally, another time-consuming step is the construction of the transform operator that maps phasor data from a Cartesian layout to a ring-based layout. In the original implementation, this operator is built as a sparse matrix of size $M^2 \times M$ with $M \gets \max(x,y)$ and $M^2$ non-zero entries. We index samples from zero throughout, $i = 0,\dots,M^2-1$, so that the one-based radial index of sample $i$ is $\lceil\sqrt{i+1}\rceil$. The reference implementation is written in one-based MATLAB indexing, where the same quantity reads $\lceil\sqrt{i}\rceil$. Each sample contributes a value of $1$ at position $(i,\lceil\sqrt{i+1}\rceil)$, and its row is then scaled by $1/\sqrt{i+1}$, which accounts for the number of samples that fall on the same ring. The matrix is then iteratively collapsed over $\log_2(\max(x,y))$ iterations; in each iteration, pairs of non-contiguous rows are averaged. After $\log_2 M$ iterations, the matrix is reduced to size $M \times M$, and the accumulated weights have been scaled by a further factor of $(\tfrac{1}{2})^{\log_2 M} = \tfrac{1}{M}$.

Composing the two scalings, this hierarchical construction is equivalent to directly creating an $M \times M$ matrix and mapping each sample index $i$ to its corresponding ring index $\lceil\sqrt{i+1}\rceil$, with a final weight of $\tfrac{1}{M\sqrt{i+1}}$: the $1/\sqrt{i+1}$ comes from the per-row scaling and the $1/M$ from the repeated averaging. In other words, the hierarchical averaging performed in the original implementation can be collapsed into a single kernel. Constructing the transform operators as in the original formulation required building large sparse matrices (e.g., using the Eigen library) and repeatedly collapsing them on the CPU. The simplified construction removes this sparse-matrix setup from the critical path and makes the offline implementation easier to run on the GPU.

Finally, note that it is not necessary to construct the inverse operator explicitly; the kernel can access it by simply using transposed indices of the forward transform operator. In our offline implementation, the forward and backward matrix multiplications are solved with the \texttt{cuBLAS} module on top of CUDA.

\section{Additional results and benchmarks}

We provide four additional experiments to support our main findings. First, a breakdown of the dynamic capture from \citet{nam_low-latency_2021} across different configurations (specifically varying the algorithm, precision, and padding). Second, more per-frame error maps (extending Section 5.6, which only shows the mean absolute difference). Third, a sweep of our Mixture of Experts' (MoE) pooling weight $\beta$. Finally, the complete offline benchmark tables expanding on Section 5.8.

\subsection{Dynamic reconstruction across configurations}

\begin{figure*}[t]
\centering

\begin{subfigure}[t]{0.98\textwidth}
    \centering
    \includegraphics[width=\linewidth]{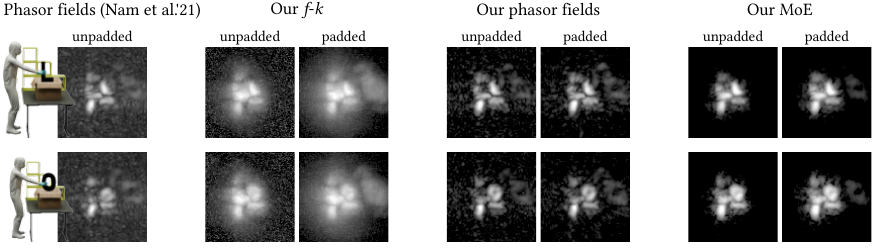}
    \caption{Qualitative comparison of reconstruction results. For visualization only, the displayed intensities are gamma-corrected ($\gamma=2$) to make low-intensity background noise more visible.}
    \label{fig:perf-comparison-qualitative}
\end{subfigure}

\vspace{0.8em}

\noindent
\begin{subfigure}[t]{0.6\textwidth}
    \vspace{0pt}
    \centering
    \includegraphics[width=\linewidth]{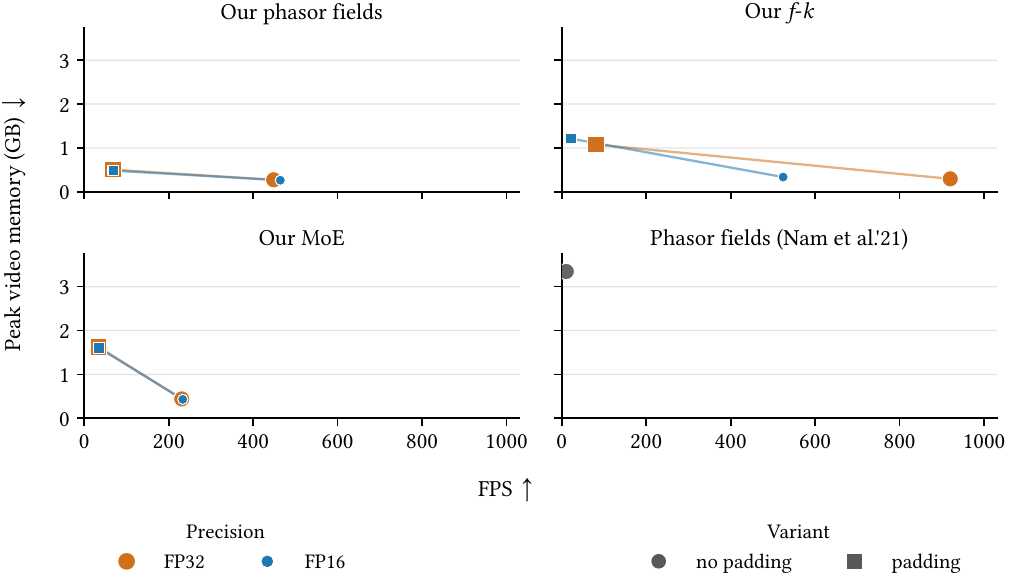}
    \caption{Response time and memory trade-off. Markers show FPS and peak VRAM.}
    \label{fig:perf-comparison-plot}
\end{subfigure}%
\hfill%
\begin{subfigure}[t]{0.38\textwidth}
\vspace{0pt}
\centering

\tiny

\resizebox{\linewidth}{!}{%
\begin{tabular}{l | c c r r}
\toprule
Method & Prec. & Padding & $\uparrow$ FPS & $\downarrow$ GB \\
\midrule
\multirow{4}{*}{\textcolor{symgreen}{Our phasor fields}}%
& \multirow{2}{*}{\texttt{FP16}} & no & 464.34 & \good{\strut \textbf{0.26}} \\
&  & yes & 68.36 & 0.48 \\
& \multirow{2}{*}{\texttt{FP32}} & no & 448.48 & 0.27 \\
&  & yes & 67.89 & 0.50 \\
\midrule
\multirow{4}{*}{\textcolor{fkblue}{Our \textit{\fkb}}}%
& \multirow{2}{*}{\texttt{FP16}} & no & 524.16 & 0.34 \\
&  & yes & 22.00 & 1.21 \\
& \multirow{2}{*}{\texttt{FP32}} & no & \good{\strut \textbf{919.85}} & 0.30 \\
&  & yes & 81.42 & 1.08 \\
\midrule
\multirow{4}{*}{\textcolor{emorange}{Our MoE}}%
& \multirow{2}{*}{\texttt{FP16}} & no & 233.31 & 0.43 \\
&  & yes & 34.98 & 1.60 \\
& \multirow{2}{*}{\texttt{FP32}} & no & 230.90 & 0.44 \\
&  & yes & 34.21 & 1.62 \\
\midrule
\multirow{1}{*}{\textcolor{refgray}{Phasor fields (Nam et al.'21)}}%
& \texttt{FP32} & no & \bad{\strut 10.69} & \bad{\strut 3.34} \\
\bottomrule
\end{tabular}
}

\vspace{0.35em}

\tiny
FPS: from mean frame time. GB: peak VRAM.\\
Pad.: padded reconstruction volume.

\caption{Numerical runtime and memory summary.}
\label{fig:perf-comparison-table}

\end{subfigure}

\caption{Per-configuration real-time comparison on the \texttt{nlosbox1} dynamic capture. The top row shows a representative reconstruction, while the bottom row summarizes the runtime-memory trade-off across methods. Frame rates are derived from mean frame times, as in Table~2 of the main paper.}
\label{fig:performance-comparison-combined}
\end{figure*}

\fref{fig:performance-comparison-combined} breaks the \texttt{nlosbox1} capture of Section~5.1 of the main paper down across every configuration. The comparison includes \fk{}, our \rsd{} and MoE, each with and without padding and at both precisions, together with the \rsd{} implementation of \citet{nam_low-latency_2021}, which does not offers a half-precision path nor a padded reconstruction volume and therefore appears as a single configuration. Frame rates here are derived from mean frame times, the same statistic as Table~2 of the main paper, so the configurations the two share report identical values. As in the main results, unpadded \fk{} gives the highest frame rate, but it is also the configuration most exposed to padding artifacts. Our \rsd{} produces sharper letters in this sequence than \fk{} while using far less GPU memory than the reference implementation.

The useful reading of this figure is the speed-memory frontier. Unpadded \fk{} sits at the high-FPS end, our \rsd{} keeps most of the visual benefit in the low-memory corner, and the reference implementation sits an order of magnitude away on both axes. MoE is therefore better interpreted as a visual-combination option than as the best point on the performance frontier. One configuration runs against the trend: \fk{} in \textsc{fp16} is both slower and slightly larger than its \textsc{fp32} counterpart, so half precision is not a useful operating point for that branch.

\subsection{Additional image-space quality}
\label{sec:appendix:quality}

\begin{figure}
    \centering
    \includegraphics[width=\linewidth]{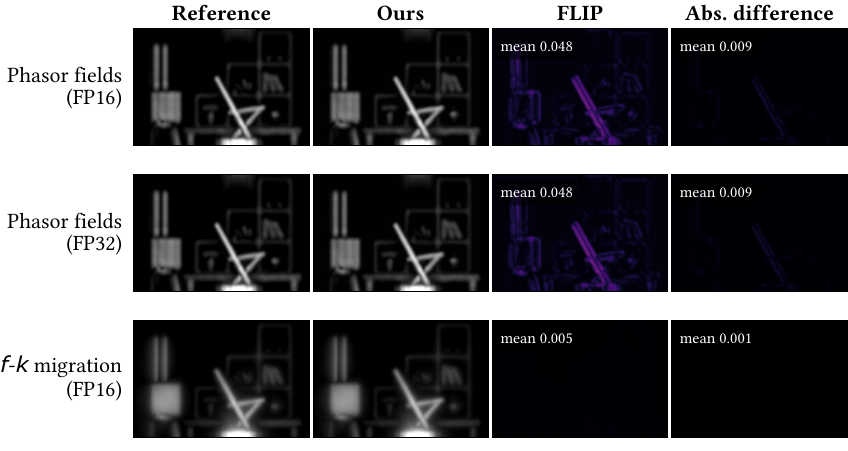}
    \caption{Image-space error on the office scene. Each row compares our reconstruction with its reference and reports FLIP and absolute error maps. The reference for the \rsd{} rows is \cite{nam_low-latency_2021}'s \rsd{} implementation, so the comparison isolates the radial kernel representation. For \fkb{}, the reference is \fkb{} using \texttt{FP32}.}
    \label{fig:image_error_office}
\end{figure}

Section~5.6 of the main paper reports the sequence-wide mean absolute difference. Here, we show the per-frame error maps, together with the corresponding FLIP errors. \fref{fig:image_error_office} shows the office scene, which contains several structures at different depths and is therefore useful for stress testing, although that also makes local error patterns harder to read.

The \texttt{FP16} and \texttt{FP32} rows are indistinguishable, so the residual difference is the radial kernel representation rather than the lack of precision from \texttt{FP16}. Our \rsd{} reaches a peak signal-to-noise ratio of \SI{37.0}{\decibel} against the dense-kernel reference, and \fkb{} reaches \SI{57.0}{\decibel} against its \texttt{FP32} reference.

\subsection{Pooling weight of the mixture}
\label{sec:appendix:beta}

\tref{tab:beta_sweep} sweeps the pooling weight $\beta$ (Equation~10) on the \texttt{nlosbox1} capture using the post-processing setup from Section~5.7 of the main paper. As $\beta$ increases, background noise decreases exponentially by roughly $2.3\times$ per step, whereas object intensity drops linearly. Image sharpness increases slightly over the same range.


Because the response is monotonic without a distinct inflection point, $\beta$ controls a direct tradeoff between background suppression and object signal retention. We select $\beta = 0.3$ across all experiments: this setting reduces background noise by three orders of magnitude relative to either standalone branch while preserving 73\% of the \fkb{} object intensity, reproducing the MoE row in Table~6 of the main paper. Higher weights yield additional background suppression at the expense of object visibility: at $\beta = 1$, where pooling simplifies to a plain product, object intensity drops to 36\% of the \fkb{} baseline.


\begin{table}
\centering
\caption{Impact of the pooling weight $\beta$ on the 100-frame \texttt{nlosbox1} sequence (FP32, unpadded). Because every metric changes monotonically with $\beta$, there is no single optimal setting: $\beta = 0.5$ provides the strongest background rejection but also the weakest object signal. Bold text highlights our chosen operating point ($\beta = 0.3$), which matches the MoE row in Table~6 of the main paper.}
\label{tab:beta_sweep}
\resizebox{\linewidth}{!}{
\begin{tabular}{|l|r|r|r|r|}
\toprule
$\beta$ & $\uparrow$ Object & $\downarrow$ Background & $\uparrow$ Contrast & $\uparrow$ Sharpness\\
\cmidrule{1-5}
0.15 & 0.520 & $8.7\cdot10^{-5}$ & 6,008 & 0.050\\
0.20 & 0.495 & $3.8\cdot10^{-5}$ & 12,993 & 0.051\\
0.25 & 0.471 & $1.7\cdot10^{-5}$ & 28,186 & 0.052\\
\textbf{0.30} & \textbf{0.449} & \textbf{$7.4\cdot10^{-6}$} & \textbf{60,981} & \textbf{0.053}\\
0.35 & 0.428 & $3.2\cdot10^{-6}$ & 132,666 & 0.054\\
0.40 & 0.408 & $1.4\cdot10^{-6}$ & 291,566 & 0.055\\
0.50 & 0.372 & $2.5\cdot10^{-7}$ & 1,478,556 & 0.056\\
\bottomrule
\end{tabular}
}
\end{table}

\subsection{Offline reconstruction}

The remaining tables give the full offline reconstruction timings. They cover several spatial and temporal resolutions, starting from the original dataset sizes and then downsampling the spatial or temporal dimensions. The \fk{} datasets from \citet{lindell_wave-based_2019} are $512^3$, while the largest Zaragoza datasets \citep{galindo_dataset_2019} are $256\times256\times4096$. We limit the temporal dimension to 2048 bins in these experiments because 4096 bins put substantial memory pressure on both the original and optimized implementations.

Besides reconstruction time, we report throughput in millions of voxels per second. For the exhaustive captures, the dimensions column lists the original measurement grid. Normal moveout correction turns it into $31\times31\times T$ confocal data, which is the format used by the throughput figures. The last two columns report the maximum observed memory footprint of each implementation's main buffers. We label the metric \emph{tracked memory} rather than VRAM because the two sides measure different resources: the original MATLAB implementations allocate their main buffers in CPU memory, whereas for our implementations the reported value is the peak of a process-scoped counter over live CUDA allocations, which counts pinned host staging buffers alongside device memory. Since that counter is not restricted to dedicated video memory, the largest configurations report totals above the 16~GB of our GPU.

Across these tables, the largest speed-ups occur when the original implementations allocate and transform large intermediate volumes. The optimized versions are still affected by setup overhead at very small sizes and by memory traffic at the largest sizes, but the throughput remains high across confocal, simulated, and exhaustive inputs. This supports our claim from the main paper that the speed-up comes from dataflow and layout changes rather than from a dataset-specific shortcut.

\begin{table*}
\scriptsize
\centering
\caption{Offline performance and memory comparison between our accelerated \fkb{} and phasor-fields implementations and their original MATLAB counterparts, across every measured resolution. Throughput is the reconstructed volume divided by reconstruction time. For exhaustive captures, the dimensions column lists the original measurement grid before normal moveout correction, which produces a $31\times31\times T$ confocal transient volume. Max. tracked memory reports the peak footprint of each implementation's main buffers: host RAM for the original MATLAB code, and the peak of all tracked CUDA allocations for ours, counting device buffers and pinned host staging together (see the discussion preceding these tables). Bold marks our value when it beats the baseline.}
\label{table:offline_table_1}
\begin{tabular}{|l|l|l|l|l|l|l|l|l|}
\toprule
& & & \multicolumn{2}{|c|}{$\downarrow$ Time (\si{\second})} & \multicolumn{2}{c|}{$\uparrow$ Throughput (Mvox/\si{\second})} & \multicolumn{2}{c}{$\downarrow$ Max. tracked memory (\si{\giga\byte})}\\
\cmidrule{4-9}
Dataset & Dimensions & Algorithm & Ours & Original & Ours & Original & Ours & Original \\
\cmidrule{1-9}
\multicolumn{9}{|c|}{Confocal -- \fk\ dataset \citep{lindell_wave-based_2019}}\\
\cmidrule{1-9}
\multirow{18}{*}{\parbox{1.8cm}{\centering bike\\[3pt]\includegraphics[width=1cm]{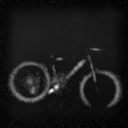}}} & \multirow{2}{*}{512$\times$512$\times$512} & \fk & \textbf{2.114 $\pm$ 0.03} & 146.624 $\pm$ 8.47 & \textbf{63.50} & 0.92 & \textbf{17.000} & 52.430 \\
&  & Phasor fields & \textbf{3.791 $\pm$ 0.03} & 30.872 $\pm$ 0.67 & \textbf{35.40} & 4.35 & \textbf{22.001} & 33.579 \\
\cmidrule{2-9}
& \multirow{2}{*}{512$\times$512$\times$256} & \fk & \textbf{0.282 $\pm$ 0.00} & 20.815 $\pm$ 1.51 & \textbf{237.72} & 3.22 & \textbf{8.500} & 26.214 \\
&  & Phasor fields & \textbf{0.425 $\pm$ 0.00} & 14.793 $\pm$ 0.35 & \textbf{157.83} & 4.54 & \textbf{11.000} & 16.778 \\
\cmidrule{2-9}
& \multirow{2}{*}{256$\times$256$\times$512} & \fk & \textbf{0.143 $\pm$ 0.00} & 10.695 $\pm$ 0.66 & \textbf{234.41} & 3.14 & \textbf{4.250} & 13.107 \\
&  & Phasor fields & \textbf{0.217 $\pm$ 0.00} & 6.894 $\pm$ 0.20 & \textbf{154.81} & 4.87 & \textbf{5.501} & 8.393 \\
\cmidrule{2-9}
& \multirow{2}{*}{512$\times$512$\times$128} & \fk & \textbf{0.141 $\pm$ 0.00} & 10.364 $\pm$ 0.37 & \textbf{237.99} & 3.24 & \textbf{4.250} & 13.107 \\
&  & Phasor fields & \textbf{0.213 $\pm$ 0.00} & 7.537 $\pm$ 0.27 & \textbf{157.66} & 4.45 & \textbf{5.500} & 8.389 \\
\cmidrule{2-9}
& \multirow{2}{*}{256$\times$256$\times$256} & \fk & \textbf{0.073 $\pm$ 0.00} & 5.243 $\pm$ 0.14 & \textbf{230.17} & 3.20 & \textbf{2.125} & 6.554 \\
&  & Phasor fields & \textbf{0.110 $\pm$ 0.00} & 3.428 $\pm$ 0.15 & \textbf{152.22} & 4.89 & \textbf{2.750} & 4.194 \\
\cmidrule{2-9}
& \multirow{2}{*}{128$\times$128$\times$512} & \fk & \textbf{0.038 $\pm$ 0.00} & 2.552 $\pm$ 0.11 & \textbf{219.35} & 3.29 & \textbf{1.062} & 3.277 \\
&  & Phasor fields & \textbf{0.057 $\pm$ 0.00} & 1.607 $\pm$ 0.02 & \textbf{147.91} & 5.22 & \textbf{1.376} & 2.101 \\
\cmidrule{2-9}
& \multirow{2}{*}{256$\times$256$\times$128} & \fk & \textbf{0.038 $\pm$ 0.00} & 2.650 $\pm$ 0.18 & \textbf{222.97} & 3.17 & \textbf{1.062} & 3.277 \\
&  & Phasor fields & \textbf{0.057 $\pm$ 0.00} & 1.751 $\pm$ 0.06 & \textbf{146.93} & 4.79 & \textbf{1.375} & 2.097 \\
\cmidrule{2-9}
& \multirow{2}{*}{128$\times$128$\times$256} & \fk & \textbf{0.020 $\pm$ 0.00} & 1.253 $\pm$ 0.03 & \textbf{207.55} & 3.35 & \textbf{0.531} & 1.638 \\
&  & Phasor fields & \textbf{0.030 $\pm$ 0.00} & 0.801 $\pm$ 0.03 & \textbf{138.92} & 5.24 & \textbf{0.688} & 1.049 \\
\cmidrule{2-9}
& \multirow{2}{*}{128$\times$128$\times$128} & \fk & \textbf{0.012 $\pm$ 0.00} & 0.652 $\pm$ 0.02 & \textbf{181.34} & 3.22 & \textbf{0.266} & 0.819 \\
&  & Phasor fields & \textbf{0.016 $\pm$ 0.00} & 0.408 $\pm$ 0.02 & \textbf{128.48} & 5.14 & \textbf{0.344} & 0.524 \\
\cmidrule{1-9}
\multirow{18}{*}{\parbox{1.8cm}{\centering teaser\\[3pt]\includegraphics[width=1cm]{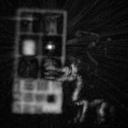}}} & \multirow{2}{*}{512$\times$512$\times$512} & \fk & \textbf{2.016 $\pm$ 0.01} & 154.730 $\pm$ 14.77 & \textbf{66.59} & 0.87 & \textbf{17.000} & 52.433 \\
&  & Phasor fields & \textbf{3.916 $\pm$ 0.08} & 31.458 $\pm$ 2.65 & \textbf{34.28} & 4.27 & \textbf{22.001} & 33.579 \\
\cmidrule{2-9}
& \multirow{2}{*}{512$\times$512$\times$256} & \fk & \textbf{0.280 $\pm$ 0.00} & 21.723 $\pm$ 0.88 & \textbf{239.68} & 3.09 & \textbf{8.500} & 26.214 \\
&  & Phasor fields & \textbf{0.427 $\pm$ 0.00} & 15.453 $\pm$ 0.38 & \textbf{157.32} & 4.34 & \textbf{11.000} & 16.777 \\
\cmidrule{2-9}
& \multirow{2}{*}{256$\times$256$\times$512} & \fk & \textbf{0.144 $\pm$ 0.00} & 11.462 $\pm$ 0.65 & \textbf{232.97} & 2.93 & \textbf{4.250} & 13.109 \\
&  & Phasor fields & \textbf{0.218 $\pm$ 0.00} & 7.155 $\pm$ 0.27 & \textbf{153.91} & 4.69 & \textbf{5.501} & 8.393 \\
\cmidrule{2-9}
& \multirow{2}{*}{512$\times$512$\times$128} & \fk & \textbf{0.142 $\pm$ 0.00} & 10.528 $\pm$ 0.08 & \textbf{236.05} & 3.19 & \textbf{4.250} & 13.107 \\
&  & Phasor fields & \textbf{0.212 $\pm$ 0.00} & 7.910 $\pm$ 0.17 & \textbf{158.02} & 4.24 & \textbf{5.500} & 8.389 \\
\cmidrule{2-9}
& \multirow{2}{*}{256$\times$256$\times$256} & \fk & \textbf{0.073 $\pm$ 0.00} & 5.597 $\pm$ 0.20 & \textbf{228.38} & 3.00 & \textbf{2.125} & 6.554 \\
&  & Phasor fields & \textbf{0.111 $\pm$ 0.00} & 3.568 $\pm$ 0.11 & \textbf{151.14} & 4.70 & \textbf{2.750} & 4.194 \\
\cmidrule{2-9}
& \multirow{2}{*}{128$\times$128$\times$512} & \fk & \textbf{0.039 $\pm$ 0.00} & 2.713 $\pm$ 0.21 & \textbf{217.19} & 3.09 & \textbf{1.062} & 3.277 \\
&  & Phasor fields & \textbf{0.057 $\pm$ 0.00} & 1.621 $\pm$ 0.06 & \textbf{146.98} & 5.17 & \textbf{1.376} & 2.101 \\
\cmidrule{2-9}
& \multirow{2}{*}{256$\times$256$\times$128} & \fk & \textbf{0.038 $\pm$ 0.00} & 2.857 $\pm$ 0.07 & \textbf{218.52} & 2.94 & \textbf{1.062} & 3.277 \\
&  & Phasor fields & \textbf{0.057 $\pm$ 0.00} & 1.917 $\pm$ 0.07 & \textbf{146.45} & 4.38 & \textbf{1.375} & 2.097 \\
\cmidrule{2-9}
& \multirow{2}{*}{128$\times$128$\times$256} & \fk & \textbf{0.021 $\pm$ 0.00} & 1.435 $\pm$ 0.18 & \textbf{200.98} & 2.92 & \textbf{0.531} & 1.638 \\
&  & Phasor fields & \textbf{0.031 $\pm$ 0.00} & 0.822 $\pm$ 0.06 & \textbf{136.51} & 5.10 & \textbf{0.688} & 1.049 \\
\cmidrule{2-9}
& \multirow{2}{*}{128$\times$128$\times$128} & \fk & \textbf{0.012 $\pm$ 0.00} & 0.672 $\pm$ 0.04 & \textbf{180.10} & 3.12 & \textbf{0.266} & 0.819 \\
&  & Phasor fields & \textbf{0.017 $\pm$ 0.00} & 0.409 $\pm$ 0.01 & \textbf{125.27} & 5.13 & \textbf{0.344} & 0.524 \\
\cmidrule{1-9}
\multirow{18}{*}{\parbox{1.8cm}{\centering statue\\[3pt]\includegraphics[width=1cm]{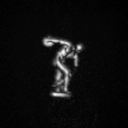}}} & \multirow{2}{*}{512$\times$512$\times$512} & \fk & \textbf{2.018 $\pm$ 0.01} & 152.567 $\pm$ 8.22 & \textbf{66.51} & 0.88 & \textbf{17.000} & 52.430 \\
&  & Phasor fields & \textbf{3.758 $\pm$ 0.02} & 32.195 $\pm$ 1.86 & \textbf{35.72} & 4.17 & \textbf{22.001} & 33.579 \\
\cmidrule{2-9}
& \multirow{2}{*}{512$\times$512$\times$256} & \fk & \textbf{0.280 $\pm$ 0.00} & 21.298 $\pm$ 0.59 & \textbf{239.83} & 3.15 & \textbf{8.500} & 26.214 \\
&  & Phasor fields & \textbf{0.427 $\pm$ 0.00} & 14.895 $\pm$ 0.40 & \textbf{157.20} & 4.51 & \textbf{11.000} & 16.778 \\
\cmidrule{2-9}
& \multirow{2}{*}{256$\times$256$\times$512} & \fk & \textbf{0.144 $\pm$ 0.00} & 11.118 $\pm$ 0.87 & \textbf{233.54} & 3.02 & \textbf{4.250} & 13.107 \\
&  & Phasor fields & \textbf{0.217 $\pm$ 0.00} & 7.063 $\pm$ 0.17 & \textbf{154.43} & 4.75 & \textbf{5.501} & 8.393 \\
\cmidrule{2-9}
& \multirow{2}{*}{512$\times$512$\times$128} & \fk & \textbf{0.141 $\pm$ 0.00} & 10.614 $\pm$ 0.21 & \textbf{237.14} & 3.16 & \textbf{4.250} & 13.107 \\
&  & Phasor fields & \textbf{0.213 $\pm$ 0.00} & 7.774 $\pm$ 0.14 & \textbf{157.64} & 4.32 & \textbf{5.500} & 8.389 \\
\cmidrule{2-9}
& \multirow{2}{*}{256$\times$256$\times$256} & \fk & \textbf{0.073 $\pm$ 0.00} & 5.340 $\pm$ 0.21 & \textbf{229.77} & 3.14 & \textbf{2.125} & 6.554 \\
&  & Phasor fields & \textbf{0.111 $\pm$ 0.00} & 3.476 $\pm$ 0.10 & \textbf{151.73} & 4.83 & \textbf{2.750} & 4.195 \\
\cmidrule{2-9}
& \multirow{2}{*}{128$\times$128$\times$512} & \fk & \textbf{0.038 $\pm$ 0.00} & 2.747 $\pm$ 0.16 & \textbf{221.81} & 3.05 & \textbf{1.062} & 3.277 \\
&  & Phasor fields & \textbf{0.057 $\pm$ 0.00} & 1.668 $\pm$ 0.04 & \textbf{147.39} & 5.03 & \textbf{1.376} & 2.101 \\
\cmidrule{2-9}
& \multirow{2}{*}{256$\times$256$\times$128} & \fk & \textbf{0.038 $\pm$ 0.00} & 2.856 $\pm$ 0.25 & \textbf{220.54} & 2.94 & \textbf{1.062} & 3.277 \\
&  & Phasor fields & \textbf{0.057 $\pm$ 0.00} & 1.820 $\pm$ 0.06 & \textbf{146.61} & 4.61 & \textbf{1.375} & 2.097 \\
\cmidrule{2-9}
& \multirow{2}{*}{128$\times$128$\times$256} & \fk & \textbf{0.020 $\pm$ 0.00} & 1.316 $\pm$ 0.06 & \textbf{204.85} & 3.19 & \textbf{0.531} & 1.638 \\
&  & Phasor fields & \textbf{0.030 $\pm$ 0.00} & 0.830 $\pm$ 0.04 & \textbf{139.19} & 5.05 & \textbf{0.688} & 1.049 \\
\cmidrule{2-9}
& \multirow{2}{*}{128$\times$128$\times$128} & \fk & \textbf{0.011 $\pm$ 0.00} & 0.686 $\pm$ 0.14 & \textbf{182.86} & 3.06 & \textbf{0.266} & 0.819 \\
&  & Phasor fields & \textbf{0.016 $\pm$ 0.00} & 0.392 $\pm$ 0.00 & \textbf{128.49} & 5.35 & \textbf{0.344} & 0.524 \\
\bottomrule
\end{tabular}
\end{table*}

\begin{table*}
\scriptsize
\centering
\caption{Continuation of \tref{table:offline_table_1}.}
\label{table:offline_table_2}
\begin{tabular}{|l|l|l|l|l|l|l|l|l|}
\toprule
& & & \multicolumn{2}{|c|}{$\downarrow$ Time (\si{\second})} & \multicolumn{2}{c|}{$\uparrow$ Throughput (Mvox/\si{\second})} & \multicolumn{2}{c}{$\downarrow$ Max. tracked memory (\si{\giga\byte})}\\
\cmidrule{4-9}
Dataset & Dimensions & Algorithm & Ours & Original & Ours & Original & Ours & Original \\
\cmidrule{1-9}
\multicolumn{9}{|c|}{Confocal -- Zaragoza dataset \citep{galindo_dataset_2019}}\\
\cmidrule{1-9}
\multirow{18}{*}{\parbox{1.8cm}{\centering usaf\\[3pt]\includegraphics[width=1cm]{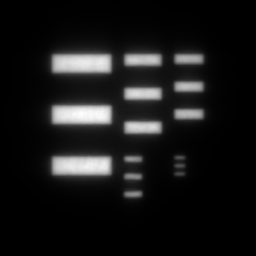}}} & \multirow{2}{*}{256$\times$256$\times$1964} & \fk & \textbf{1.747 $\pm$ 0.00} & 139.592 $\pm$ 22.22 & \textbf{73.70} & 0.92 & \textbf{16.303} & 52.429 \\
&  & Phasor fields & \textbf{4.172 $\pm$ 0.01} & 31.655 $\pm$ 1.26 & \textbf{30.85} & 4.07 & \textbf{21.112} & 33.621 \\
\cmidrule{2-9}
& \multirow{2}{*}{256$\times$256$\times$982} & \fk & \textbf{0.280 $\pm$ 0.00} & 20.740 $\pm$ 1.52 & \textbf{230.05} & 3.10 & \textbf{8.151} & 26.214 \\
&  & Phasor fields & \textbf{0.428 $\pm$ 0.00} & 14.870 $\pm$ 0.24 & \textbf{150.21} & 4.33 & \textbf{10.552} & 16.794 \\
\cmidrule{2-9}
& \multirow{2}{*}{128$\times$128$\times$1964} & \fk & \textbf{0.142 $\pm$ 0.00} & 10.361 $\pm$ 0.48 & \textbf{226.13} & 3.11 & \textbf{4.076} & 13.107 \\
&  & Phasor fields & \textbf{0.222 $\pm$ 0.00} & 8.209 $\pm$ 0.05 & \textbf{144.67} & 3.92 & \textbf{5.289} & 8.454 \\
\cmidrule{2-9}
& \multirow{2}{*}{256$\times$256$\times$491} & \fk & \textbf{0.141 $\pm$ 0.00} & 10.714 $\pm$ 0.29 & \textbf{228.16} & 3.00 & \textbf{4.076} & 13.107 \\
&  & Phasor fields & \textbf{0.220 $\pm$ 0.01} & 6.665 $\pm$ 0.28 & \textbf{146.26} & 4.83 & \textbf{5.275} & 8.393 \\
\cmidrule{2-9}
& \multirow{2}{*}{128$\times$128$\times$982} & \fk & \textbf{0.073 $\pm$ 0.00} & 4.976 $\pm$ 0.32 & \textbf{221.74} & 3.23 & \textbf{2.038} & 6.554 \\
&  & Phasor fields & \textbf{0.109 $\pm$ 0.00} & 3.424 $\pm$ 0.17 & \textbf{147.40} & 4.70 & \textbf{2.641} & 4.211 \\
\cmidrule{2-9}
& \multirow{2}{*}{128$\times$128$\times$491} & \fk & \textbf{0.039 $\pm$ 0.00} & 2.468 $\pm$ 0.18 & \textbf{206.95} & 3.26 & \textbf{1.019} & 3.277 \\
&  & Phasor fields & \textbf{0.057 $\pm$ 0.00} & 1.611 $\pm$ 0.04 & \textbf{140.95} & 4.99 & \textbf{1.320} & 2.101 \\
\cmidrule{2-9}
& \multirow{2}{*}{64$\times$64$\times$1964} & \fk & \textbf{0.039 $\pm$ 0.00} & 2.590 $\pm$ 0.13 & \textbf{206.38} & 3.11 & \textbf{1.019} & 3.277 \\
&  & Phasor fields & \textbf{0.059 $\pm$ 0.00} & 3.285 $\pm$ 0.09 & \textbf{136.22} & 2.45 & \textbf{1.333} & 2.163 \\
\cmidrule{2-9}
& \multirow{2}{*}{64$\times$64$\times$982} & \fk & \textbf{0.022 $\pm$ 0.00} & 1.509 $\pm$ 0.23 & \textbf{181.21} & 2.67 & \textbf{0.509} & 1.638 \\
&  & Phasor fields & \textbf{0.031 $\pm$ 0.00} & 0.987 $\pm$ 0.03 & \textbf{131.32} & 4.08 & \textbf{0.663} & 1.065 \\
\cmidrule{2-9}
& \multirow{2}{*}{64$\times$64$\times$491} & \fk & \textbf{0.014 $\pm$ 0.00} & 0.605 $\pm$ 0.05 & \textbf{148.94} & 3.32 & \textbf{0.255} & 0.819 \\
&  & Phasor fields & \textbf{0.018 $\pm$ 0.00} & 0.449 $\pm$ 0.04 & \textbf{110.21} & 4.48 & \textbf{0.331} & 0.528 \\
\cmidrule{1-9}
\multirow{18}{*}{\parbox{1.8cm}{\centering bunny\\[3pt]\includegraphics[width=1cm]{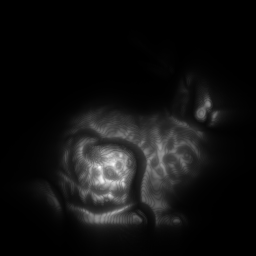}}} & \multirow{2}{*}{256$\times$256$\times$1964} & \fk & \textbf{1.734 $\pm$ 0.00} & 138.419 $\pm$ 8.19 & \textbf{74.24} & 0.93 & \textbf{16.303} & 52.429 \\
&  & Phasor fields & \textbf{4.169 $\pm$ 0.02} & 33.418 $\pm$ 0.67 & \textbf{30.88} & 3.85 & \textbf{21.112} & 33.621 \\
\cmidrule{2-9}
& \multirow{2}{*}{256$\times$256$\times$982} & \fk & \textbf{0.280 $\pm$ 0.00} & 22.058 $\pm$ 0.97 & \textbf{229.92} & 2.92 & \textbf{8.151} & 26.214 \\
&  & Phasor fields & \textbf{0.430 $\pm$ 0.00} & 14.709 $\pm$ 0.33 & \textbf{149.75} & 4.38 & \textbf{10.552} & 16.793 \\
\cmidrule{2-9}
& \multirow{2}{*}{128$\times$128$\times$1964} & \fk & \textbf{0.143 $\pm$ 0.00} & 10.740 $\pm$ 0.57 & \textbf{225.81} & 3.00 & \textbf{4.076} & 13.107 \\
&  & Phasor fields & \textbf{0.223 $\pm$ 0.00} & 8.290 $\pm$ 0.14 & \textbf{144.02} & 3.88 & \textbf{5.289} & 8.454 \\
\cmidrule{2-9}
& \multirow{2}{*}{256$\times$256$\times$491} & \fk & \textbf{0.140 $\pm$ 0.00} & 10.384 $\pm$ 0.18 & \textbf{229.36} & 3.10 & \textbf{4.076} & 13.107 \\
&  & Phasor fields & \textbf{0.213 $\pm$ 0.00} & 7.006 $\pm$ 0.26 & \textbf{151.12} & 4.59 & \textbf{5.275} & 8.393 \\
\cmidrule{2-9}
& \multirow{2}{*}{128$\times$128$\times$982} & \fk & \textbf{0.072 $\pm$ 0.00} & 5.020 $\pm$ 0.15 & \textbf{222.54} & 3.20 & \textbf{2.038} & 6.554 \\
&  & Phasor fields & \textbf{0.109 $\pm$ 0.00} & 3.410 $\pm$ 0.07 & \textbf{147.34} & 4.72 & \textbf{2.641} & 4.211 \\
\cmidrule{2-9}
& \multirow{2}{*}{128$\times$128$\times$491} & \fk & \textbf{0.038 $\pm$ 0.00} & 2.552 $\pm$ 0.03 & \textbf{209.20} & 3.15 & \textbf{1.019} & 3.277 \\
&  & Phasor fields & \textbf{0.057 $\pm$ 0.00} & 1.556 $\pm$ 0.06 & \textbf{141.99} & 5.17 & \textbf{1.320} & 2.101 \\
\cmidrule{2-9}
& \multirow{2}{*}{64$\times$64$\times$1964} & \fk & \textbf{0.039 $\pm$ 0.00} & 2.633 $\pm$ 0.23 & \textbf{207.03} & 3.06 & \textbf{1.019} & 3.277 \\
&  & Phasor fields & \textbf{0.059 $\pm$ 0.00} & 3.293 $\pm$ 0.10 & \textbf{136.40} & 2.44 & \textbf{1.333} & 2.163 \\
\cmidrule{2-9}
& \multirow{2}{*}{64$\times$64$\times$982} & \fk & \textbf{0.021 $\pm$ 0.00} & 1.232 $\pm$ 0.08 & \textbf{187.23} & 3.26 & \textbf{0.509} & 1.638 \\
&  & Phasor fields & \textbf{0.031 $\pm$ 0.00} & 1.026 $\pm$ 0.04 & \textbf{129.02} & 3.92 & \textbf{0.663} & 1.065 \\
\cmidrule{2-9}
& \multirow{2}{*}{64$\times$64$\times$491} & \fk & \textbf{0.014 $\pm$ 0.00} & 0.679 $\pm$ 0.08 & \textbf{148.54} & 2.96 & \textbf{0.255} & 0.819 \\
&  & Phasor fields & \textbf{0.018 $\pm$ 0.00} & 0.420 $\pm$ 0.01 & \textbf{110.09} & 4.79 & \textbf{0.331} & 0.528 \\
\cmidrule{1-9}
\multirow{18}{*}{\parbox{1.8cm}{\centering z\\[3pt]\includegraphics[width=1cm]{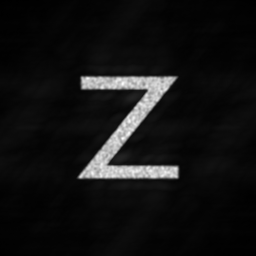}}} & \multirow{2}{*}{256$\times$256$\times$1964} & \fk & \textbf{1.780 $\pm$ 0.08} & 142.930 $\pm$ 6.27 & \textbf{72.29} & 0.90 & \textbf{16.303} & 52.429 \\
&  & Phasor fields & \textbf{4.208 $\pm$ 0.02} & 32.743 $\pm$ 1.42 & \textbf{30.58} & 3.93 & \textbf{21.112} & 33.636 \\
\cmidrule{2-9}
& \multirow{2}{*}{256$\times$256$\times$982} & \fk & \textbf{0.290 $\pm$ 0.01} & 23.453 $\pm$ 0.80 & \textbf{222.01} & 2.74 & \textbf{8.151} & 26.214 \\
&  & Phasor fields & \textbf{0.430 $\pm$ 0.00} & 14.576 $\pm$ 0.38 & \textbf{149.71} & 4.42 & \textbf{10.552} & 16.794 \\
\cmidrule{2-9}
& \multirow{2}{*}{128$\times$128$\times$1964} & \fk & \textbf{0.147 $\pm$ 0.00} & 10.414 $\pm$ 0.36 & \textbf{218.17} & 3.09 & \textbf{4.076} & 13.107 \\
&  & Phasor fields & \textbf{0.223 $\pm$ 0.00} & 8.356 $\pm$ 0.20 & \textbf{144.47} & 3.85 & \textbf{5.289} & 8.454 \\
\cmidrule{2-9}
& \multirow{2}{*}{256$\times$256$\times$491} & \fk & \textbf{0.144 $\pm$ 0.00} & 10.379 $\pm$ 0.20 & \textbf{222.93} & 3.10 & \textbf{4.076} & 13.107 \\
&  & Phasor fields & \textbf{0.212 $\pm$ 0.00} & 6.895 $\pm$ 0.36 & \textbf{151.78} & 4.67 & \textbf{5.275} & 8.400 \\
\cmidrule{2-9}
& \multirow{2}{*}{128$\times$128$\times$982} & \fk & \textbf{0.077 $\pm$ 0.00} & 5.169 $\pm$ 0.11 & \textbf{208.14} & 3.11 & \textbf{2.038} & 6.554 \\
&  & Phasor fields & \textbf{0.109 $\pm$ 0.00} & 3.464 $\pm$ 0.04 & \textbf{147.30} & 4.64 & \textbf{2.641} & 4.211 \\
\cmidrule{2-9}
& \multirow{2}{*}{128$\times$128$\times$491} & \fk & \textbf{0.041 $\pm$ 0.00} & 2.933 $\pm$ 0.21 & \textbf{196.97} & 2.74 & \textbf{1.019} & 3.277 \\
&  & Phasor fields & \textbf{0.057 $\pm$ 0.00} & 1.567 $\pm$ 0.06 & \textbf{140.21} & 5.13 & \textbf{1.320} & 2.101 \\
\cmidrule{2-9}
& \multirow{2}{*}{64$\times$64$\times$1964} & \fk & \textbf{0.039 $\pm$ 0.00} & 2.525 $\pm$ 0.16 & \textbf{206.03} & 3.19 & \textbf{1.019} & 3.277 \\
&  & Phasor fields & \textbf{0.059 $\pm$ 0.00} & 3.268 $\pm$ 0.04 & \textbf{135.43} & 2.46 & \textbf{1.333} & 2.163 \\
\cmidrule{2-9}
& \multirow{2}{*}{64$\times$64$\times$982} & \fk & \textbf{0.022 $\pm$ 0.00} & 1.270 $\pm$ 0.08 & \textbf{185.57} & 3.17 & \textbf{0.509} & 1.638 \\
&  & Phasor fields & \textbf{0.031 $\pm$ 0.00} & 1.028 $\pm$ 0.01 & \textbf{128.08} & 3.91 & \textbf{0.663} & 1.065 \\
\cmidrule{2-9}
& \multirow{2}{*}{64$\times$64$\times$491} & \fk & \textbf{0.014 $\pm$ 0.00} & 0.642 $\pm$ 0.07 & \textbf{144.58} & 3.13 & \textbf{0.255} & 0.819 \\
&  & Phasor fields & \textbf{0.018 $\pm$ 0.00} & 0.416 $\pm$ 0.01 & \textbf{111.45} & 4.83 & \textbf{0.331} & 0.528 \\
\bottomrule
\end{tabular}
\end{table*}

\begin{table*}
\scriptsize
\centering
\caption{Continuation of \tref{table:offline_table_2}.}
\label{table:offline_table_3}
\begin{tabular}{|l|l|l|l|l|l|l|l|l|}
\toprule
& & & \multicolumn{2}{|c|}{$\downarrow$ Time (\si{\second})} & \multicolumn{2}{c|}{$\uparrow$ Throughput (Mvox/\si{\second})} & \multicolumn{2}{c}{$\downarrow$ Max. tracked memory (\si{\giga\byte})}\\
\cmidrule{4-9}
Dataset & Dimensions & Algorithm & Ours & Original & Ours & Original & Ours & Original \\
\cmidrule{1-9}
\multicolumn{9}{|c|}{Tal-generated dataset \citep{royo_diegoroyotal_2024}, inspired by a Phasor Fields scene \citep{liu_phasor_2020}}\\
\cmidrule{1-9}
\multirow{18}{*}{\parbox{1.8cm}{\centering office\\[3pt]\includegraphics[width=1cm]{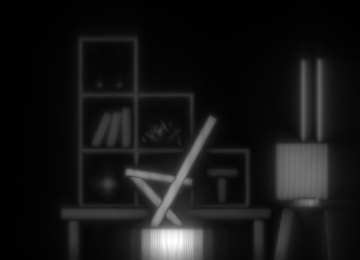}}} & \multirow{2}{*}{360$\times$260$\times$1500} & \fk & \textbf{2.945 $\pm$ 0.05} & 507.808 $\pm$ 47.68 & \textbf{47.67} & 0.28 & \textbf{17.784} & 75.938 \\
&  & Phasor fields & \textbf{5.414 $\pm$ 0.02} & 138.281 $\pm$ 13.87 & \textbf{25.93} & 1.02 & \textbf{23.023} & 66.424 \\
\cmidrule{2-9}
& \multirow{2}{*}{360$\times$260$\times$750} & \fk & \textbf{0.308 $\pm$ 0.01} & 64.483 $\pm$ 7.22 & \textbf{227.77} & 1.09 & \textbf{8.893} & 37.969 \\
&  & Phasor fields & \textbf{0.458 $\pm$ 0.00} & 28.830 $\pm$ 1.34 & \textbf{153.27} & 2.43 & \textbf{11.510} & 33.194 \\
\cmidrule{2-9}
& \multirow{2}{*}{180$\times$130$\times$1500} & \fk & \textbf{0.150 $\pm$ 0.00} & 15.883 $\pm$ 1.75 & \textbf{234.54} & 2.21 & \textbf{4.446} & 18.984 \\
&  & Phasor fields & \textbf{0.232 $\pm$ 0.00} & 14.810 $\pm$ 0.46 & \textbf{151.47} & 2.37 & \textbf{5.762} & 16.654 \\
\cmidrule{2-9}
& \multirow{2}{*}{360$\times$260$\times$375} & \fk & \textbf{0.153 $\pm$ 0.00} & 14.011 $\pm$ 0.08 & \textbf{229.12} & 2.51 & \textbf{4.447} & 18.984 \\
&  & Phasor fields & \textbf{0.228 $\pm$ 0.00} & 13.549 $\pm$ 0.21 & \textbf{153.85} & 2.59 & \textbf{5.755} & 16.593 \\
\cmidrule{2-9}
& \multirow{2}{*}{180$\times$130$\times$750} & \fk & \textbf{0.076 $\pm$ 0.00} & 7.340 $\pm$ 0.39 & \textbf{229.82} & 2.39 & \textbf{2.223} & 9.492 \\
&  & Phasor fields & \textbf{0.115 $\pm$ 0.00} & 6.564 $\pm$ 0.12 & \textbf{152.06} & 2.67 & \textbf{2.879} & 8.311 \\
\cmidrule{2-9}
& \multirow{2}{*}{180$\times$130$\times$375} & \fk & \textbf{0.040 $\pm$ 0.00} & 3.350 $\pm$ 0.07 & \textbf{221.61} & 2.62 & \textbf{1.112} & 4.746 \\
&  & Phasor fields & \textbf{0.059 $\pm$ 0.00} & 3.101 $\pm$ 0.02 & \textbf{148.09} & 2.83 & \textbf{1.439} & 4.151 \\
\cmidrule{2-9}
& \multirow{2}{*}{90$\times$65$\times$1500} & \fk & \textbf{0.041 $\pm$ 0.00} & 3.818 $\pm$ 0.14 & \textbf{214.11} & 2.30 & \textbf{1.112} & 4.746 \\
&  & Phasor fields & \textbf{0.061 $\pm$ 0.00} & 4.814 $\pm$ 0.02 & \textbf{142.68} & 1.82 & \textbf{1.447} & 4.213 \\
\cmidrule{2-9}
& \multirow{2}{*}{90$\times$65$\times$750} & \fk & \textbf{0.022 $\pm$ 0.00} & 1.809 $\pm$ 0.04 & \textbf{203.46} & 2.43 & \textbf{0.556} & 2.373 \\
&  & Phasor fields & \textbf{0.032 $\pm$ 0.00} & 1.782 $\pm$ 0.02 & \textbf{137.10} & 2.46 & \textbf{0.721} & 2.090 \\
\cmidrule{2-9}
& \multirow{2}{*}{90$\times$65$\times$375} & \fk & \textbf{0.012 $\pm$ 0.00} & 0.912 $\pm$ 0.06 & \textbf{176.53} & 2.41 & \textbf{0.278} & 1.187 \\
&  & Phasor fields & \textbf{0.018 $\pm$ 0.00} & 0.806 $\pm$ 0.01 & \textbf{122.81} & 2.72 & \textbf{0.360} & 1.041 \\
\cmidrule{1-9}
\multicolumn{9}{|c|}{Exhaustive -- Zaragoza dataset \citep{galindo_dataset_2019}}\\
\cmidrule{1-9}
\multirow{6}{*}{\parbox{1.8cm}{\centering concavities\\[3pt]\includegraphics[width=1cm]{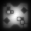}}} & \multirow{2}{*}{16$\times$16$\times$16$\times$16$\times$2048} & \fk & \textbf{0.018 $\pm$ 0.00} & 0.588 $\pm$ 0.04 & \textbf{107.02} & 3.21 & \textbf{0.352} & 0.770 \\
&  & Phasor fields & \textbf{0.027 $\pm$ 0.00} & 2.107 $\pm$ 0.01 & \textbf{68.74} & 0.90 & \textbf{0.436} & 0.558 \\
\cmidrule{2-9}
& \multirow{2}{*}{16$\times$16$\times$16$\times$16$\times$1024} & \fk & \textbf{0.011 $\pm$ 0.00} & 0.305 $\pm$ 0.02 & \textbf{88.41} & 3.09 & \textbf{0.176} & 0.384 \\
&  & Phasor fields & \textbf{0.016 $\pm$ 0.00} & 0.479 $\pm$ 0.01 & \textbf{60.09} & 1.97 & \textbf{0.215} & 0.263 \\
\cmidrule{2-9}
& \multirow{2}{*}{16$\times$16$\times$16$\times$16$\times$512} & \fk & \textbf{0.007 $\pm$ 0.00} & 0.157 $\pm$ 0.02 & \textbf{67.26} & 3.01 & \textbf{0.088} & 0.192 \\
&  & Phasor fields & \textbf{0.012 $\pm$ 0.00} & 0.153 $\pm$ 0.00 & \textbf{38.93} & 3.08 & \textbf{0.106} & 0.127 \\
\cmidrule{1-9}
\multirow{6}{*}{\parbox{1.8cm}{\centering t (in a box)\\[3pt]\includegraphics[width=1cm]{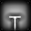}}} & \multirow{2}{*}{16$\times$16$\times$16$\times$16$\times$2048} & \fk & \textbf{0.015 $\pm$ 0.00} & 0.744 $\pm$ 0.05 & \textbf{122.69} & 2.53 & \textbf{0.351} & 0.773 \\
&  & Phasor fields & \textbf{0.024 $\pm$ 0.00} & 4.263 $\pm$ 0.15 & \textbf{78.85} & 0.44 & \textbf{0.436} & 0.562 \\
\cmidrule{2-9}
& \multirow{2}{*}{16$\times$16$\times$16$\times$16$\times$1024} & \fk & \textbf{0.008 $\pm$ 0.00} & 0.372 $\pm$ 0.09 & \textbf{112.12} & 2.53 & \textbf{0.176} & 0.384 \\
&  & Phasor fields & \textbf{0.015 $\pm$ 0.00} & 0.491 $\pm$ 0.00 & \textbf{63.70} & 1.92 & \textbf{0.214} & 0.269 \\
\cmidrule{2-9}
& \multirow{2}{*}{16$\times$16$\times$16$\times$16$\times$512} & \fk & \textbf{0.008 $\pm$ 0.00} & 0.159 $\pm$ 0.01 & \textbf{59.02} & 2.97 & \textbf{0.088} & 0.192 \\
&  & Phasor fields & \textbf{0.010 $\pm$ 0.00} & 0.158 $\pm$ 0.01 & \textbf{48.51} & 2.99 & \textbf{0.106} & 0.127 \\
\cmidrule{1-9}
\multirow{6}{*}{\parbox{1.8cm}{\centering bunny\\[3pt]\includegraphics[width=1cm]{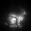}}} & \multirow{2}{*}{16$\times$16$\times$16$\times$16$\times$2048} & \fk & \textbf{0.019 $\pm$ 0.00} & 0.685 $\pm$ 0.10 & \textbf{100.31} & 2.76 & \textbf{0.352} & 0.769 \\
&  & Phasor fields & \textbf{0.039 $\pm$ 0.02} & 2.187 $\pm$ 0.03 & \textbf{48.33} & 0.86 & \textbf{0.436} & 0.560 \\
\cmidrule{2-9}
& \multirow{2}{*}{16$\times$16$\times$16$\times$16$\times$1024} & \fk & \textbf{0.012 $\pm$ 0.00} & 0.348 $\pm$ 0.03 & \textbf{76.51} & 2.71 & \textbf{0.176} & 0.384 \\
&  & Phasor fields & \textbf{0.016 $\pm$ 0.00} & 0.490 $\pm$ 0.01 & \textbf{60.86} & 1.93 & \textbf{0.215} & 0.262 \\
\cmidrule{2-9}
& \multirow{2}{*}{16$\times$16$\times$16$\times$16$\times$512} & \fk & \textbf{0.008 $\pm$ 0.00} & 0.159 $\pm$ 0.02 & \textbf{56.80} & 2.97 & \textbf{0.088} & 0.192 \\
&  & Phasor fields & \textbf{0.010 $\pm$ 0.00} & 0.158 $\pm$ 0.00 & \textbf{45.54} & 2.99 & \textbf{0.106} & 0.127 \\
\bottomrule
\end{tabular}
\end{table*}





\clearpage
\addtocounter{page}{-1}
\end{document}